\documentclass[twoside,onecolumn,prd,amsmath,amssymb,nofootinbib]{revtex4}

\usepackage{dcolumn}
\usepackage{bm}
\usepackage[dvips]{graphicx}
\usepackage{hyperref}
\usepackage{wasysym}
\usepackage{hyperref}
\usepackage{latexsym}
\usepackage{fancyheadings}

\def\aap  {A\&A}
\def\apj  {ApJ}
\def\apjl {ApJ}
\def\apjs {ApJ}

\def\mnras{MNRAS}

\def\aj   {AJ}

\def\nat  {Nature}

\newcommand{\planss}{Planetary and Space Science}
\newcommand{\araa}{Annual Review of Astronomy \& Astrophysics}

\newcommand{\be}{\begin{equation}}
\newcommand{\ee}{\end{equation}}
\newcommand{\ba}{\begin{array}}
\newcommand{\ea}{\end{array}}
  
  \newcommand{\lirs}{$\mbox L_{\mbox{\scriptsize IR}}$~}

  \newcommand{\mip}{24~$\mu$m}
  \newcommand{\mips}{24~$\mu$m~}

\begin{document}

\begin{center}
\begin{figure}
\includegraphics[height=3.cm]{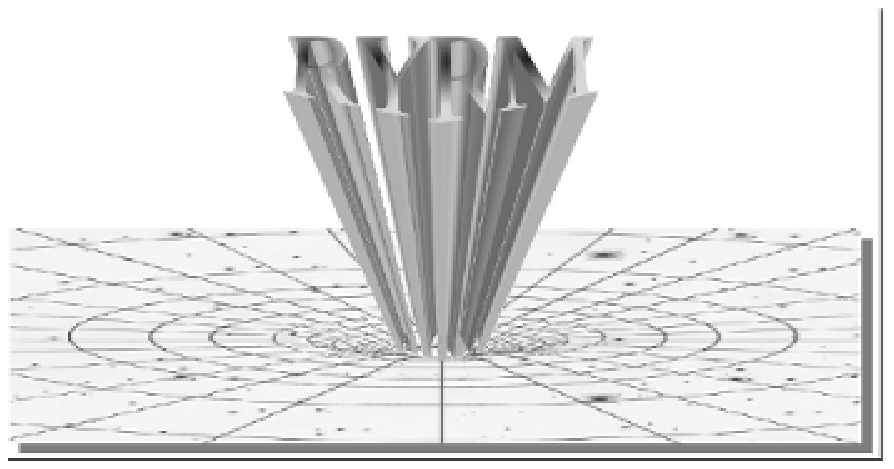}
\end{figure}
\end{center}

\title{$1^{\rm st}$ Roman Young Researchers Meeting Proceedings}
\author{E.~Cannuccia$^1$, M.~Migliaccio$^1$, D.~Pietrobon$^{1,3}$, F.~Stellato$^{1,2}$,
M.~Veneziani$^4$}\email{r.young.res.meet@gmail.com}
\affiliation{
$^1$ \mbox{Dipartimento di Fisica dell'Universit\`a di Roma ``Tor Vergata", Via della Ricerca
Scientifica 1 00133 Roma - Italy}\\
$^2$ \mbox{I.N.F.N.~- Sezione di Roma ``Tor Vergata'' Via della Ricerca
Scientifica 1 00133 Roma - Italy}\\
$^3$ \mbox{Institute of Cosmology and Gravitation, Dennis Sciama
Building, Burnaby Road Portsmouth, PO1 3FX - United Kingdom}\\
$^4$ \mbox{Dipartimento di Fisica, Universit\`a di Roma ``La
Sapienza'' P.~le A.~Moro 2 00185 Roma - Italy}}




%
%
\begin{abstract}
During the last few decades scientists have been able to test the bases of the physics paradigms, where the
quantum mechanics has to match the cosmological scales. Between the
extremes of this scenario, biological phenomena and their complexity take
place, challenging the laws we observe in the atomic and sub-atomic
world. In order to explore the details of this world, new huge
experimental facilities are under
construction. These projects involve people coming from several
countries and give physicists the opportunity to work together with
chemists, biologists and other scientists. The Roman Young Researchers
Meeting is a conference, organised by Ph.~D. students and young
post-docs connected to the Roman area. It is aimed primarily at graduate
students and post-docs, working in physics. The 1$^{\rm st}$
conference has been held on the 21$^{\rm st}$ of July 2009 at the
University of Roma ``Tor Vergata''. It was organised in three
sessions, devoted to Astrophysics and Cosmology, Soft and Condensed
Matter Physics and Theoretical and Particle Physics. In this
proceeding we collect the
contributions which have been presented and discussed during the
meeting, according to the specific topics treated.
\end{abstract}
\maketitle

\section*{Preface}

Undergraduate students in science face a fascinating and challenging future, particularly in physics. Scientists have proposed very interesting theories, which describe fairly well the 
microscopic world as well the macroscopic one, trying to match the quantum regime
with cosmological scales; the complexity which governs the biological
phenomena has been the following target addressed by scientists, led by the productive studies on
the matter behaviour. More and more accurate experiments have been
planned and are now going to test the bases of the physics paradigms,
such as the Large Hadronic Collider (LHC), which is going to shed
light on the physics of the Standard Model of Particles and its extensions, the
Planck-Herschel satellites, which target a very precise  measurement
of the properties of our Universe, and the Free Electron Lasers
facilities, which produce
high-brilliance, ultrafast X-ray pulses, thus allowing to investigate the fundamental processes of solid state physics, chemistry, and biology. These
projects are the result of intense collaborations spread across the world,
involving people belonging to different and complementary
fields: physicists, chemists, biologists and other scientists, keen to
make the best of these extraordinary laboratories.

In this context, in which each branch of science becomes more and more
focused on the details, it is very important to keep an eye on the global
picture, being aware of the possible interconnections between inherent
fields. This is even more crucial for students, who are approaching
the research.

With this in mind, Ph.~D. students and young post-docs
connected to the Roman area have felt the need for an event able to
establish the background and the network, necessary for interactions
and collaborations. This resulted in the 1$^{\rm st}$ Roman Young Researchers
Meeting~\footnote{\href{http://ryrm.roma2.infn.it/ryrm/index1.htm}{http://ryrm.roma2.infn.it/ryrm/index1.htm}}, a one day conference aimed primarily at graduate students and
post-docs, working in physics. In its first edition, the meeting
has been held at the University of Roma ``Tor Vergata'', and organised
in three sections dedicated to up-to-date topics spanning broad
research fields: Astrophysics-Cosmology, Soft-Condensed Matter
Physics and Theoretical-Particle Physics.

\vspace{.25cm}
Our journey through the realm of physics started with the astrophysics and
cosmology session, which opened the meeting with the 
description of two experiments,
 COCHISE and PILOT, the former dedicated to the measurements of the SZ signal on the
Cosmic Microwave Background radiation, the latter focused on the
polarised light emitted from
interstellar dust. A very important study in cosmology is
related to the high redshift Universe, and the amount of information
we can extract from quasars and galaxy formation: this has been
discussed from the point of view of the power spectra we measure from
distant quasars and the formation of massive galaxies. The Universe
displays a variety of objects, which are very interesting \emph{per se},
in addition to the useful information they bring on cosmogony, and
represent indeed a genuine laboratory where we can observe processes
at very high energy,
unachievable elsewhere. An overview of the properties of blazars
detected by the AGILE experiment, the BL lacs and the Active Galactic
Nuclei observed by SWIFT and XMM has been presented.

The second session covered an extremely rich and vast topic, condensed
matter, whose research is fed and propulsed also by its applications
to different fields. An example has been discussed focusing on the
study of surgical infections, and how this process is modelled and
simulated in order to guarantee the safety of the patient. Two other
challenging topics have been addressed, in particular the simulations
of gels and nanocrystals and the application of spectroscopical tools
to the study of nanomaterials.

The last session shed light on theoretical and particle physics,
starting from a description of a section of the ATLAS experiment built
at LHC, followed by an overview of a possible anomaly-free supersymmetric extension of the
Standard Model of Particles and its signature at the LHC. The last talk
discussed one of the most outstanding problem in cosmology, the
so-called \emph{cosmological constant} one, from a new interesting prospective which
finds its root in the neutrino flavour mixing mechanism. 

The 1$^{\rm st}$ Roman Young Researchers Meeting has been a great
success mainly thanks to the high quality of the scientists who
participated and gave rise to interesting discussions, stimulated by
excellent presentations. Encouraged by this result, the next
appointment has already been set: the 2$^{\rm nd}$ Young Researchers
Meeting in Rome will take place in February 2010 at the university ``La
Sapienza'' in Roma. Further details will appear soon on the website of
the meeting, where the presentations of the talks and more detailed
information are already available.

The contributions of each speaker who attended the meeting follow,
organised in sessions which resemble the schedule of the meeting.

\begin{flushright}
The organisers

\end{flushright}

Elena Cannuccia graduated in physics at the University of Rome "Tor Vergata".
She went on working in the field of the ab-inito optical properties of matter
taking the opportunity to do the Ph.~D. in the same research group.
During this year she has focused on the investigation of the role played by the
electron-phonon coupling in the optical properties of conjugated polymers.
In order to do that she is giving a contribution to the development of YAMBO~\footnote{\href{http://www.yambo-code.org/}{http://www.yambo-code.org/}},
a FORTRAN/C code for Many-Body calculations in solid state and molecular physics.

Marina Migliaccio graduated in Universe Science at the University of Rome
Tor Vergata in May 2008. Her degree thesis was dedicated to the analysis
of the cosmic microwave background maps produced by the BOOMERanG
experiment, investigating the presence of non-Gaussian signatures which
could shed light on the mechanism of cosmic inflation. As Ph.~D. student in
Astronomy at the University of Rome Tor Vergata, she is now involved in
the Planck mission Core Cosmology project.

Davide Pietrobon graduated in Astronomy, sharing the Ph.~D. between the University of Roma
``Tor Vergata'' and the Institute of Cosmology and Gravitation at the
University of Portsmouth, within the context of the European
\emph{Cotutela} project. His thesis represents a detailed analysis
of the cosmological perturbations through needlets, a statistical tool he developed together
with his colleagues in Rome. In particular he focused on two main open
questions in cosmology: dark energy and non-Gaussianity. He took the bachelor in physics at the
University of Modena and Reggio Emilia and the master in physics at
the University of Roma ``Tor Vergata''. He spent three months at the
University of California Irvine as a visiting student and he is now going to
start a postdoctoral fellowship at Jet Propulsion Laboratory. 

Francesco Stellato has studied during his Ph.~D. the role of metals in
the pathogenesis of neurodegenerative diseases such as Parkinson and
Alzheimer. To this purpose, he mainly used synchrotron radiation based
techniques, e.g. the X-ray Absorption Spectroscopy.
He is interested in the development of new generation light sources
such as high-brilliance synchrotron and Free Electron Lasers and in
their  application to the structural and dynamical study of
biomolecules.

Marcella Veneziani is a postdoc fellow in the cosmology group of the University of Rome La Sapienza.
 She is involved in the HiGal Herschel key project, working with the scientific team at the Institute of Physics of Interstellar Space (IFSI), and she is a member of the Planck High Frequency Instrument (HFI) Core team. 
She graduated on February 2009 with a joint project between the Astroparticle and Cosmology group of the University Paris Diderot and the University La Sapienza. 
During her education she worked on two important cosmological surveys: the Planck-HFI satellite, focusing on the instrumental calibration, and the BOOMERanG Balloon, measuring the Galactic emission in the microwave band and the level of its contamination on the cosmic microwave background radiation. Part of her work has been performed in collaboration with the University of California Irvine, where she has been a visiting student for 4 months.

\begin{acknowledgments}
The Roman Young Researchers Meeting organisers would like to thank the
speakers and the scientists who attended the meeting, and the
University of Roma Tor Vergata for hosting the first edition of the
meeting. We are grateful to Marco Veneziani, Rossella Cossu and Paolo
Cabella for technical support and
useful discussions.
\end{acknowledgments}

\newpage
\section*{Contents}

\begin{tabular}{ll}
\multicolumn{2}{l}{\bf Astrophysics and Cosmology} \\
& \\
\hspace{.25cm}COHISE: Cosmological Observations from Concordia, Antarctica & Sec.~\ref{sabbatini} \\
& \\
\hspace{.25cm}What can we learn from quasars absorption spectra? & Sec.~\ref{gallerani}\\
& \\
\hspace{.25cm}How do galaxies accrete their mass? Quiescent and star-forming massive galaxies at high redshift & Sec.~\ref{santini}\\
& \\
\hspace{.25cm}Multiwavelength observations of the gamma-ray blazars detected by AGILE & Sec.~\ref{dammando}\\
& \\
\hspace{.25cm}The power from BL Lacs & Sec.~\ref{paggi}\\
& \\
\hspace{.25cm}Studying the X-ray/UV Variability of Active Galactic Nuclei with data from Swift and XMM archives & Sec.~\ref{turriziani}\\
& \\
\multicolumn{2}{l}{\bf Soft and Condensed Matter} \\
& \\
\hspace{.25cm}The problem of surgical wound infections: air flow simulation in operating room & Sec.~\ref{abundo}\\
& \\
\multicolumn{2}{l}{\bf Theoretical and Particle Physics} \\
& \\
\hspace{.25cm}The Resistive Plate Chambers of the ATLAS experiment: performance studies & Sec.~\ref{cattani}\\
& \\
\hspace{.25cm}Anomalous U(1)' Phenomelogy: LHC and Dark Matter & Sec.~\ref{racioppi}\\
& \\
\hspace{.25cm}Flavour mixing in an expanding universe & Sec.~\ref{tarantino}\\
\end{tabular}

\newpage

\pagestyle{fancy}
\lhead[\fancyplain{}{}]{\itshape \small{COHISE: Cosmological Observations from Concordia, Antarctica}}
\rhead[\itshape \large{1$^\mathrm{st}$ RYRM Proceedings}]{\fancyplain{}{L. Sabbatini}}





\begin{center}

\setcounter{section}{0}
\section{\large{COCHISE: Cosmological Observations from Concordia, Antarctica}}
\label{sabbatini}

\normalsize{L.~ Sabbatini$^*$, G.~ Dall'Oglio, L.~ Pizzo}

\emph{\small{University of Roma Tre, Department of Physics}}

\vspace{.25cm}
\normalsize{F.~ Cavaliere}

\emph{\small{University of Milano, Department of Physics}}

\vspace{.25cm}
\normalsize{A.~ Miriametro}

\emph{\small{University of Rome "La Sapienza", Department of
Physics}}

\vspace{.5cm}
\begin{minipage}{.8\textwidth} \small{COCHISE is a 2.6 meter millimetric telescope
devoted to cosmological observations. It is located near the
Concordia Station, on the high Antarctic plateau, probably the
best site in the world for (sub)millimetric observations. At
present time, COCHISE is the largest telescope installed at
Concordia: besides the scientific expectations, it is of great
interest as a pathfinder for future Antarctic telescopes. The main
characteristics of the telescope will be presented, including the
scientific goals and the technical aspects related to the use of
such an instrument at the extreme conditions of the Antarctic
environment. Key aspects of the atmospheric transmission will be
also discussed, by showing the preliminary results of site testing
experiments.}

\vspace{.25cm}
$*$ e-mail: \href{mailto:sabbatini@fis.uniroma3.it}{sabbatini@fis.uniroma3.it}
\end{minipage}

\end{center}

\normalsize
\setcounter {section} {0}
\setcounter {subsection} {0}
\setcounter {figure} {0}
\setcounter {equation} {0}

\section{Scientific context}
Observations at (sub)millimeter wavelengths are extremely
important since they provide many useful information about
different astrophysical problems, such as cosmological studies,
early stages of stellar evolution, properties of cold interstellar
medium, infrared galaxies, cosmic structures at high redshift and
so on. We are particularly interested in the study of the
so-called Sunyaev-Zeldovich Effect (SZE). This is the process by
which the photons of the Cosmic Microwave Background (CMB)
radiation undergo inverse Compton effect on the high energy
electrons contained in clusters of galaxies. This effect changes
the spectral shape of the CMB by increasing on average the energy
of the photons (see for example Carlstrom et al. 2002
\cite{carlstrom02}). SZE surveys can be used to extract values of
the main cosmological parameters. The main features of the SZE are
found in the wavelength range between 850 $\mu$m and 2 mm, hence
the exact comprehension of this effect requires the observation of
clusters of galaxies in this wavelength range in order to
constraint the theory and retrieve physical parameters.

\section{The site}
\begin{figure}[h]
\includegraphics[width=.6\columnwidth]{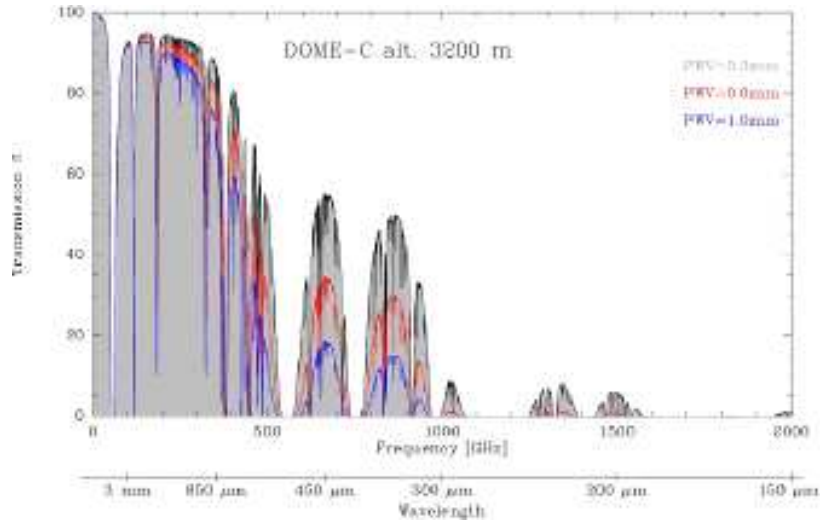}
\caption{\small{The plot shows the atmospheric transmission in
percent for values of {\it pwv}=0.3mm (grey), 0.6mm (red), and 1mm
(blue) against frequency/wavelength for a frequency range 0-2000
GHz transmission (Schneider, Urban \& Baron 2009
\cite{schneider09} http://transmissioncurves.free.fr).}}
\label{figATMO}
\end{figure}

A major problem when performing ground-based millimetric
observations is the presence of water vapor in the atmosphere.
Indeed, in this wavelength range water vapor shows absorption
bands that attenuate the intensity of the incoming signal. Since
the vertical distribution of water vapor in the atmosphere has a
typical scale height of 3 Km, a site located at high altitude is
required to perform (sub)millimetric observations. Preliminary
site testing measurements have revealed that from this point of
view the Antarctic plateau (and Dome C in particular) is one of
the best site in the world; similar atmospheric conditions are
found only in the Chilean desert. In Figure \ref{figATMO}, a
typical atmospheric transmission curve for the Dome C site is
reported as a function of wavelength between 100 and 2000 GHz
(from Schneider et al. 2009 \cite{schneider09}). The exact shape
of the curve depends on the local vertical profiles of pressure
and temperature, and strongly on the integrated amount of water
vapor on the column. This quantity is called {\it precipitable
water vapor} ({\it pwv}): it is the height (measured in
millimetre) of the column of liquid water that would be on the
ground if all the vapor on the vertical is condensed. As evident
from Figure \ref{figATMO}, the amount of {\it pwv} is critical in
particular for the window at 200 $\mu$m, that is very interesting
for cosmologists. The Dome C site seems to allow the observations
at this wavelength. In that site, during the last few years Italy
and French have realized a common scientific station called
Concordia: it is located at 3200 meter above sea level, at 1200 km
from the coast (Lat. 75$^\circ$S, Long. 123$^\circ$E); typical
temperatures vary during the year from -35$^\circ$C to
-80$^\circ$C. The Station is completely isolated from the rest of
the world for almost 10 months each year, from February to
November. The location and the prohibitive working conditions,
both for people and for instrumentation, make the site
logistically difficult. Anyway, there are many advantages, from
astronomical point of view, that made this site of extreme
interest. First of all, the atmosphere is dry, hence the
millimetric transmission is exceptionally good. Then, the
atmospheric temperature is very low, hence the sky emissivity is
at minimum. Moreover, the atmosphere is basically stable, with
laminar wind motion and very low turbulence levels. Due to the
lack of natural or human obstacles, the horizon is completely
available for observations. The site is very far from human or
natural activities, that means that the air is free from aerosol
particles and pollution. The latitude of the site makes a large
part of the sky observable without interruption for an indefinite
time. Finally, the duration of the polar night, the distance from
other continents and the lack of significative human activities
make Antarctica an ideal site for astronomical observations.

\section{The COCHISE telescope}
\begin{figure}[hbt]
\begin{center}
\includegraphics[width=.5\columnwidth]{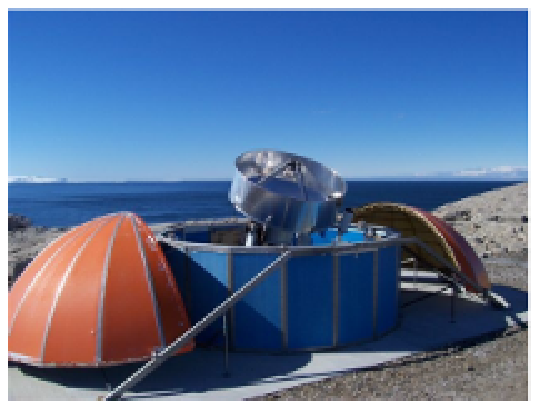}
\includegraphics[width=.5\columnwidth]{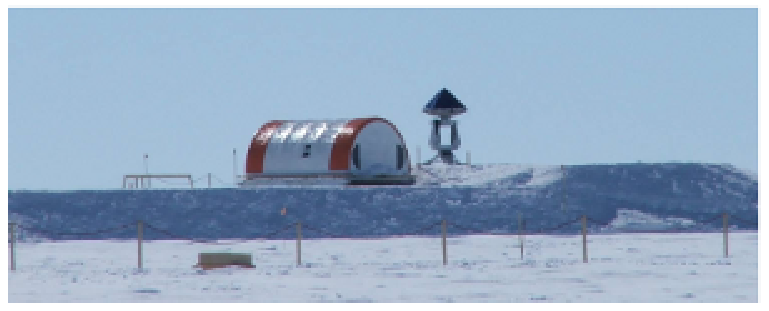}
\caption{\small{A view of the OASI telescope (top) located at the
Italian Mario Zucchelli Station on the coast, and the COCHISE
telescope (bottom) located at the Italian and French Concordia
Station, on the Antarctic plateau. Also the laboratory tent is
visible on the astrophysical platform.}}
 \label{figCOCHISE}
\end{center}
\end{figure}

For the reasons exposed so far, Concordia as been chosen to
install astronomical instrumentation. COCHISE is a Cassegrain 2.6m
millimetric telescope, with a wobbling secondary mirror, a field
of view of few arcminutes. A detailed description can be found in
Sabbatini et al. 2009 (submitted). The COCHISE telescope is very
similar to the OASI one (Infrared and Submillimetric Antarctic
Observatory), installed at the Italian Mario Zucchelli Station by
the same Group and described in Dall'Oglio et al. 1992
\cite{dalloglio92}; the work performed at the OASI telescope
provided to the Group a deep experience in the working conditions
at Antarctic sites. Images of OASI and COCHISE are shown in Figure
\ref{figCOCHISE}. The main scientific objective of COCHISE is the
SZE, even though also very interesting galactic observations
devoted to study the presence of cold dust can be performed with
this instrument, as the ones carried out from OASI telescope
(Sabbatini et al. 2005 \cite{sabbatini05}). Moreover, in the
perspective of very big telescopes to be installed at Concordia in
the future (see for instance Minier et al. 2008 \cite{minier08}),
COCHISE will provide further site testing measurements, and the
possibility to study and solve the technological aspects related
to the use of a telescope at the Antarctic conditions.

The telescope is located on the astrophysical platform, about 400
meters far from the Station; the laboratory tent used for
maintenance is hosted on the same platform. The installation of
COCHISE has been accomplished during two summer campaigns. Up to
now, COCHISE has not performed astrophysical observations yet,
while it has been used for preliminary operations and
technological developments. Indeed, particular attention has to be
posed on some technical aspects, in order to adapt the telescope
at the harsh Antarctic environment. For example, it is mandatory
to take into account the strong thermal variations (temperature at
Concordia can change of more than 20$^\circ$C in few hours) that
cause differential contractions on the materials; it has to be
avoided the sinking into the ice of the structures. The
electronics must be properly heated and insulated; the functioning
of all the parts (mechanics and electronics) has to be tested at
various temperatures. Cables and connectors must be suitable for
operations at -80$^\circ$C.

During the preliminary operations, a cryogenic photometer has been
used in order to to arrange the optics, including the alignment
and the focus. The photometer requires the use of liquid Helium
and $^3$He refrigerators (Graziani et al. 2003 \cite{graziani03},
Pizzo et al. 2006 \cite{pizzo06}) in order to keep the detectors
at their operating temperature, about 0.3 K for the bolometers
used. In the next future, the liquid Helium will be substituted by
the use of a cryocooler, more suitable to the difficult conditions
of the Antarctic environment.

\begin{figure}[!thb]
\begin{center}
\includegraphics[height=5cm]{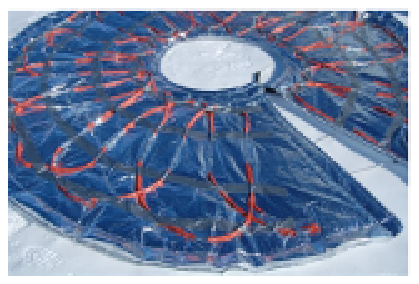}
\includegraphics[height=5cm]{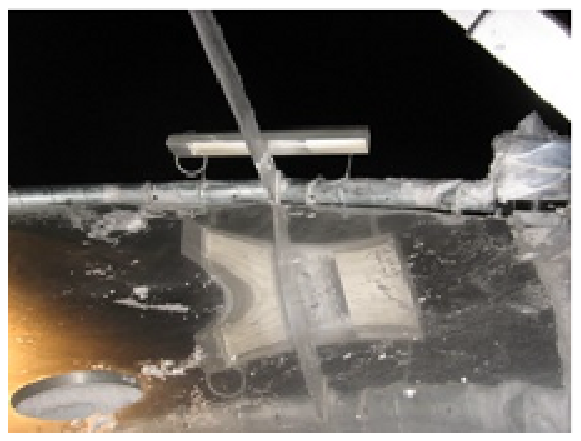}
\caption{\small{The defrosting system. [Left] The preparation of
the heating system: the heating cables have been uniformly
distributed over an insulating coverture and fixed on the rear of
the primary mirror. [Right] A view of one of the infrared lamps
located on the edge of the primary mirror; the lamp is installed
on the tube of the blowing system and its position and inclination
can be easily arranged.}} 
\label{fig:defrosting}
\end{center}
\end{figure}

\begin{figure}[!htb]
\begin{center}
\includegraphics[height=5cm]{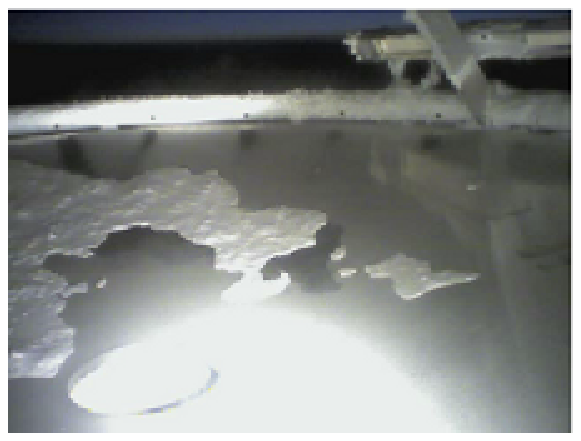}
\includegraphics[height=5cm]{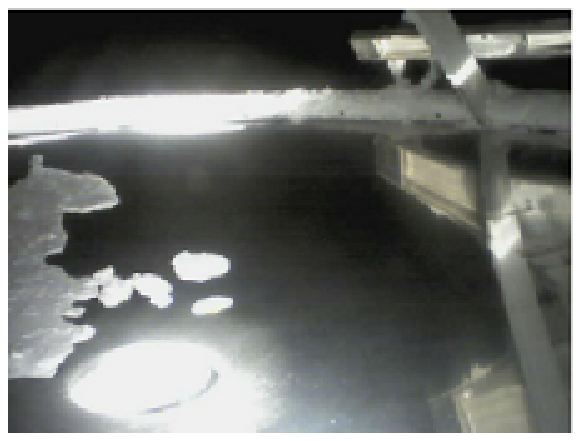}
\caption{\small{Images taken by the webcam that monitors the
primary mirror. The effect of the heating system is shown: on the left 
image the mirror appears opaque because it is covered by a
layer of frost; after 12 hours of heating, at a set point
temperature of few degrees higher than the air temperature, the
mirror appears shiny: the frost has been removed.}}
\label{fig:effheat}
\end{center}
\end{figure}

One of the major problem in the functioning of a telescope on the
high Antarctic plateau during the winter is the formation of
frost, due to the condensation of the water vapor, and the
accumulation of snow. An experimental defrosting system has been
designed and realized for COCHISE by the CEA-Saclay team, within
an international collaboration under the responsibility of Gilles
Durand. The goal of this system is to keep the primary mirror free
from frost during winter. The system consists of three subsystems
based on the three main thermodinamical processes: conduction,
convection, irradiation. For what concern the conduction, the
primary mirror is heated from the rear with heating cables
properly insulated, to force frost sublimation (see the left side
of Figure \ref{fig:defrosting}). The convection has been applied
by means of a blowing system that provides dry, cold air at the
edges of the primary mirror, directed toward the center. The
irradiation includes infrared lamps located all around the primary
mirror pointing in a direction transverse to the beam (see the
right side of Figure \ref{fig:defrosting}). The different
subsystems can be used independently or in combination and all the
system is controlled by remote; the primary mirror is monitored by
a webcam that takes regularly images. This system has been tested
during the whole year in order to find the good configuration to
keep the surfaces clean; as a first preliminary result, the
combined use of heating and blowing has been found very
interesting. It has been found that keeping the mirror for few
hours at a temperature of few degrees higher than the air one can
remove the frost formation (see Figure \ref{fig:effheat}). The
heating also assures a slippery surface that allows the removal of
snow by the mechanical action of the blowing system.

This experiment is still in progress, with few changes that have
been performed at the end of the first season of tests. The
telescope has been left in tilted position to avoid snow
accumulation and have a better understanding of the frost
formation alone; the blowing system has been replaced by a
compressed air system. The correlation with meteorological
conditions will be better studied.

\section{Intraday measurements of water vapor content at Dome C}
Although the Concordia site has been chosen for its extremely cold
and dry air conditions, the water vapor content in the atmosphere
can still great affect the incoming radiation. Therefore, routine
measurements of {\it pwv} are needed to evaluate its effects on
transmission.

For that purpose, during the installation of COCHISE, measurements
of {\it pwv} have been performed by using a solar hygrometer,
devoted to the intraday monitoring of the atmospheric
transmission. This is a simple and robust instrument, accurate and
reliable, designed to work at harsh conditions; it performs the
simultaneous measurements of the solar radiation in two infrared
spectral intervals chosen the first within an absorption band and
the second in a transparency window. Since the solar spectrum is
very well known in these bands, the ratio between the two measured
values, appropriately calibrated, is directly related to the
content of water vapor. The prototype of this instrument has been
realized by Tomasi \& Guzzi 1974 \cite{tomasi74}. The same
instrument has been already used in Antarctica by Valenziano et
al. 1998 \cite{valenziano98}; improvements have been adopted,
including a better procedure of measurements, a more refined
analysis of the radiosoundings and a more accurate calibration.

The calibration of the solar hygrometer is attained by performing
measurements at the same time of the radiosoundings that are daily
launched at Concordia. The radiosoundings provide the vertical
profiles of temperature $T$, dew point temperature $T_d$, pressure
$p$, relative humidity $RH$, wind speed $w_s$ and wind direction
$w_d$. The data for the period of interest have been kindly
provided by PNRA - Osservatorio Meteo Climatologico
(http://www.climantartide.it/). The vertical profiles are used to
evaluate the integrated columnar content of water vapor, after an
accurate correction procedure needed to remove the systematic
effects, as pointed out by the work of Tomasi et al. 2006
\cite{tomasi06}.

The square-root calibration curve usually adopted for this kind of
instrument (Volz 1974 \cite{volz74}) is inappropriate for the
Antarctic atmosphere, since it represents a strong water vapor
absorption regime law, inadequate to the low content of water
vapor usually measured in Antarctica. A more realistic analytical
form for the hygrometric calibration curve has been proposed by
Tomasi et al. 2008 \cite{tomasi08}, using radiative transfer codes
to simulate the weak absorption by water vapor, taking into
account the spectral near-IR curves of extraterrestrial solar
irradiance, instrumental responsivity parameters, atmospheric
transmittance and the field measurements taken at Concordia to
determine empirically some shape parameters of the calibration
curve.

\begin{figure}
\begin{center}
\includegraphics[width=.6\columnwidth]{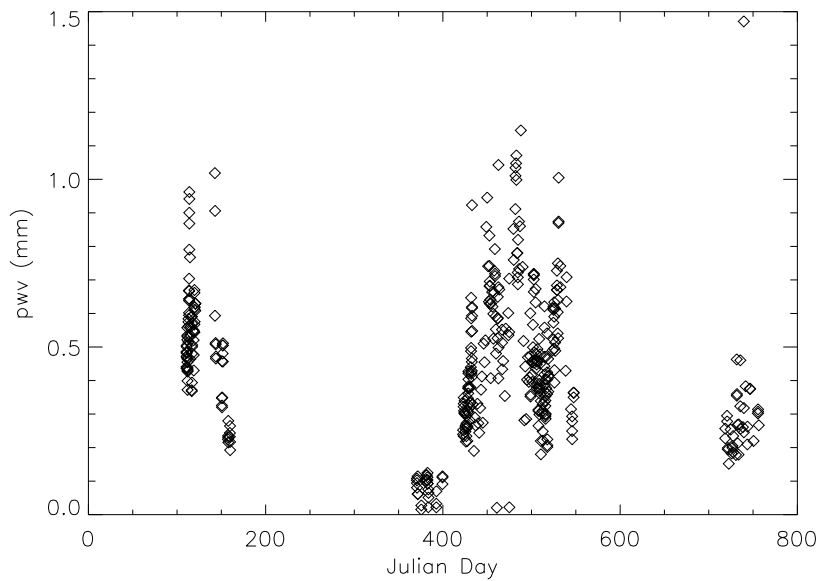}
\caption{\small{Measurements of {\it pwv} in the period December
2007-November 2008. The two main groups of data corresponds to the
two summer periods covered in this work, while the gap between
them is due to the winter periods, where the Sun elevation is not
enough to perform good measurements (or the Sun is above the
horizon).}} 
\label{fig:hygropwv}
\end{center}
\end{figure}

The whole set of data of {\it pwv} is reported in Figure
\ref{fig:hygropwv}, plotted as a function Julian day, starting
from December 2007. So far, this is the largest dataset collected
with this instrument, since more than 400 observations have been
performed; they allow the evaluation on daily and seasonal
fluctuations. In that figure, there are two evident gaps
corresponding to the polar nights, when the solar hygrometer is
not usable due to the elevation of the Sun. As preliminary
results, the average lower values (and also the lower
fluctuations) are present during September and October (JD around
400), that are the coldest months at Concordia, with typical
temperature at -75$^\circ$C. Looking at the first block of data, a
quick look reveals that February values (around the Julian Day
150) are sensibly lower than the previous ones, corresponding at
the December-January period, showing a steep decrease. Very low
values are also measured at the end of the winter season, October
2007 (corresponding in the plot to Julian Day about 380). A second
peak is also evident, showing the increasing of water vapor
content during summer period, when the humidity and the
temperature are both higher. In any case, with few exceptions, the
value is always less than 1 mm. It has to be underlined that this
value constitutes an upper limit, since the measurements have been
taken during the worst conditions: summer period and daytime, when
the temperatures get higher and the Sun is always above the
horizon. Also the dispersion of the values shows a trend with the
period: we found lower fluctuations in the same period in which we
have the lower values of $pwv$.

\section{Conclusions}
There is an enormous interest in performing millimetric and
sub-millimetric astronomical observations from Antarctica, since
the site testing has revealed exceptional conditions, especially
for the low water vapor content, that originates a very high
atmospheric transmission. Therefore, it is of extreme interest the
realization of cosmological observations at Concordia; that will
be possible in the next future thanks to the COCHISE telescope,
that will allow the exploitation of the spectral range between 850
$\mu$m and 3 mm of wavelength.

Besides the astronomical observations that will be carried out
from COCHISE, the work performed is important also with respect to
the technological issues related to the realization and running of
this kind of structures in the Antarctic environment. In
particular, the application of the defrosting system on COCHISE
has answered to many questions regarding the problem of frost
formation on the structures and its removal. This work is of
considerable importance for future large telescopes to be
installed at Concordia.

The measurements realized with the solar hygrometer represent the
first systematic monitoring of water vapor content and its daily
and monthly fluctuations. The results from this work are
particularly interesting for the astronomical community, since
even a small advantage in transparency may constitute a critical
issue in site selection for future instruments. Therefore a
complete analysis to avoid bias and systematics is very delicate
and it is still in progress.

\begin{acknowledgments}
COCHISE is founded by PNRA - Programma
Nazionale di Ricerche in Antartide. We are grateful to PNRA and
IPEV people for the technical and logistic support provided during
the installation phases. We are also grateful to the DC4 team, and
especially to Laurent Fromont, for the great help during winter
operations.
\end{acknowledgments}



\newpage

\lhead[\fancyplain{}{}]{\itshape \small{What can we learn from quasars absorption spectra?}}
\rhead[\itshape \large{1$^\mathrm{st}$ RYRM Proceedings}]{\fancyplain{}{S. Gallerani}}
%
%
%
%
%
%
%







\begin{center}

\setcounter{section}{1}
\section{\large{What can we learn from quasar absorption spectra?}}
\label{gallerani}

\normalsize{Simona Gallerani$^*$}

\emph{\small{INAF - Osservatorio Astronomico di Roma, Via Frascati 33,
00040 Monteporzio (RM).}}


\vspace{.5cm}
\begin{minipage}{.8\textwidth}
\small{We analyze optical-near infrared spectra of a large sample of quasars at high redshift with the aim of investigating both the cosmic reionization history at $z\sim 6$ and the properties of dust extinction at $z>4$. 

In order to constrain cosmic reionization, we study the transmitted flux in the region blueward the Ly$\alpha$ emission line in a sample of 17 quasars spectra at $5.7\leq z_{em}\leq 6.4$. By comparing the properties of the observed spectra with the results of a semi-analytical model of the Ly$\alpha$ forest we find that actual data favor a model in which the Universe reionizes at $z\geq 7$, thus being consistent with an highly ionized intergalactic medium at $z\sim 6$.
 
For what concerns the study of the high-z dust, we focus our attention on the region redward the Ly$\alpha$ emission line of 33 quasars at $4\leq z_{em}\leq 6.4$. We compute simulated dust-absorbed quasar
 spectra by taking into account a large grid of extinction curves. We find that the SMC extinction curve, which has been shown to reproduce the dust reddening of most quasars at $z<4$, is not a good prescription for describing dust
 extinction also at higher redshifts.}

\vspace{.25cm}
$*$ e-mail: \href{mailto:gallerani@oa-roma.inaf.it}{gallerani@oa-roma.inaf.it}
\end{minipage}

\end{center}

\setcounter {section} {0}
\setcounter {subsection} {0}
\setcounter {figure} {0}
\setcounter{equation}{0}

\section{Introduction}
Quasar absorption spectra retain a huge amount of information on the ionization level of the intergalactic medium (IGM), therefore being exquisite tools for studying the cosmic reionization process.\\ 
Moreover, the dust present in the host galaxy of a quasar affects its spectrum, preferentially absorbing the blue part of the rest frame ultraviolet quasar continuum, effect generally called ``reddening''.\\
In what follows we discuss the interpretation of a sample of almost 50 quasars observed at $z>4$ in terms of the neutral hydrogen and dust content in the early Universe.     

\section{Cosmic Reionization}
After the recombination epoch at $z\sim 1100$, the Universe remained almost neutral until the first generation of luminous sources (stars, accreting black holes, etc...) were formed. The photons from these sources ionized the surrounding neutral medium and once these individual ionized regions started overlapping, the global ionization and thermal state of the intergalactic gas changed drastically. This is known as the reionization of the Universe, which has been an important subject of research over the last few years, especially because of its strong impact on the formation and evolution of the first cosmic structures (for a comprehensive review on the subject of reionization and first cosmic structures, see \cite{CiardiFerrara2005}.\\
Although observations of cosmic epochs closer to the present have indisputably 
shown that the IGM is in an ionized state, it is yet unclear when the phase transition from the neutral state to the ionized one started. Thus, the redshift of reionization, $z_{rei}$, is still very uncertain.\\
In the last few years a possible tension has been identified between WMAP5 data \cite{Dunkley2009} and SDSS observations of quasar absorption 
spectra \cite{Fan2006}, the former being consistent with an epoch of reionization 
$z_{\rm rei}\approx 11$, the latter suggesting $z_{\rm rei}\approx 6$. Long Gamma Ray 
Bursts may constitute a complementary way to study the reionization 
process, possibly probing $z>6$ (e.g. \cite{Salvaterra2009}). Moreover, an increasing number of Lyman Alpha 
Emitters are routinely found at $z>6$ (e.g. \cite{Stark2007}).\\ 
In this work, we analyze statistically the transmitted flux of 17 quasar absorption spectra observed at $5.7\leq z_{em}\leq 6.4$ in order to understand whether current data of quasars absorption spectra strongly require a sudden change in the global properties (temperature, ionization level, etc...) of the Universe at $z\sim 6$, or they are still compatible with a highly ionized IGM at these redshifts.
\begin{figure*}
\centerline{
\includegraphics[width=6. cm]{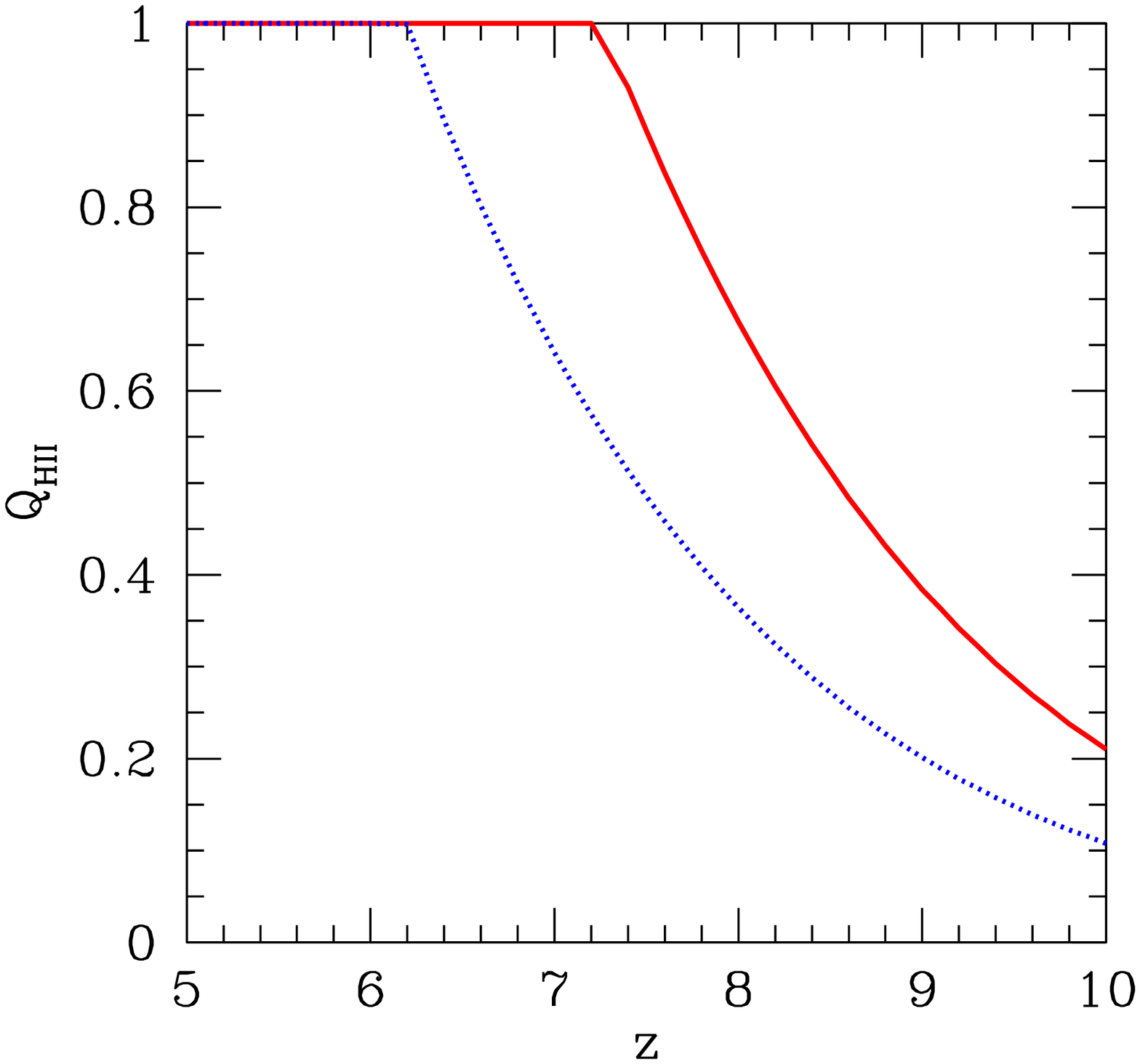}
\includegraphics[width=6. cm]{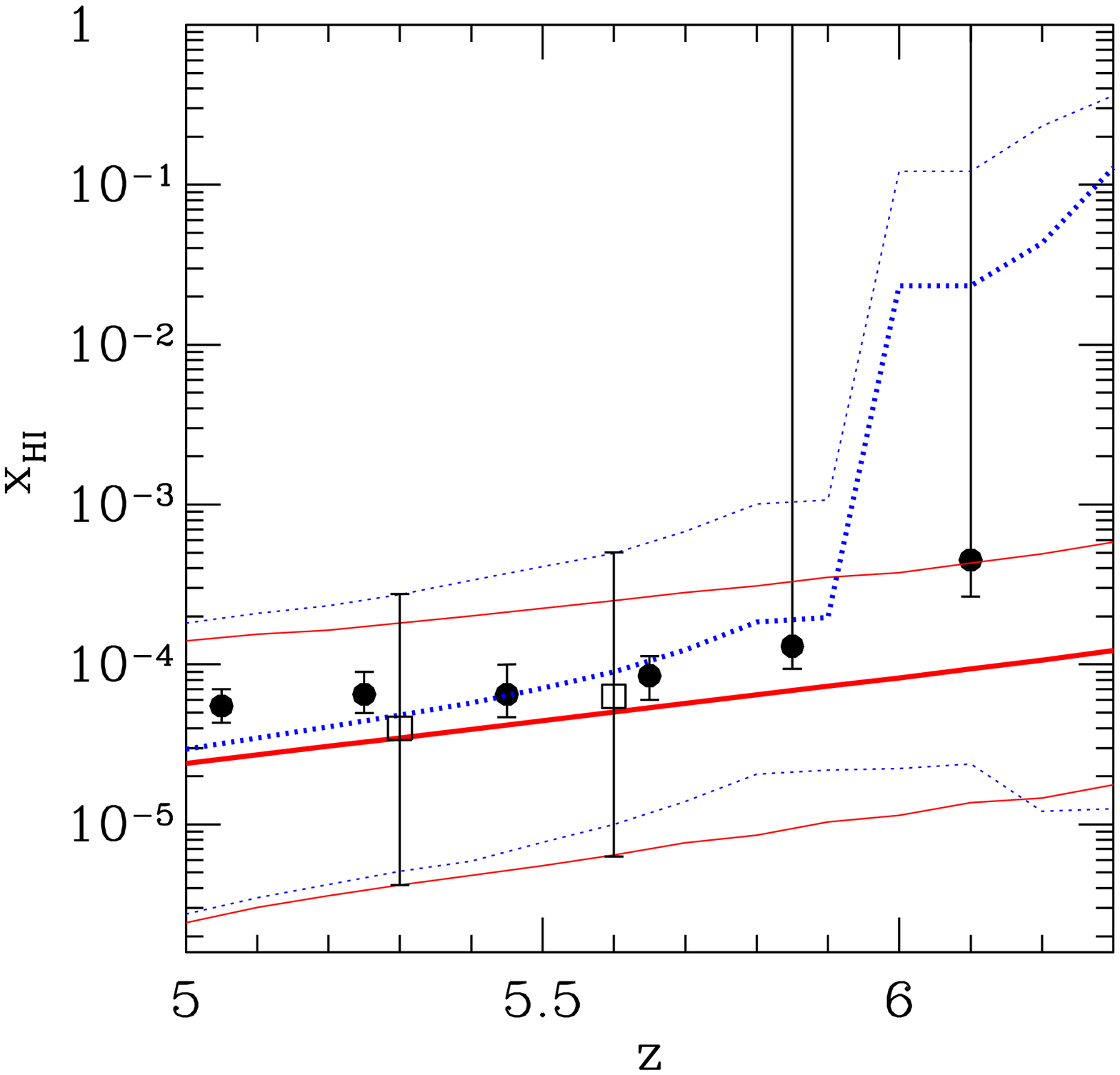}
\includegraphics[width=6. cm]{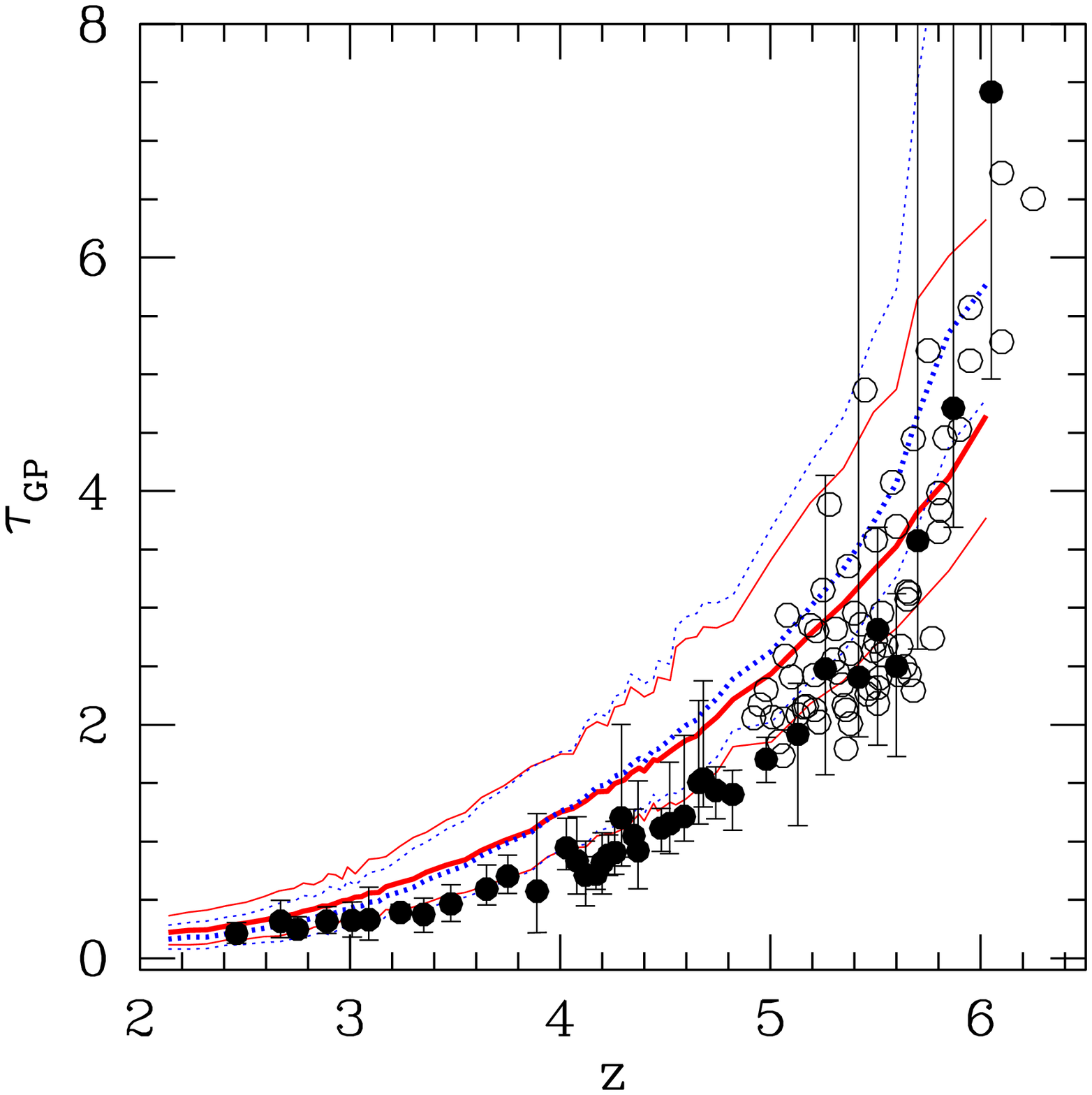}
}
\caption{{\it Left panel}: Evolution of the volume filling 
factor of ionized regions for the early (red solid lines) and late
 (blue dotted lines) reionization models. 
{\it Middle panel}: Evolution of the neutral hydrogen fraction. 
Thick lines represent average results over 100 LOS, while the thin lines 
denote the upper and lower neutral hydrogen fraction extremes in each redshift 
interval. Solid circles represent neutral hydrogen fraction estimates by \citet{Fan2006}; 
empty squares denote the results obtained in this work. {\it Right panel}: Evolution of the Gunn-Peterson 
optical depth for early (ERM, solid red line)  and late (LRM, blue dotted). 
Thick lines represent average results on 100 LOS for each emission redshift, 
while the thin lines denote the upper and lower transmission extremes in 
each redshift bin, weighted on 100 LOS. Filled and empty circles are 
observational 
data from \citet{Songaila2004} and \citet{Fan2006}, respectively.}
\label{prop}
\end{figure*} 
\begin{figure*}
\includegraphics[width=12 cm]{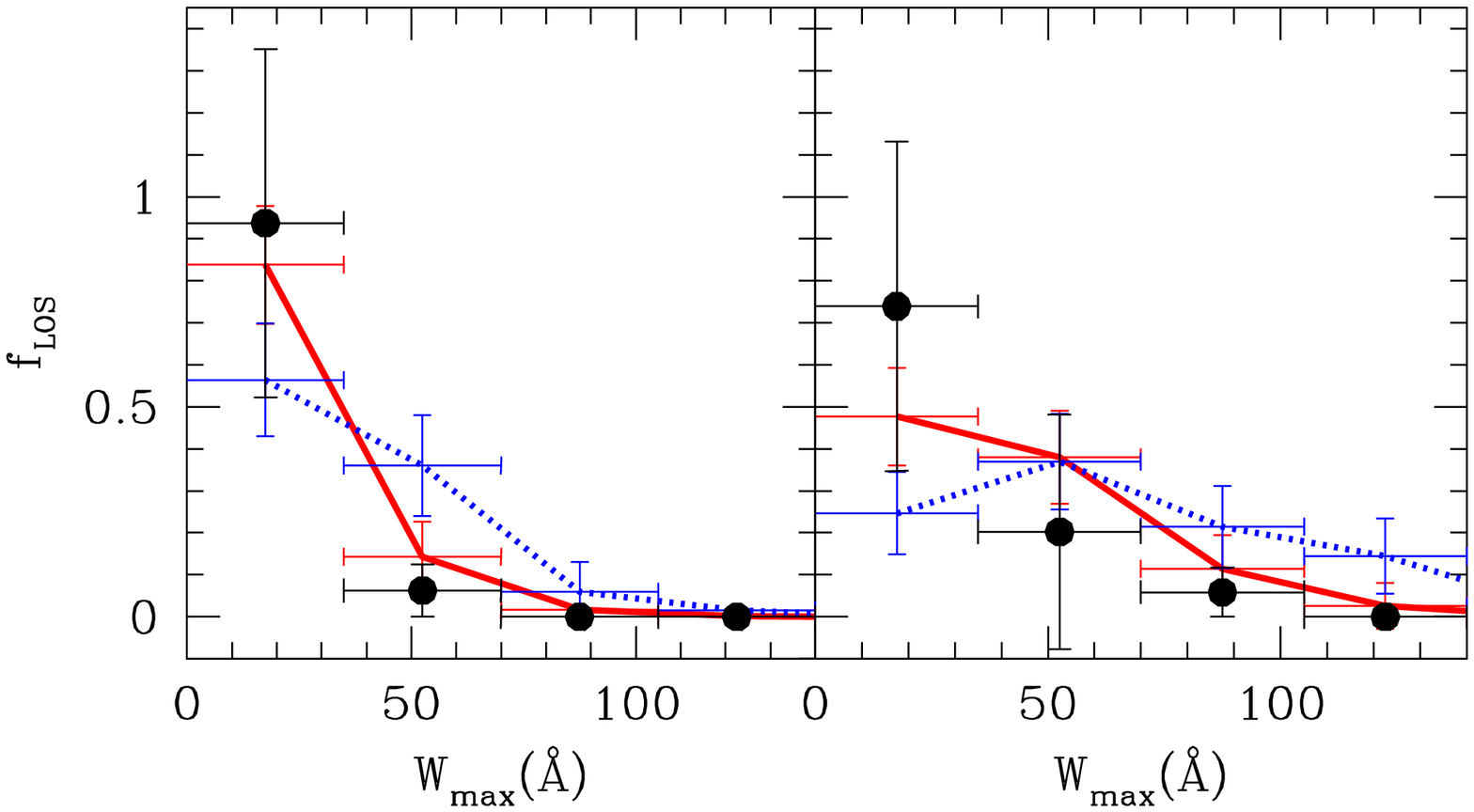}
\caption{Largest Gap Width distribution for the LR and the 
HR cases (left and right, respectively). 
Filled circles represent the result of the analysis of the 17 quasars observed 
spectra.
Solid red (dotted blue) lines show the 
results obtained by the semi-analytical modeling implemented for the ERM 
(LRM). Vertical error bars measure poissonian noise, horizontal errors define 
the bin for the gap widths.}
\label{larg}
\end{figure*}

\subsection{The Ly$\alpha$ forest}
The radiation emitted by quasars could be absorbed through Ly$\alpha$ transition 
by the neutral hydrogen intersecting the line of sight, the so-called 
Gunn-Peterson (GP) effect \cite{GunnPeterson1965}. 
The Ly$\alpha$ forest arises from absorption by low amplitude-fluctuations in 
the underlying 
baryonic density field. To simulate the GP optical depth ($\tau_{GP}$) distribution we use the method 
described by \citet{Gallerani2006}, whose
main features are recalled in the following. The spatial distribution of the 
baryonic density field and its correlation with the peculiar velocity field are taken into account adopting the formalism introduced by Bi \& Davidsen (1997). To enter the mildly non-linear regime 
which characterizes the Ly$\alpha$ forest  absorbers we use a Log-Normal 
model, firstly introduced by \citet{ColesJones1991}.
 
\subsection{Reionization models}
For a given IGM temperature, the neutral hydrogen fraction, $x_{\rm HI}$, can be computed from the
photoionization equilibrium as a function of the baryonic density field and photoionization rate 
due to the ultraviolet background radiation field. For all these quantities we 
follow the approach introduced by \cite{ChoudhuryFerrara2006}, hereafter CF06. By assuming as ionizing sources quasars, PopII and PopIII stars, 
their model provides excellent fits to a large number of observational data, namely the redshift evolution of 
Lyman-limit systems, Ly$\alpha$ and Ly$\beta$ optical depths, electron scattering optical depth, cosmic star formation history, and the number counts of high redshift sources. 
In the CF06 model, a reionization scenario is defined by the product of two free 
parameters: (i) the star-formation efficiency $f_*$, and (ii) the escape fraction $f_{esc}$ 
of ionizing photons of PopII and PopIII stars. We select two sets of free parameters values yielding two different reionization histories: (i) an Early Reionization Model (ERM), for $f_{*,PopII}=0.1; f_{esc,PopII}=0.07$, and (ii) a Late Reionization Model (LRM), for $f_{*,PopII}=0.08; f_{esc,PopII}=0.04$. The properties of the two models considered are shown in Fig. 1. The right panel shows the evolution of the volume filling factor of ionized regions, from which it results that in the LRM (blue dotted line) the epoch of reionization is $z_{\rm rei}\sim 6$, while in the ERM (red solid line) $z_{\rm rei}\sim 7$, meaning that in this case the Universe is highly ionized at $z\sim 6$. In the middle panel the volume averaged neutral hydrogen fraction  $x_{\rm{HI}}$ is plotted for the two models, and in the right panel the corresponding optical depth evolution is shown. 

\subsection{Observational constraints on cosmic reionization}
Since at $z\approx 6$ regions with high transmission in the Ly$\alpha$ forest 
become rare, an appropriate method to analyze the statistical properties 
of the transmitted flux is the distribution of gaps. A gap is defined as a contiguous 
region of the spectrum characterized by a transmission above a given flux threshold ($F_{th}=0.08$ in this work). In particular 
\citet{Gallerani2006} suggested that the Largest Gap Width statistics are suitable tools to study the ionization state of the IGM at high redshift. 
The LGW distribution quantifies the fraction of LOS 
which are characterized by the largest gap of a given width. 
We apply the LGW statistics both 
to simulated and observed spectra with the aim of measuring  the evolution of 
$x_{\rm HI}$ with redshift. \\
We use observational data including 17 quasars obtained by \citet{Fan2006}.
We divide the observed spectra into two redshift-selected sub-samples: 
the ``Low-Redshift'' (LR) sample (8 emission redshifts $5.7 < z_{em} < 6$), and
the ``High-Redshift'' (HR) one (9 emission redshifts $ 6 < z_{em} < 6.4$). The comparison between the simulated and the observed results for the LGW statistics is shown in Fig. 2. In the LR (HR) sample, the quasars emission redshifts used and the wavelength interval analyzed in the spectra are such that the mean redshift of the 
absorbers is $\langle z\rangle=5.3$ ($\langle z\rangle=5.6$). From the good agreement between the simulated and the observed LGW distribution it results that the $x_{\rm{HI}}$ evolves smoothly from $10^{-4.4}$ at $z=5.3$ to $10^{-4.2}$ at $z=5.6$. However, it is also clear from the figure that the ERM is in better agreement with observations. We find that actual data favor a highly ionized Universe at $z\sim 6$, with a robust upper limit $x_{\rm{HI}} < 0.36$ at $z=6.3$ \citep{Gallerani2008a}.

\section{Dust at High Redshift}
Dust represents one of the key ingredients of the Universe by playing a crucial
role both in the formation
and evolution of the stellar populations in galaxies as well as in their observability. The presence of dust is shown by several observational evidences covering a large redshift interval, ranging from our Galaxy to the frontiers of the observable Universe.\\ 
\citet{Richards2003} and \citet{Hopkins2004} have found that, at $z<2$, reddened quasars (including Broad Absorption Line, BAL, quasars) are characterized by SMC-like extinction curves.\\   
This result is in disagreement with the analysis of the reddened quasar SDSSJ1048+46 at $z=6.2$ \cite{Maiolino2004} and of the spectral energy distribution of the Gamma Ray Burst GRB050904 afterglow at $z=6.3$ \cite{Stratta2007}. These studies show that the inferred dust extinction curves are different with respect to those observed at low redshift,
while being in very good agreement with the \citet{TodiniFerrara2001} predictions of dust formed in SNe ejecta.\\
Here, we apply a similar analysis to a sample of 33 observed quasars at $4\leq z_{em}\leq 6.4$ to understand whether the SMC extinction curve is a good prescription for describing dust extinction at these redshifts.\\
Observational data are from \citet{Juarez2009,Jiang2006,Willott2007,Mortlock2008}.
\begin{figure*}
\centerline{
\includegraphics[width=6 cm]{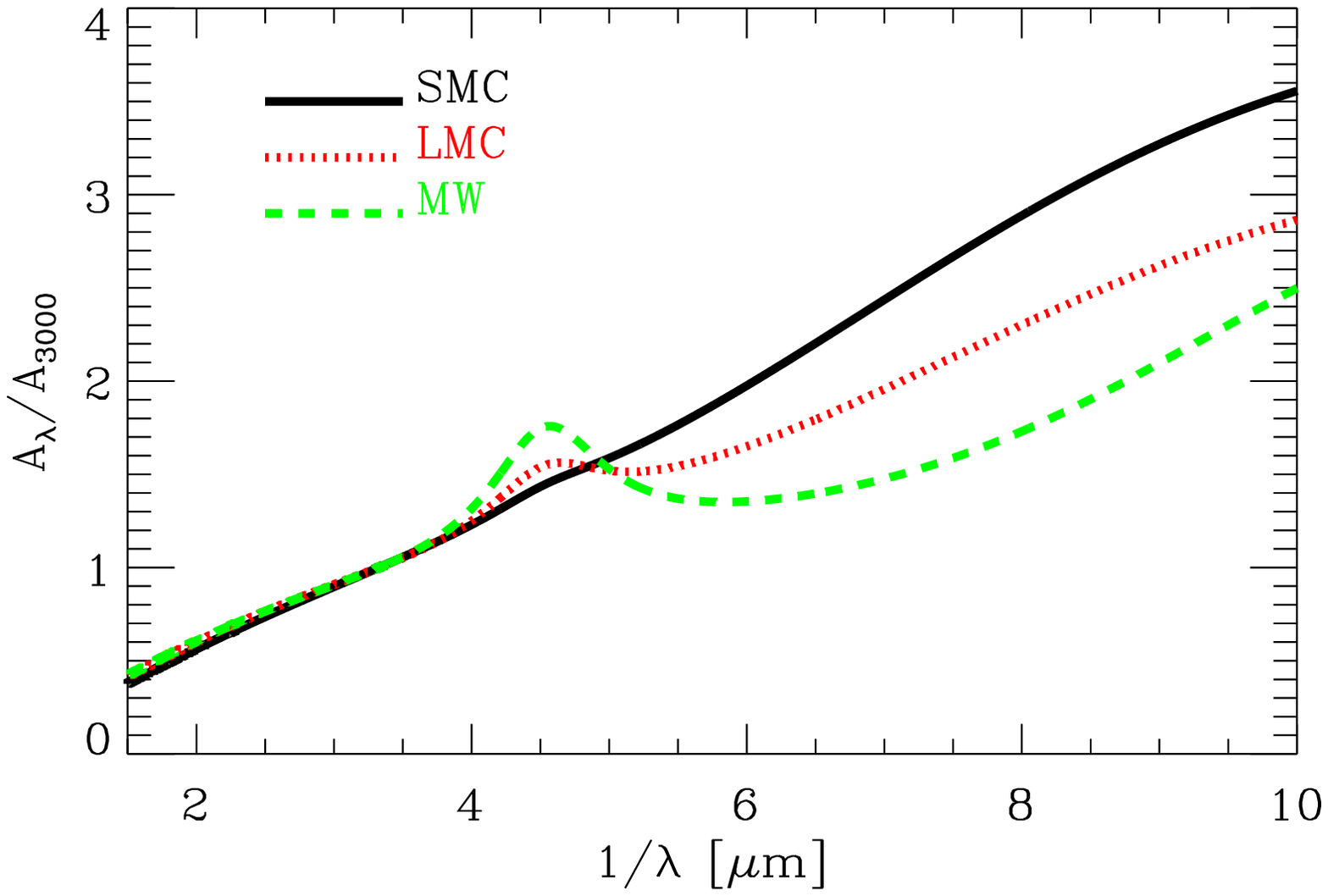}
\includegraphics[width=6 cm]{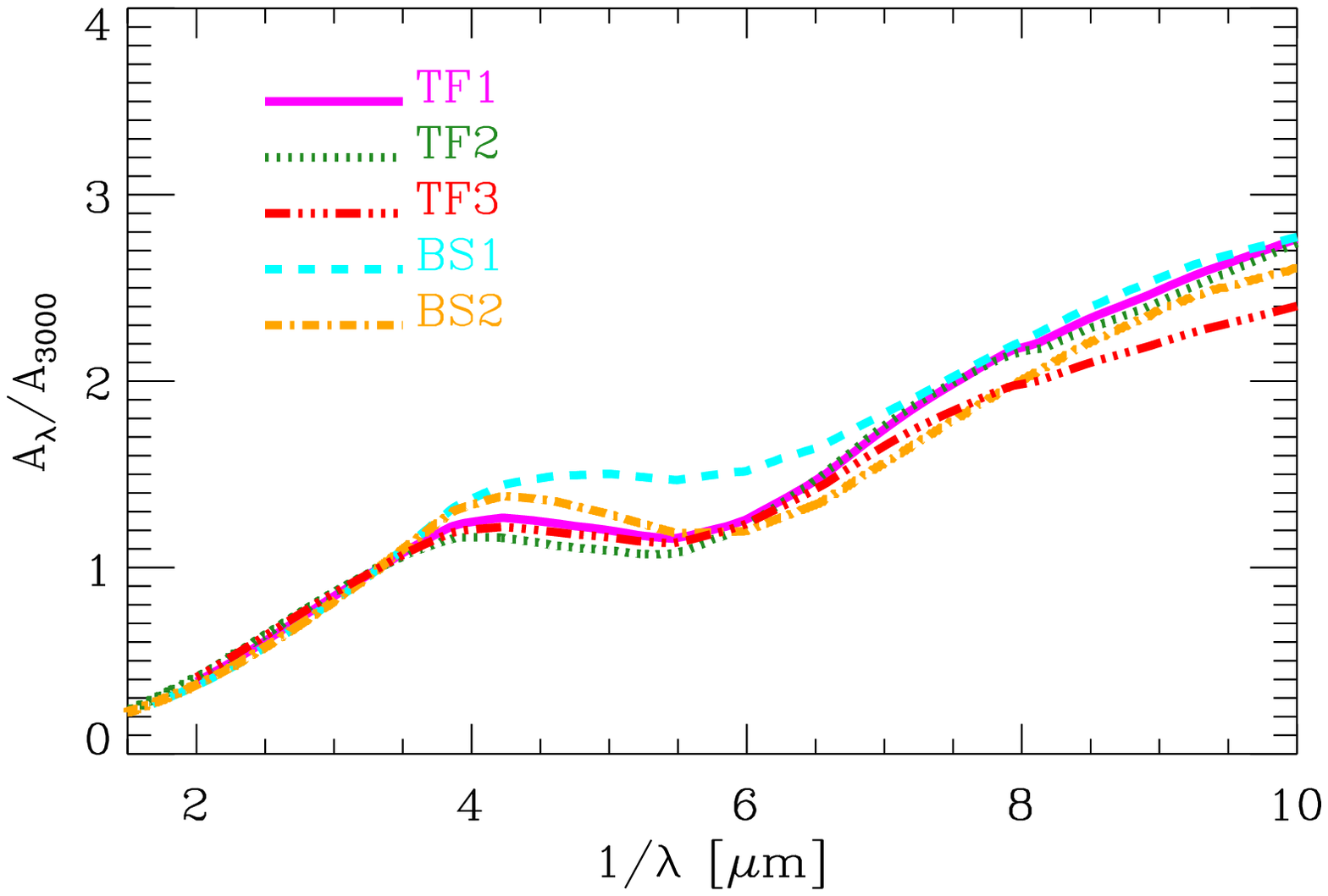}
\includegraphics[width=6 cm]{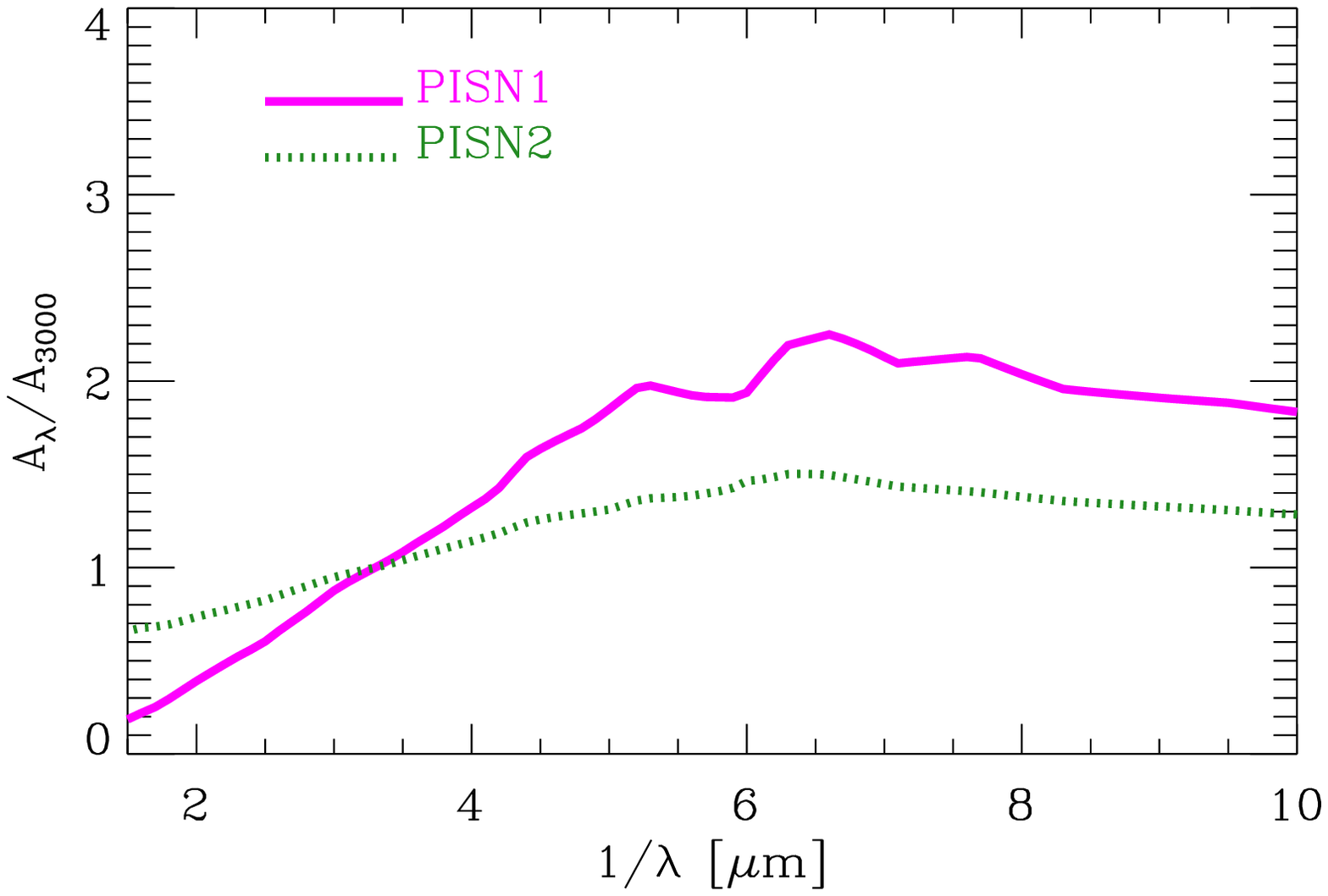}
}
\caption{{\it Left panel}: Empirical curves for dust extinction in the local Universe \cite{Pei1992}. {\it Middle panel}: Theoretical extinction curves predicted by Type II SNe dust models proposed by citet{TodiniFerrara2001} and \citet{BianchiSchneider2007}. {\it Right panel}: Predictions for extinction curves produced by PISNe dust \cite{Hirashita2008}.}
\label{ec}
\end{figure*} 
\begin{figure*}
\centerline{
\includegraphics[width=10 cm]{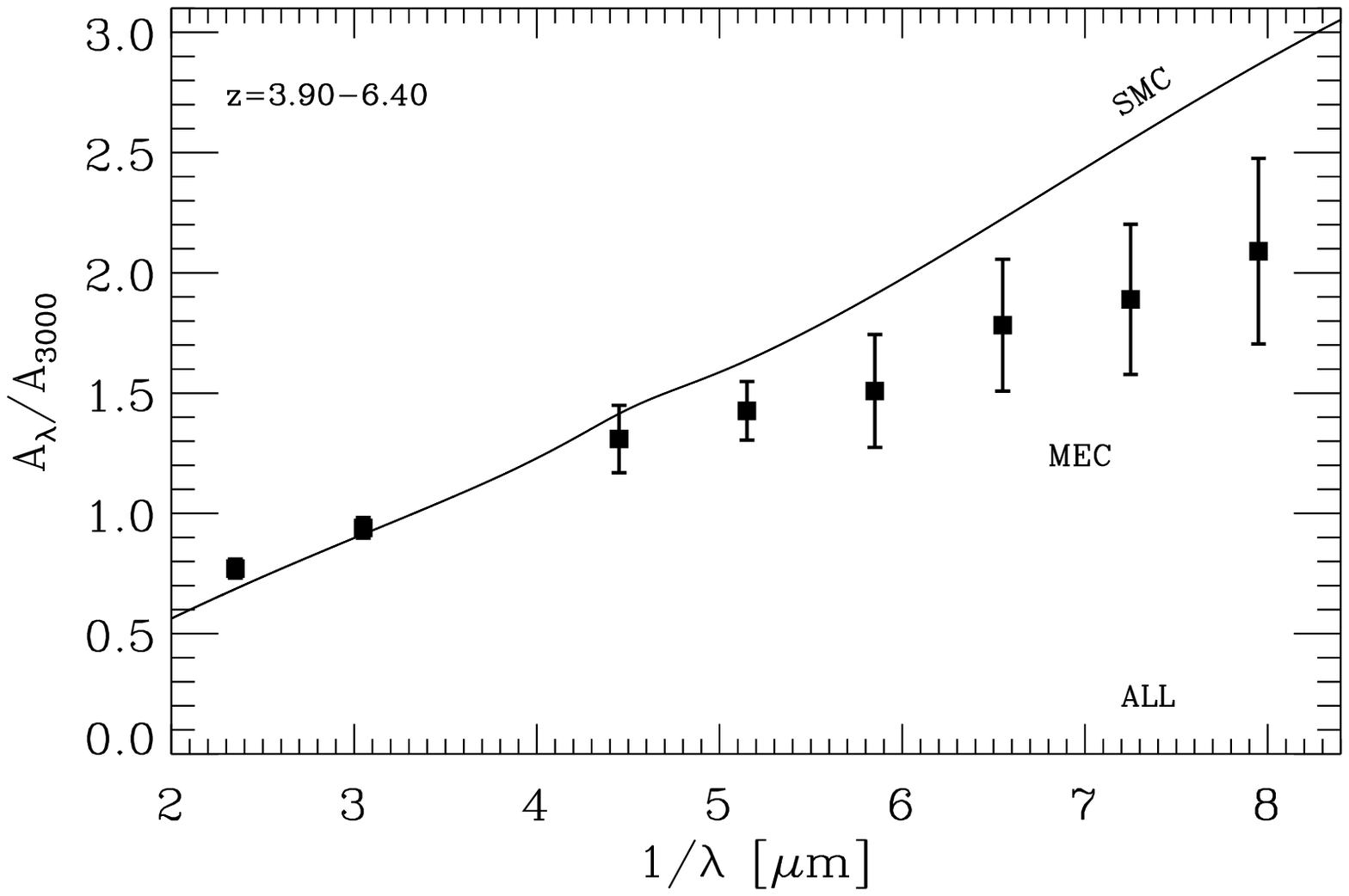} 
}
\caption{The mean extinction curve (MEC) computed by combining the results obtained for individual quasars is shown through black filled squares, along with the $1\sigma$ dispersion denoted by the error bars.}
\label{MEC}
\end{figure*}
 
\subsection{Extinction curves}\label{secec}
In order to investigate the evolution of the dust properties across cosmic times, we consider a grid of
extinction curves to characterize the extinction produced by dust in the rest-frame wavelength range
$0.1\leq \lambda \leq 0.5~\rm{\mu m}$.
First, we consider the empirical curves which describe the dust extinction in the local Universe \cite{Pei1992}: the Milky Way (MW) extinction curve, characterized by a a prominent bump at 2175 \AA; the
featureless Small Magellanic Cloud (SMC) extinction curve, which steeply rises with inverse wavelength
from near infrared to far ultraviolet ($A_{\lambda}/A_{3000}\sim \lambda^{-1.2}$); the Large Magellanic Clouds (LMC)
extinction curve, being intermediate between the MW and the SMC. We show the SMC, LMC and MW extinction
curves in Fig. \ref{ec}, left panel (black solid, red dotted, green dashed lines, respectively). 
\\
We also use the extinction curves expected by Type II SNe dust models as
predicted by \citet{TodiniFerrara2001,BianchiSchneider2007}. The extinction curves by \citet{TodiniFerrara2001} depend on the metallicity of the SNe progenitors; the middle panel of Fig.~\ref{ec}
shows the cases of $Z=0$ (TF1, magenta solid line), $Z=Z_{sun}$ (TF2, green dotted line), and
$Z=10^{-2}Z_{sun}$ (TF3, red dotted-long-dashed line).
For what concerns the models by \citet{BianchiSchneider2007}, these authors take
into account the possibility that the freshly formed dust in the SNe envelopes can be destroyed and/or
reprocessed by the passage of the reverse shock. 

The middle panel of Fig.~\ref{ec} shows the extinction curve predicted by the model
before (BS1, cyan dotted line) and after (BS2, orange dotted-short-dashed line) the reverse shock,
assuming solar metallicity for the SNe progenitor.\\ 
In the right panel of Fig. \ref{ec}, we show the results of the models by \citet{Hirashita2008} for
dust produced by Pair Instability SNe (PISN) having progenitor mass of 170~$M_{sun}$. \citet{Hirashita2008} study
two extreme cases for the mixing of the elements constituting the dust grains: the first one
is the unmixed
case, in which the original onion-like structure of elements is preserved in the helium core;
the second one
is the mixed case, characterized by a uniform mixing of the elements.  
In the bottom right panel of Fig. \ref{ec}, we show the results of their analysis in the mixed (PISN1, solid magenta line) and unmixed (PISN2,
dotted green line) cases.

\subsection{Observational constraint on the high-z dust}
We fit the spectral region redward the Ly$\alpha$ emission line with the following equation:
\begin{equation}
F_{\lambda}=C~F_{\lambda}^{t}~\left ( \frac{\lambda}{3000}\right )^{(1.62-\alpha)}~10^{-A_{3000}\frac{A_{\lambda}}{2.5}},
\label{bfeq}
\end{equation}
where C is a normalization constant, $F_{\lambda}^{t}$ is a quasar template spectrum, $\alpha$ the slope
of the unreddened spectrum,
$A_{3000}$ the absolute extinction at 3000~\AA~, and $A_{\lambda}$ the extinction curve
normalized at 3000~\AA. In this study we adopt the template by \citet{Reichard2003}, obtained
considering 892 quasars classified as non-BAL. The spectral index of the averaged non-BAL spectrum
($F_{\lambda}^{t}\propto \lambda^{-\alpha_{t}}$) is $\alpha_{t}=1.62$. Therefore, the term
$(\lambda/3000)^{(1.62-\alpha)}$ allows us to force the slope of the template to the value $\alpha$. We
let $\alpha$ to vary in the interval [0.2; 3.0], which is the range encompassed by more than 95\% of the quasars.  For
each quasar spectra, in the fitting procedure, we avoid the emission features characterizing the
following spectral regions in the rest frame of the source: Ly$\alpha$+NV [1215.67; 1280]~\AA; OI+SiII
[1292; 1320]~\AA; SiIV+OIV] [1380; 1418]~\AA; CIV [1515; 1580]~\AA; AlIII+CIII] [1845; 1945]~\AA. We also
exclude the region [2210; 3000]~\AA, characterized by a prominent FeII bump. 
After having
selected the spectral regions to be included in the analysis, we rebin the observed spectra to a
resolution $R\sim 50$, which is about the spectral resolution delivered by the Amici prism in the
NICS-TNG observations.\\
The $\chi^2$ analysis reveals that 8 quasars require substantial ($0.2\geq A_{3000}\geq 1.9$) reddening and that
these reddened spectra favor extinction curves which differ from the
SMC. This study mostly aims at investigating whether the properties of the extinction curves at $z\gtrsim 4$ deviate
or not from the SMC, which has been shown to reproduce the dust reddening of quasars at $z<4$. Therefore, we compute the mean of the inferred extinction curves to provide an empirical,
average extinction
law at $z\gtrsim 4$, called MEC. We find that the MEC deviates from the SMC
 extinction curve at a confidence level $\gtrsim 68\%$ for $\lambda<2000$ \cite{Gallerani2009}.
  
\section{Conclusions}
We analyze optical-near infrared spectra of a large sample of quasars at high redshift with the aim of investigating both the cosmic reionization history at $z\sim 6$ and the properties of dust extinction at $z>4$. 

In order to investigate cosmic reionization, we study the transmitted flux in the region blueward the Ly$\alpha$ emission line in a sample of 17 quasars spectra at $5.7\leq z_{em}\leq 6.4$. 
We analyze the wide dark portions (gaps) in the observed absorption spectra and we compare the statistics of these spectral features with a semi-analytical model of the Ly$\alpha$ forest. We consider two physically motivated reionization models: (ii) a Late Reionization Model LRM in which the epoch of reionization is $z_{\rm rei}\sim 6$; (ii) an Early Reionization Model in which $z_{\rm rei}\sim 7$, meaning that in this case the Universe is highly ionized at $z\sim 6$. We find that the volume-averaged neutral hydrogen fraction $x_{\rm{HI}}$ evolves smoothly from $10^{-4.4}$ at $z=5.3$ to $10^{-4.2}$ at $z=5.6$, with a robust upper limit $x_{\rm{HI}} < 0.36$ at $z=6.3$. The frequency and physical sizes of the gaps favor the ERM, thus being consistent with an highly ionized IGM at $z\sim 6$. This result is also confirmed by the analysis of the optical afterglow spectrum of the Gamma Ray Burst GRB050904 at $z=6.3$ \cite{Gallerani2008b} and by the evolution of the luminosity function of Ly$\alpha$ emitters between $z=5.7$ and $z=6.6$ \cite{Dayal2008}.
 
For what concerns the study of the high-z dust, we focus our attention on the region redward the Ly$\alpha$ emission line of 33 quasars observed at $4\leq z_{em}\leq 6.4$. We compute simulated absorbed quasar spectra by taking into account a large grid of extinction curves which includes well-known empirical laws describing the dust extinction of the local Universe, namely the SMC, the LMC, and the MW extinction curves, as well as several theoretical extinction curves predicted by supernova dust models. We apply a $\chi^2$ analysis to the observed spectra, by comparing them with the synthetic absorbed ones. We find that 8 quasars in our sample require substantial ($0.2\geq A_{3000}\geq 1.9$) reddening. Since the SMC extinction curve has been shown to reproduce the dust reddening of most quasars at $z<4$, the main goal of this work is to investigate whether this curve provides a good prescription for describing dust extinction also at higher redshifts. Starting from the results obtained for individual quasars, we compute an empirical mean extinction curve (MEC) with the corresponding standard deviation. We find that the MEC deviates from the SMC extinction curve at a confidence level $\gtrsim 68\%$ for $\lambda<2000$. This result suggests that the properties of dust (chemical composition and/or grain size distribution) may differ at $z>4$ with respect to the local Universe \cite{Gallerani2009}.

\begin{acknowledgments}
This work has been done in collaboration with: A. Ferrara, R. Maiolino, S. Bianchi, T. Roy Choudhury, P. Dayal, X. Fan, L. Jiang, Y. Juarez, F. Mannucci, A. Marconi, A. Maselli, D. Mortlock, T. Nagao, T. Oliva, R. Salvaterra, R. Schneider, R. Valiante, C. Willott.    
\end{acknowledgments}

\newpage

\lhead[\fancyplain{}{}]{\itshape \small{How do galaxies accrete their mass? Quiescent and star-forming massive galaxies at high redshift }}
\rhead[\itshape \large{1$^\mathrm{st}$ RYRM Proceedings}]{\fancyplain{}{P. Santini}}

  \newcommand{\kmips}{F(24$\mu$m)/F($K_s$)~}
  \newcommand{\kmip}{F(24$\mu$m)/F($K_s$)}



\begin{center}

\setcounter{section}{2}
\section{\large{How do galaxies accrete their mass? Quiescent and star-forming massive galaxies at high redshift}}
\label{santini}

\normalsize{Paola Santini$^*$}

\emph{\small{INAF - Osservatorio Astronomico di Roma,\\Via Frascati 33,
00040 Monteporzio (RM)}}


\vspace{.5cm}
\begin{minipage}{.8\textwidth}
\small{In recent years, several surveys have shown that massive galaxies have undergone a major evolution during the epoch corresponding to the redshift range 1.5-3, assembling a significant fraction of their stellar mass in this epoch. To understand the origin of this rapid rise, a closer scrutiny on the nature and physical properties of massive galaxies at high redshift is needed. I will present our recent results based on the analysis of the \mips MIPS data of the GOODS-S field, that allow to trace star formation (or the lack of) in high redshift galaxies without biases due to dust extinction.
I will show the results of our analysis focusing in particular on: {\it a)} the fraction of quiescent galaxies as a function of redshift; {\it b)} the evolution of the specific star formation rate as a function of redshift and stellar mass.
The scenario emerging from these data will be compared with recent predictions of theoretical models, to discuss the validity of their physical ingredients.}

\vspace{.25cm}
$*$ e-mail: \href{mailto:santini@oa-roma.inaf.it}{santini@oa-roma.inaf.it}
\end{minipage}



\end{center}
\setcounter {section} {0}
\setcounter {subsection} {0}
\setcounter {figure} {0}
\setcounter{equation}{0}

\section{Introduction}

According to several independent lines of evidence, the population of
massive galaxies has undergone major evolution during the very short epoch
corresponding to the redshift range $1.5-3$. Many previous works measured a rapid evolution in the stellar mass density within this redshift range (e.g., \cite{dickinson03}, \cite{fontana06}, \cite{papovich06} and references therein) and demonstrated that a substantial fraction (30-50\%) of the stellar mass formed during this epoch. 

The nature of the physical processes responsible for this rapid rise
remains unclear.  A large number of massive ($\simeq
10^{11}M_\odot$) actively star-forming galaxies is clearly in place
at $z\simeq 2$ \cite{daddi04}, \cite{papovich07}. These galaxies have been demonstrated to
experience an extremely active phase in the same redshift range (e.g., \cite{daddi07a}).  

At the same time, galaxies with very low levels of
star formation rates (SFR) at $z\simeq 1.5-2$ have been detected by imaging
surveys based on color criteria (e.g., \cite{daddi04}) or SED fitting
\cite{grazian07},  and by spectroscopic observations of red galaxy samples
(e.g., \cite{cimatti04}, \cite{kriek06}. These results have motivated the
inclusion of efficient methods for providing a rapid assembly of
massive galaxies at high $z$ (such as starburst during interactions) 
as well as quenching of the SFR, most notably via AGN feedback
(e.g., \cite{menci06}, \cite{bower06}).

In the first part of this work we will focus on quiescent galaxies, while star-forming ones will be the object of the second part. 
Once presented our data sample in section \ref{sec:data}, in section \ref{sec:red} we attempt to compile a statistically well defined, mass-selected sample of galaxies with very low levels of star formation at high redshift ($0.4-4$), with the aim of comparing their abundance with theoretical model predictions. Such a comparison can shed light upon the nature of the star formation suppression. 
In section \ref{sec:sfr}, we investigate the star formation properties of our mass-selected sample between $z\sim 0.3$ and $z\sim 2.5$ in order to understand whether the rapid growth of the stellar mass density is due to star formation episodes inside the galaxies or to merging events. Once again, we will compare our observations with theoretical expectations. Conclusions will follow in section \ref{sec:concl}.

Throughout this work, unless  stated otherwise,  
we assume a Salpeter initial mass function (IMF) and adopt the $\Lambda$-CDM concordance cosmological model (H$_0$ = 70 km/s/Mpc, $\Omega_{\small M}$ = 0.3 and $\Omega_{\Lambda}$ = 0.7).

\section{\label{sec:data}The data sample}
For this study, we used data from an updated version of the multicolour
GOODS-MUSIC sample \cite{santini09}, \cite{grazian06}.  
The catalog covers an area of about 143 arcmin$^2$ located in the Chandra Deep Field South and consists of 15\,208 sources. After culling Galactic stars, it contains 14\,999 objects selected in either the $z$ band or the $K_s$ band or at 4.5 $\mu$m. 
The 15-bands multiwavelength coverage ranges from 0.35 to \mip, as a result of the combination of images from different instruments (2.2ESO, VLT-VIMOS, ACS-HST, VLT-ISAAC, Spitzer-IRAC,
Spitzer-MIPS). 
The whole catalog has been cross-correlated with spectroscopic
catalogs available to date, and a spectroscopic redshift has been
assigned to $\sim$12 \% of all sources.  For all other objects, we have
computed well-calibrated photometric redshifts ($|\Delta z/(1+z)|<0.06$). 
Physical properties of each object, such as total stellar mass, SFR, age and dust obscuration, have been obtained through a standard SED fitting technique to the overall photometric data from 0.3 to 8 $\mu$m using the synthetic templates of  \citet{bc03}. 

Our work is mainly based on the analysis of the \mips MIPS data, which is described in details in \cite{santini09}. 
Notable contaminants affecting galaxy mid-IR emission are represented by highly obscured AGNs, where the IR emission is generated by matter accretion onto a central black hole rather than dust heating by young stars. In order to avoid bias in our IR-based SFR estimates as well in the selection of quiescent galaxies, we remove highly obscured AGNs candidates from our sample by following the approach described by \cite{fiore08}.

\section{\label{sec:red}Quiescent galaxies}

\subsection{Selection criterium}

It is difficult to isolate passively evolving galaxies from the wider population of intrinsically red galaxies at high redshift, which include also a (probably larger) contribution of star-forming galaxies reddened by a large amount of dust. The two classes are indeed indistinguishable when selected by means of traditional single colour criteria, since a dusty stellar population may have similar UV-optical characteristics of an old population. 
The problem can be overcome by considering the mid-IR emission of these optically red galaxies, which  appears to be clearly different for the two populations. 
Star-forming galaxies are bright because the UV light released by their young stars is absorbed by dust and re-emitted at mid-IR wavelengths. By contrast, passively evolving galaxies are faint, as the starlight from evolved populations peaks in the near-IR. 
We separated the quiescent and the star-forming population by means of their \kmips flux ratio, which we demonstrated to have a clear bimodal distribution \cite{fontana09}.  

However, we were not simply interested to quiescent galaxies, but rather we aimed to explore the very quiescent tail of the red galaxy population, which can be used to investigate the mechanisms which shut down the star formation. We therefore combined the information derived from the mid-IR photometry with the SED fitting analysis, and we selected a sample of ``red and dead'' galaxies by requiring that $SFR/M<10^{-11} yr^{-1}$, which ensures negligible levels of star formation in these galaxies (see \cite{fontana09}).

In the following, we adopt a mass-selected sample, obtained by applying a mass threshold at $M\geq 7\times 10^{10}M_\odot$ to our photometric catalog based on a combined selection
$K_s<23.5$ {\it or} $m_{45}<23.5$. This photometric sample is complete
at this mass limit to $z\simeq 4$,  also for dust-absorbed star-forming
galaxies \cite{fontana06}.

\subsection{\label{sec:redfrac}The fraction of ``red and dead'' galaxies and the comparison with theoretical predictions}

\begin{figure}
\includegraphics[width=9cm]{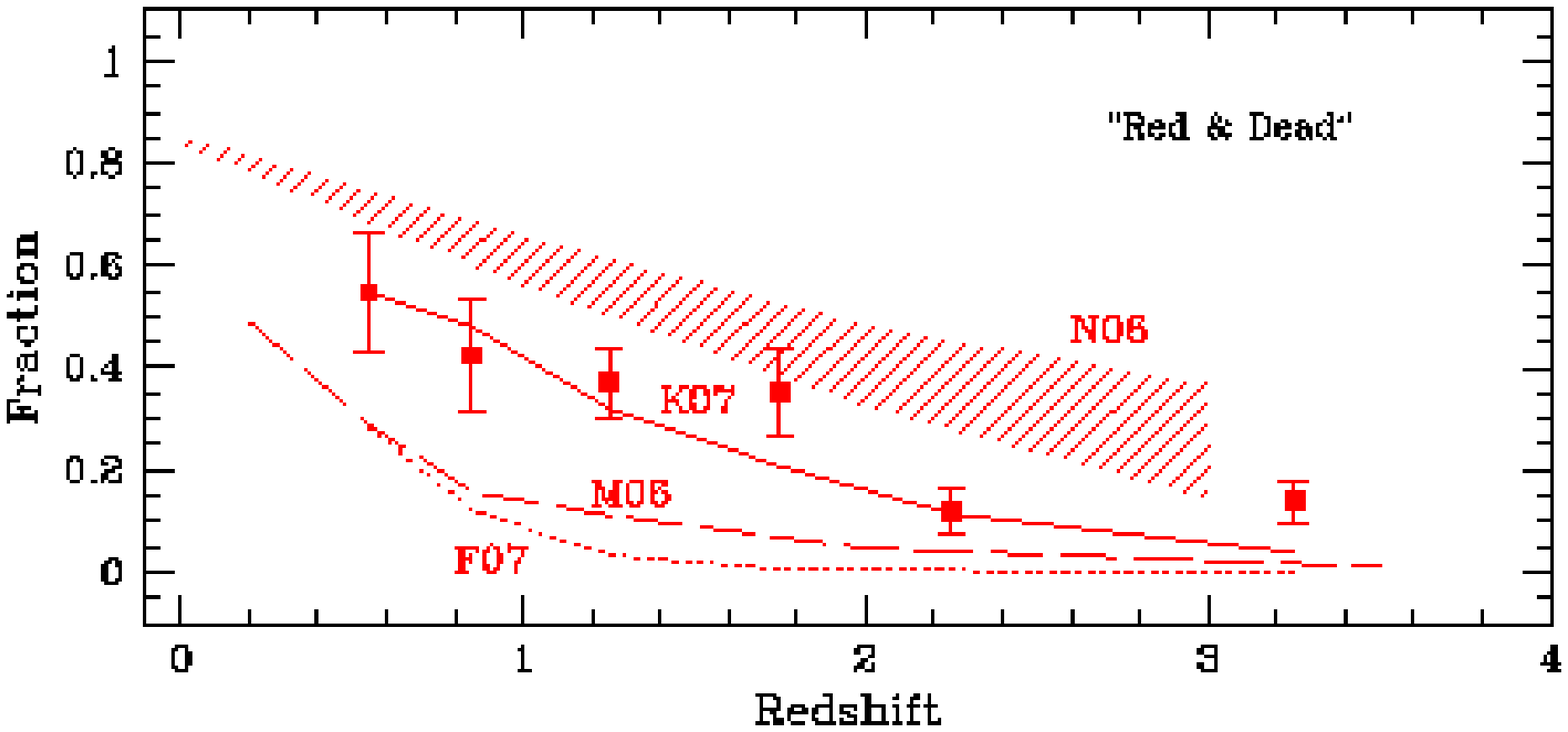}
\caption{Fraction of ``red and dead'' galaxies (defined as described in the text) as a function of redshift, in the  $M\geq 7\times10^{10}M_\odot$ mass-selected sample. Points
  represent the observed values. Error bars include Poisson and cosmic variance errors.  Lines
  refer to the predictions of theoretical models, as described by the
  legend (see text for references). }
\label{fig:redfrac}
\end{figure}

The fraction of ``red and dead'' is shown in Fig.\ref{fig:redfrac} as a function of redshift from $z=0.4$ to $z=4$. A detailed analysis of the highest redshift ($z>2.5$) candidates, where the selection is affected by large uncertainties, is presented in \cite{fontana09}. Error bars were computed by summing (in quadrature) the Poisson and cosmic variance error. The latter was computed by measuring the relative variance
within 200 samples bootstrapped from the Millennium Simulation
\cite{kitzbichler07}, using an area as large as GOODS-South and applying
the same selection criteria.

Our analysis confirms the cosmological decrease in the number density of massive early-type
galaxies at high redshifts: ``red and dead'' galaxies make up more than 50\% of the
population of massive galaxies at $z<<1$, and become progressively
less common at higher $z$. 
However, we note that a sizeable fraction of galaxies with extremely low levels of SFR is already in
place at $z>1.5$ and up to the highest redshifts sampled here ($z\simeq 3.5$) with
a fraction of about 15\% at $z>2$. This implies that the star formation episodes in these galaxies must be quenched either by efficient feedback mechanism and/or by the
stochastic nature of the hierarchical merging process.

It is interesting to determine whether theoretical models agree with these
observational results.  In Fig. \ref{fig:redfrac}, we plot the predictions
of several models, applying the  same selection criteria that we apply on the data. 
We consider purely semi-analytical models
(\citet{menci06}, M06, and MORGANA \cite{monaco07}, F07), a semi-analytical
rendition of the Millennium N-body dark matter Simulation
(\citet{kitzbichler07}, K07), and purely hydrodynamical simulations (\citet{nagamine06}, N06). 
The final are presented for three
different timescales $\tau$ of the star formation rate (ranging from
$2\times10{^7}$ yrs to $2\times10{^8}$ yrs), and represented with a
shaded area. 

All these models agree in predicting a gradual decline
with redshift in the fraction of galaxies with very low SFR. 
However, the predicted fraction of ``red and dead'' galaxies
varies significantly at all redshifts.
For this reason, it turns out to be 
a particularly sensitive quantity, which provides a powerful
way of highlighting the differences between the models.  Some
models (M06, F07) underpredict the fraction of ``red and dead'' galaxies
at all redshifts, and in particular predict virtually no object at
$z>2$, in contrast to what observed. The Millennium-based model agrees
with the observed quantities, while the hydro model appears to
overpredict them.

\section{\label{sec:sfr}Star-forming galaxies}

\subsection{\label{sec:sfrestimate}The estimate of the SFR}

Since the most intense star formation episodes are expected to occur in dusty regions, and
based on the assumption that most of the photons originating in newly formed
stars are absorbed and re-emitted by dust, the mid-IR emission is in
principle the most sensitive tracer of the star formation rate. 
Moreover, it has the great advantage of being unaffected by dust obscuration and  not depending on still uncertain dust corrections.

In addition to mid-IR emission, a small fraction of unabsorbed photons will be 
detected at UV wavelengths. A widely used SFR indicator is therefore based on a
combination of IR and UV luminosity, which supply complementary
knowledge of the star formation process. For \mips detected
sources, we estimated the instantaneous SFR using the same calibration
as \cite{papovich07}: 
\be
\mbox{SFR}_{\mbox{\scriptsize IR} + \mbox{\scriptsize UV}}/\mbox{M}_\odot \mbox{yr}^{-1} = 1.8 \times 10^{-10} \times \mbox{L}_{bol}/\mbox{L}_\odot \\
\label{eq:sfr}
\ee \be \mbox{L}_{bol} = (2.2 \times \mbox{L}_{\mbox{\scriptsize UV}}
+ \mbox{L}_{\mbox{\scriptsize IR}}) \ee 

The total IR luminosity \lirs was computed by fitting \mips emission to \citet{dh02}  (DH) synthetic templates. 
The rest-frame UV luminosity, uncorrected for extinction, was derived from the SED fitting technique, L$_{\mbox{\scriptsize UV}} = 1.5 \times \mbox{L}_{2700 \mbox{\scriptsize \AA}}$; although often negligible, this can account for the contribution from young unobscured stars. 

Following \cite{papovich07}, we then applied a lowering correction to the estimate obtained from Eq. \ref{eq:sfr}. 
They found that the \mips flux, fitted with the same DH library, overestimates the SFR for bright IR galaxies, with respect to the case where longer wavelengths (70 and 160 $\mu$m MIPS bands) are considered as well, and they corrected the trend using an empirical second-order polynomial. 

We derive the SFR using the prescriptions above to all objects with $F_{24\mu m} \geq 20\ \mu$Jy (which we estimated to be the flux limit of our \mips catalog), while we adopt the estimate derived from the SED fitting analysis for all fainter galaxies undetected at \mip.

\subsection{The evolution of the specific star formation rate and the mass assembly process}

With respect to other surveys, our sample has the distinctive
advantage of being selected by a multiwavelength approach. In
this section, we consider a subsample created by performing the following cuts: $z<26$ {\it or} $K_s<23.5$ {\it or} $m_{4.5}<23.2$. The $K_s$ and $m_{4.5}$ cuts ensure a proper sampling of
highly absorbed star-forming galaxies, and hence probably a complete
census of all galaxies with high SFR. On the other hand, the deep $z$-selected sample
contains the lower mass, fainter and bluer galaxies of both low levels of dust
extinction and low star formation rate.

\begin{figure*}[!t]
\includegraphics[width=11cm]{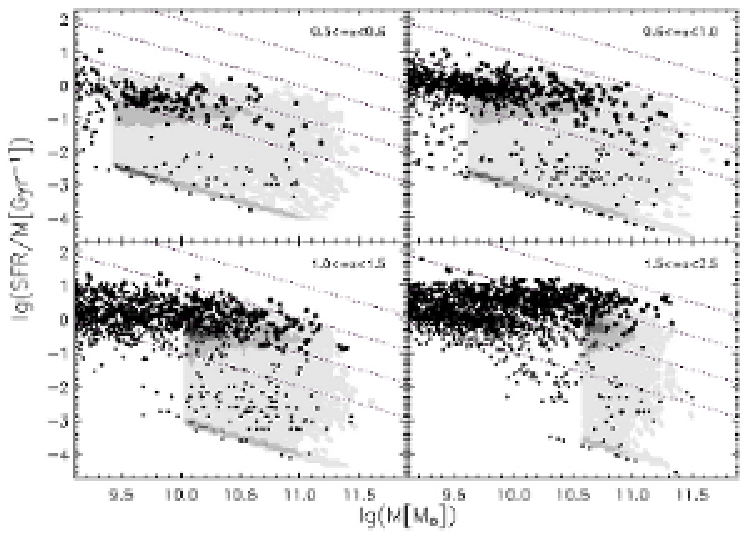}
\caption{Relation between the specific star formation rate and the stellar mass calibrated to a Chabrier IMF. 
Large dots correspond to \mips sources with F$_{24\mu m}>20 \mu$Jy, while small ones refer to \mips upper limits. Dotted lines correspond to constant SFRs of 1, 10, 100 and 1000 $M_\odot yr^{-1}$.  The horizontal dashed lines indicate the inverse of the age of the Universe at the centre of each redshift interval. 
Shaded contour levels (at 0.05\%, 10\%, 50\% and 80\% level) represent the predictions K07  \cite{kitzbichler07} model above the mass limit in each redshift bin. }
\label{fig:mdot}
\end{figure*}

In Fig. \ref{fig:mdot}, we plot the relation between the stellar mass
and the specific star formation rate (defined as the SFR per unit mass, SSFR hereafter) for all galaxies
divided into redshift bins, from $z=0.3$ to $z=2.5$. To be able to compare  our findings with the Millennium Simulation predictions, we converted our masses and SFR to the Chabrier IMF used by the Millennium Simulation.

First of all, we notice a strong bimodality in the SSFR distribution at all redshifts. 
Two distinct populations, together with some sources lying
between the two, are detectable, one of young, active and blue
galaxies  (the so-called blue cloud) and the other one consisting of old, ``red and dead'', early-type galaxies  (red sequence). 
The loci of these two populations are consistent with
the selection in \cite{salimbeni08} between early- and late-type
galaxies.  

A trend for the specific star formation rate to increase with redshift
at a given stellar mass is evident: galaxies tend to form
their stars more actively at higher redshifts, or, in other words, the bulk of
active sources shifts to higher values of SSFR with increasing
redshift. Moreover, at a given redshift, low mass galaxies are more actively star-forming than their higher mass counterparts. 
Our findings are in good agreement with
\cite{perezgonzalez05} and \cite{papovich06}.

A significant fraction of the sample, increasing with redshift, is in
an active phase (see also section \ref{sec:redfrac}). 
It is natural to compare the SSFR (which has units of the
inverse of a timescale) of these galaxies with the inverse of the age of the Universe at the
corresponding redshift ($t_U(z)$). We  define galaxies with
$M/SFR<t_U(z)$ as ``active'' in the following, since they are
experiencing a major episode of star formation, potentially building
up a substantial fraction of their stellar mass in this episode (see \cite{santini09}).
Galaxies selected following this criterium are forming stars more actively than in their recent past.

At $1.5 \leq z < 2.5$, the fraction of massive (with $M>7\cdot 10^{10} M_\odot$) active galaxies in the total sample is
66\%, and their mean SFR is $309\ M_\odot yr^{-1}$. To compute the total stellar mass produced within this redshift interval, it is necessary to know the duration of the active phase. For this purpose, we use a duty cycle argument and suppose that the active fraction of galaxies is indicative of the time interval spent in an active phase. 
We adopt the assumption that the active fraction is stable within the redshift bin considered. The time spanned in the 1.5--2.5 redshift interval corresponds to 1.5 Gyr. By multiplying the fraction of active galaxies by the time available, we derived an average duration of the active phase of 0.99 Gyr. The average amount of stellar mass assembled within each galaxy during these bursts is measured to be the product of the average SFR and the average duration of the active phase, and  equals to $3.1 \times 10^{11}M_\odot$,  representing a significant fraction of the final stellar mass of the galaxies considered.   
Although quite simplified, this analysis implies that most of the stellar mass
of massive galaxies is assembled during a long-lasting active phase at
$1.5 \leq z < 2.5$. It is important to remark that this process of
intense star formation occurs directly within already massive
galaxies, and, given its intensity, prevails throughout growth episodes due to merging events of already
formed progenitors.

\subsection{Comparison with theoretical predictions}

To provide a further physical insight in this process, we 
compared our results with the predictions of three recent theoretical
models of galaxy formation and evolution. Our sample is affected by mass incompleteness, so only galaxies above the completeness limit in each redshift bin were considered in the comparison.  We note that this limit depends on the redshift bin \cite{fontana06}, and it is equal to $5 \cdot 10^{9}$, $8 \cdot 10^{9}$, $2 \cdot 10^{10}$ and $7 \cdot 10^{10} M_\odot$, respectively (calibrating masses to a Salpeter IMF).

In Fig. \ref{fig:mdot}, we show the predictions of the K07 \cite{kitzbichler07} model based on the Millennium Simulation. 
We find that the model  predicts an overall
trend that is consistent with our findings. The SSFR
decreases with stellar mass (at given redshift) and increases with
redshift (at given stellar mass). In addition, it forecasts the
existence of quiescent galaxies even at $z > 1.5$, as already found in section \ref{sec:redfrac}.  
However, the average observed SSFR is systematically under-predicted
(at least above our mass limit) by a factor $\sim$3-5 by the
Millennium Simulation. 
A similar trend for the Millennium Simulation at $z\sim2$ 
was already shown by \cite{daddi07a}. 

We also compared our findings with the semi-analytical models of M06 \cite{menci06} and MORGANA \cite{monaco07}. They show very similar trends with respect to the Millennium Simulation, with only slightly different normalizations, up to a factor 10 discrepancy (see \cite{santini09}). 

A more comprehensive comparison between theoretical predictions and observations was presented in \cite{fontanot09}.

\section{\label{sec:concl}Conclusions}

We have studied the mass accretion process in galaxies around $z\sim 2$ by investigating the properties of both the quiescent and the star-forming population and by comparing observations with theoretical expectations. 

In the first part we considered the fraction of very quiescent, ``red and dead'' galaxies, as a function of redshift, to the total sample of massive galaxies. 
Since these galaxies are reproduced by the models by shutting down the star formation (SF), the fraction of ``red and dead'' galaxies provides information about the mechanism responsible for the SF quenching. 
A non-negligible fraction ($\sim$ 15\%) of galaxies with very low SF activity is already in place at the highest redshift sampled in this work. This motivates the inclusion in the models of very efficient SF mechanisms as well as their rapid suppression.

In the second part we studied the evolution of the specific star formation rate as a function of redshift and stellar mass. The SSFR shows a well-defined bimodal distribution, with a clear separation between actively star-forming and passively evolving galaxies. Massive star-forming galaxies at $z\simeq 2$ are vigorously forming stars, typically at a rate of $\sim 300 M_\odot$yr$^{-1}$. A simple duty-cycle argument suggests that they assemble a significant fraction of their final stellar mass during this phase, implying that star formation episodes in already massive galaxies are the main responsible for the rapid growth of the stellar mass density at $z\sim 2$.

We used our results for the quiescent and the star-forming galaxy populations to investigate
the predictions of a set of theoretical models of galaxy formation in a $\Lambda$-CDM scenario. 
All the models taken into account qualitatively reproduce the global observed trend. However, quantitatively, they predict an average specific star formation rate that is systematically lower than observed, at least in the mass regimes considered. On the other side, for what concerns the ``red and dead'' fraction, they vary to a large extent in their predictions and are unable to provide a global match to the data, making this observable very sensitive and powerful to constrain the SF quenching mechanism.

Although some hypothesis have been suggested by \cite{santini09}, the origins of the discrepancies between observations and theoretical predictions are difficult to ascertain, because of the complex interplay between all the physical processes involved in
these models, the different physical processes implemented - most notably those related to AGN feedback - and their different technical implementations. The failure of most models to reproduce simultaneously the fraction of ``red and dead'' massive galaxies in the early Universe and the star formation activity probably implies that the balance between the amount of cool gas and the star formation efficiency on the one side, and the different feedback mechanisms on the other, is still poorly understood.


\newpage

\lhead[\fancyplain{}{}]{\itshape \small{Multiwavelenght observations of the gamma-ray blazars detected by AGILE}}
\rhead[\itshape \large{1$^\mathrm{st}$ RYRM Proceedings}]{\fancyplain{}{F. D'Ammando}}


\begin{center}

\setcounter{section}{3}
\section{\large{Multiwavelength Observations of the Gamma-ray Blazars Detected by AGILE}}
\label{dammando}

\normalsize{F.~D'Ammando$^*$}

\emph{\small{Dip. di Fisica, Univ. ``Tor Vergata'', Via della Ricerca
  Scientifica 1, 00133 Roma, Italy\\
  INAF-IASF Roma, Via Fosso del Cavaliere 100, 00133 Roma, Italy}}

\vspace{.25cm}
\normalsize{S.~Vercellone}

\emph{\small{INAF-IASF Palermo, Via Ugo La Malfa 153, 90146 Palermo, Italy}}

\vspace{.25cm}
\normalsize{I.~Donnarumma, L.~Pacciani, G.~Pucella, M.~Tavani, V.~Vittorini}

\emph{\small{INAF-IASF Roma, Via Fosso del Cavaliere 100, 00133 Roma, Italy}}

\vspace{.25cm}
\normalsize{A.~Bulgarelli}

\emph{\small{INAF-IASF Bologna, Via Gobetti 101, 40129 Bologna, Italy}}

\vspace{.25cm}
\normalsize{A.~W.~Chen, A.~Giuliani}

\emph{\small{INAF-IASF Milano, Via E. Bassini 15, 20133 Milano, Italy}}

\vspace{.25cm}
\normalsize{F.~Longo}

\emph{\small{Dip.~ di Fisica and INFN, Via Valerio 2, 34127 Trieste, Italy}}

\vspace{.25cm}
\normalsize{on behalf of the AGILE Team} 

\vspace{.5cm}

\begin{minipage}{.8\textwidth} 
\small{Since its launch in April 2007, the AGILE satellite
  detected with the Gamma-Ray Imaging Detector several blazars in high
  $\gamma$-ray activity: 3C 279, 3C 454.3, PKS 1510--089, S5 0716$+$714, 3C 273, W Comae
  and Mrk 421. Thanks to the rapid dissemination of our alerts, we were able
  to obtain multiwavelength ToO data from other observatories such as $Spitzer$, $Swift$, RXTE, $Suzaku$, INTEGRAL, MAGIC,
  VERITAS, as well as radio-to-optical coverage by means of the GASP Project
  of the WEBT and the REM Telescope. This large multifrequency coverage gave us the
  opportunity to study truly simultaneous spectral energy distributions of these sources from
  radio to gamma-ray energy bands and to investigate the different mechanisms
  responsible for their emission. We present an overview of the AGILE results
  on these gamma-ray blazars and the relative multifrequency data.}

\vspace{.25cm}
$*$ e-mail: \href{mailto:filippo.dammando@iasf-roma.inaf.it}{filippo.dammando@iasf-roma.inaf.it}

\end{minipage}


\end{center}

\setcounter {section} {0}
\setcounter {subsection} {0}
\setcounter {figure} {0}
\setcounter{equation}{0}

\section{Introduction}

Among Active Galactic Nuclei (AGNs), blazars are a subclass characterized by the
emission of strong non-thermal radiation across the entire electromagnetic spectrum and in particular intense and variable
$\gamma$-ray emission above 100 MeV \cite{Har99}. The typical
observational properties of blazars include
irregular, rapid and often very large variability, apparent super-luminal
motion, flat radio spectrum, high and variable polarization at radio and
optical frequencies. These features are interpreted as the result of the
emission of electromagnetic radiation from a relativistic jet that is viewed
closely aligned to the line of sight \cite{BR}, \cite{UP}. 

Blazars emit across several decades of energy, from radio
to TeV energy bands, and thus they are the perfect candidates for simultaneous
observations at different wavelengths. Multiwavelength studies of variable
$\gamma$-ray blazars have been carried out 
since the beginning of the 1990s, thanks to the EGRET instrument onboard $Compton$ $Gamma$-$Ray$ $Observatory$
$(CGRO)$, providing the first evidence that the Spectral Energy Distributions (SEDs) of
the blazars are typically double humped with the first peak occurring in the
IR/optical band in the so-called $red$ $blazars$ (including Flat Spectrum Radio
Quasars, FSRQs, and Low-energy peaked BL Lacs, LBLs)  and in UV/X-rays in the
so-called $blue$ $blazars$ (including High-energy peaked BL Lacs, HBLs). 

The first peak is
interpreted as synchrotron radiation from high-energy electrons in a
relativistic jet. The SED second component, peaking at MeV--GeV energies in
$red$ $blazars$ and at TeV energies in $blue$ $blazars$, is commonly interpreted as inverse Compton
scattering of seed photons by highly relativistic
electrons \cite{Ul}, although other models involving hadronic
processes have been proposed (see e.g. \cite{Bo} for a recent review). 

3C 279 is the best example of multi-epoch studies at different frequencies
performed by EGRET during the period 1991--2000 \cite{Har01}. Nevertheless,
only a few objects were detected on a time scale of two
weeks or more in the $\gamma$-ray band and simultaneously monitored at different
energies in order to obtain a wide multifrequency coverage.

With the advent of the AGILE and $Fermi$ $\gamma$-ray satellites, together with
the ground based Imaging Atmospheric Cherenkov Telescopes H.E.S.S., MAGIC and VERITAS, a new
exiting era for the $\gamma$-ray extragalactic astronomy and in particular for
the study of blazars is now open. Observations in the high-energy part of the
electromagnetic spectrum in conjunction with a complete multiwavelength coverage will allow us to shed
light on the structure of the inner jet and the emission mechanisms working in this class of objects.  

\begin{table*}
  \caption{List of the AGILE flaring blazars. 
    References: 1.  Chen et al., 2008, A\&A, 489, L37;
                2.  Giommi et al., 2008, A\&A, 487, L49;
		3.  Donnarumma et al., 2009, ApJ, 691, L13;
                4.  Acciari et al., 2009, in preparation;
                5.  Pucella et al., 2008, A\&A, 491, L21;
                6.  D'Ammando et al., 2009, accepted for the publication in A\&A;
                7.  D'Ammando et al., 2009, in preparation;
                8.  Pacciani et al., 2009, A\&A, 494, 49;
                9.  Giuliani et al., 2009, A\&A, 494, 509;
                10. Vercellone et al., 2008, ApJ, 676, L13;
                11. Wehrle et al., 2009, in preparation;
                12. Vercellone et al., 2009a, ApJ, 690, 1018;
                13. Donnarumma et al., 2009, submitted to ApJ;
                14. Vercellone et al., 2009b, in preparation.
}
\begin{center}  
\begin{tabular}{|l|c|c|c|l|}
\noalign{\smallskip}
    \hline
 \bf{Name} & \bf{Period} & \bf{Sigma} & \bf{ATel $\#$} & \bf {Ref.} \\
 &  {\it start : stop} & & & \\
    \hline
    S5 0716$+$714 & 2007-09-04 : 2007-09-23 & 9.6  & 1221 & 1\\
                  & 2007-10-24 : 2007-11-01 & 6.0  &  -   & 2\\
    Mrk 421
         & 2008-06-09 : 2008-06-15 & 4.5  & 1574, 1583 & 3\\
    W Comae
         & 2008-06-09 : 2008-06-15 & 4.0  & 1582 & 4\\
    PKS 1510$-$089
         & 2007-08-23 : 2007-09-01 & 5.6  & 1199 & 5\\
         & 2008-03-18 : 2008-03-20 & 7.0  & 1436 & 6\\
         & 2009-03-01 : 2009-03-31 & 19.9  & 1957, 1968, 1976 & 7\\
    3C 273
         & 2007-12-16 : 2008-01-08 & 4.6  & -    & 8\\ 
    3C 279
         & 2007-07-09 : 2007-07-13 & 11.1 & -    & 9\\
    3C 454.3
         & 2007-07-24 : 2007-07-30 & 13.8 & 1160, 1167 & 10, 11\\
	 & 2007-11-10 : 2007-12-01 & 19.0 & 1278, 1300 & 12\\
	 & 2007-12-01 : 2007-12-16 & 21.3    & - & 13\\
	 & 2008-05-10 : 2008-06-30 & 30.3    & 1545, 1581, 1592 & 14\\
	 & 2008-07-25 : 2008-08-14 & 17.5    & 1634 & 14\\
    \hline
  \end{tabular}
\end{center}
  \label{tab:blazar_sample}
\end{table*}

\section{Blazars and AGILE}

AGILE ({\it Astrorivelatore Gamma ad Immagini LEggero}) is an Italian Space Agency (ASI) mission launched
on April 2007 and devoted to high energy astrophysics \cite{Ta}. The AGILE
satellite is capable of observing cosmic sources simultaneously in
X-ray (18--60 keV) and $\gamma$-ray (30 MeV--30 GeV) energy bands with the
coded-mask hard X-ray imager (SuperAGILE) and the Gamma-Ray Imaging Detector
(GRID), respectively. The GRID consists of a Silicon
Tracker, a non-imaging CsI Mini-Calorimeter and a segmented
anticoincidence system. 

Gamma-ray observations of blazars are a key scientific project of the AGILE
satellite. In the last two years, AGILE detected several blazars during high $\gamma$-ray activity and
extensive multiwavelength campaigns were organized for many of them. 
Table 1 shows the list of AGILE flaring blazars observed up to now.
The $\gamma$-ray activity time scale goes from a few days (e.g. S5
0716$+$714 and 3C 273) to several weeks (e.g. 3C 454.3 and PKS 1510--089) and the flux
$\gamma$-ray variability observed has been negligible (e.g. 3C 279), very rapid (e.g. PKS
1510--089) or extremely high (e.g. 3C 454.3 and PKS 1510--089). 

Even if at least one object for each blazar category (LBL, IBL, HBL and FSRQ)
was detected, we note that only a few objects were detected more than once in flaring state by AGILE and only
already known $\gamma$-ray emitting source showed flaring activity. This
evidence together with the early results from the first three months of $Fermi$-LAT $\gamma$-ray all-sky survey (\cite{Ab}) suggest possible constraint on the properties of the most intense
$\gamma$-ray emitters. 
In the following section we will present the most interesting results on multiwavelength
observations of the individual sources detected by AGILE.

\section{Individual Sources}

\subsection{3C 454.3}

3C 454.3 is the blazar which exhibited the most variable activity in the $\gamma$-ray
sky during the last two years. In the period July 2007--January 2009 the AGILE
satellite monitored intensively 3C 454.3 together with $Spitzer$, WEBT, REM, MITSuME,
$Swift$, RXTE, $Suzaku$ and INTEGRAL observatories,
yielding the longest multiwavelength coverage of this $\gamma$-ray quasar so
far (see \cite{V18}). 

AGILE detected 3C 454.3 for the first time during a dedicated Target of Opportunity (ToO)
activated immediately after an extremely bright optical flare in mid July 2007
\cite{Ver08}. The average $\gamma$-ray flux over 6 days of observation was $F_{E>100 MeV}$ = (280 $\pm$ 40) $\times$ 10$^{-8}$ photons cm$^{-2}$
 s$^{-1}$, more than a factor of two higher than the maximum value reported by
 EGRET. Moreover, the peak flux on daily time scale was of the order of 400 $\times$ 10$^{-8}$ ph cm$^{-2}$ s$^{-1}$. Since this
 detection, 3C 454.3 become the most luminous object of the $\gamma$-ray
 sky. 

Subsequently, two multiwavelength campaigns on 3C 454.3 were organized
 during November 2007 and December 2007, as reported in \cite{Ver09} and  \cite{Don09}.
The source underwent an unprecedented long period of very high
$\gamma$-ray activity, showing flux levels variable on short
timescales of 24--48 hours and reaching on daily timescale a $\gamma$-ray flux
of the order of 600$\times$10$^{-8}$ ph cm$^{-2}$ s$^{-1}$ (see Fig. 1). Also the optical flux appears
extremely variable with a brightening of several tenths of magnitude in a few
hours. A correlation analysis between the optical and $\gamma$-ray flux
variations is consistent with a time lag less than one day, as
confirmed also by the analysis of the early $Fermi$-LAT data in \cite{Bn}. As
 shown in Figure 2, the dominant emission mechanism in this source above 100
 MeV seems to be the inverse Compton scattering of relativistic electrons in the jet on the
external photons from the Broad Line Region (BLR), even if the hard $\gamma$-ray
spectrum and the high Compton dominance observed by AGILE in December 2007 seems to
suggest that in some cases also the contribution of external Compton of seed photons from a
 hot corona could be not negligible \cite{Don09}. 

Finally, the results of a extremely long $\gamma$-ray activity period between 10 May 2008 and 12 January 2009 will be
 presented in a forthcoming paper \cite{V18}.

\begin{figure}[bh!]
   \centering
   \includegraphics[width=8cm]{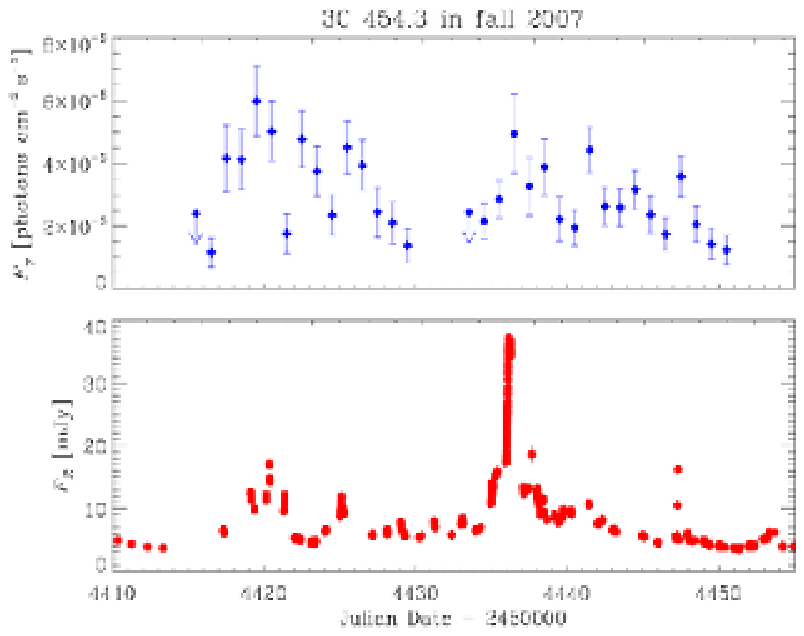}
   \caption{Light curves of 3C 454.3 in $\gamma$-ray (upper panel) and optical
   band (lower panel) acquired during
   November-December 2007 by AGILE and GASP-WEBT, respectively.}
 \end{figure}

\begin{figure}[th!]
   \centering
   \includegraphics[width=8cm]{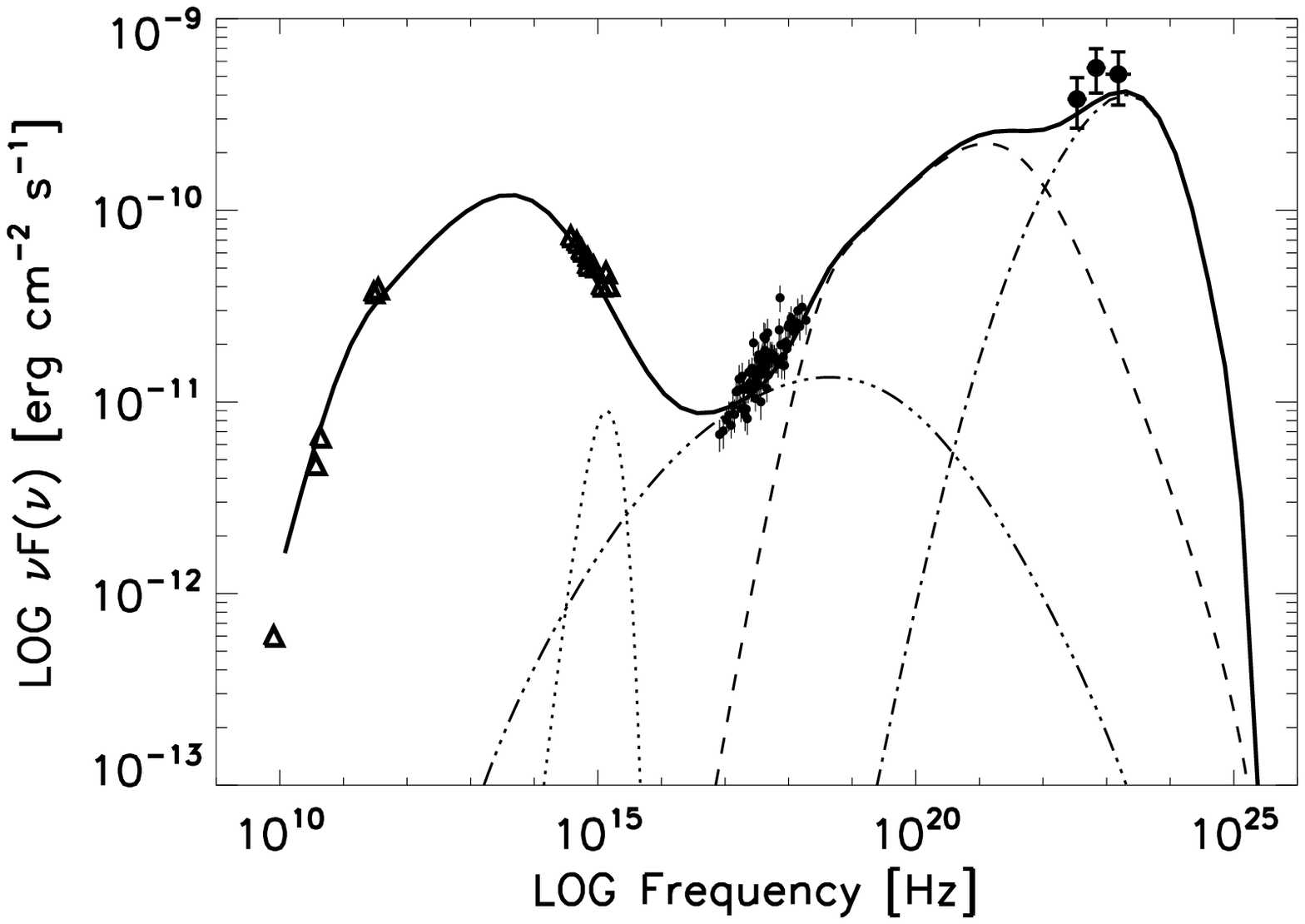}
   \caption{Spectral Energy Distribution of 3C 454.3 for the period 13--16
   November 2007 with AGILE, $Swift$ and GASP-WEBT data. The dotted, dashed,
   dot-dashed, and triple dot-dashed lines represent the accretion disk black
   body, the external Compton on the disk radiation, the external Compton on
   the BLR radiation and the SSC, respectively.}
 \end{figure}
 
\subsection{PKS 1510--089}

PKS 1510--089 showed in the last two years high variability over all the electromagnetic spectrum,
in particular a high $\gamma$-ray activity was observed by AGILE and
$Fermi$. AGILE detected two
intense flaring episodes in August 2007 (\cite{Puc08}) and March 2008 ({\cite{DAm09a}) and an extraordinary
actitivity during March 2009, with several flaring episode and a
flux reaching 500$\times$10$^{-8}$ ph cm$^{-2}$ s$^{-1}$ \cite{DAm09b}. Fig. 3 shows the SED of PKS 1510--089 relative to the rapid $\gamma$-ray flaring
episode of mid March 2008 \cite{DAm09a}. 

\begin{figure}[th!]
   \centering
   \includegraphics[width=8.5cm]{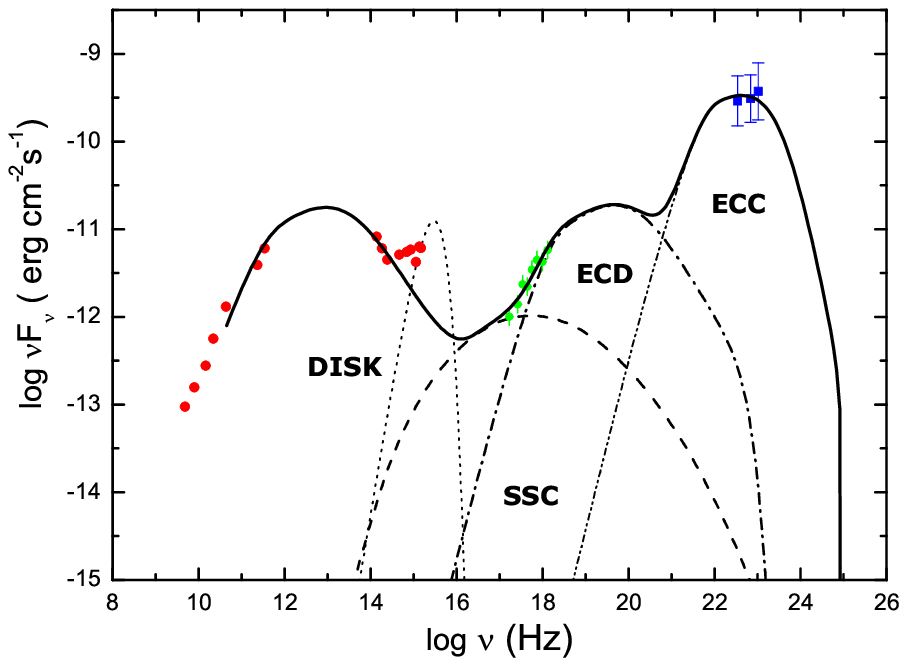}
   \caption{Spectral Energy Distribution of PKS 1510-089 on mid March
   2008 with AGILE, $Swift$ and GASP-WEBT data.The dotted, dashed,
   dot-dashed, and double dot-dashed lines represent the accretion disk black
   body, the SSC, the external Compton on the disk radiation (ECD) and the external Compton on
   the BLR radiation (ECC), respectively.}
 \end{figure}

The
multiwavelength data collected in optical and UV bands by GASP-WEBT and
$Swift$/UVOT not only in March 2008 but during the entire period 2008--2009 seem to indicate
the presence of thermal features quasar-like such as the little blue bump and
the big blue bump in the optical/UV spectrum of this source. Moreover, the $Swift$/XRT observations of mid March
2008 showed in X-ray band a
redder-when-brighter behaviour, already observed in this source by
\cite{Ka}, but not usual in FSRQs.

\subsection{3C 279}

This is the first extragalactic source detected by AGILE, in mid July 2007,
as reported in \cite{Giu09}. The average
$\gamma$-ray flux over 4 days of observation is F$_{E>100 MeV}$ = (210 $\pm$
38)$\times$10$^{-8}$ ph cm$^{-2}$ s$^{-1}$, a flux level similar to the highest
observed by EGRET and $Fermi$. The spectrum of this source during the flaring episode observed by
AGILE is soft ($\Gamma$ = 2.22 $\pm$ 0.23) and this could be an
indication of a low accretion state of the disk occurred some
months before the $\gamma$-ray observations, suggesting a dominant
contribution of the external Compton scattering of direct disk (ECD) radiation
compared to the external Compton scattering of the Broad Line Region
clouds (ECC). As a matter of fact, a strong minimum in the optical band was detected by
REM two
months before the AGILE observations and the reduction of the activity of the disk should
cause the decrease of the photon seed population produced by the disk and
then a deficit of the ECC component with respect to the ECD, an effect delayed
by the light travel time required to the photons to go from the inner disk to the BLR. 

\subsection{3C 273}

3C 273 was detected simultaneously by the GRID
and SuperAGILE detectors onboard AGILE during a pre-planned multiwavelength campaign over three
weeks between mid December 2007 and January 2008, involving also simultaneous REM,
$Swift$, RXTE and INTEGRAL observations. During this campaign, whose results
are reported in \cite{Pac09}, the average flux in the 20--60 energy
band is (23.9 $\pm$ 1.2) mCrab, whereas the source was detected by the GRID only in the second week, with an average flux of F$_{E>100 MeV}$ =
(33 $\pm$ 11)$\times$10$^{-8}$ ph cm$^{-2}$ s$^{-1}$. The analysis of
the light curves seems to indicate no optical variability during the whole campaign and a possible anti-correlation between the
$\gamma$-ray emission and the soft and hard X-rays. The SED is consistent with
a leptonic model where the soft X-ray emission is produced by the combination
of Synchrotron Self Compton (SSC) and External Compton (EC) models, while the hard X-ray and $\gamma$-ray emission is due to
ECD. The spectral variability
between the first and the second week is consistent with the acceleration
episode of the electron population responsible for the synchrotron emission.    

\begin{figure}[th!]
   \centering
   \includegraphics[width=7cm]{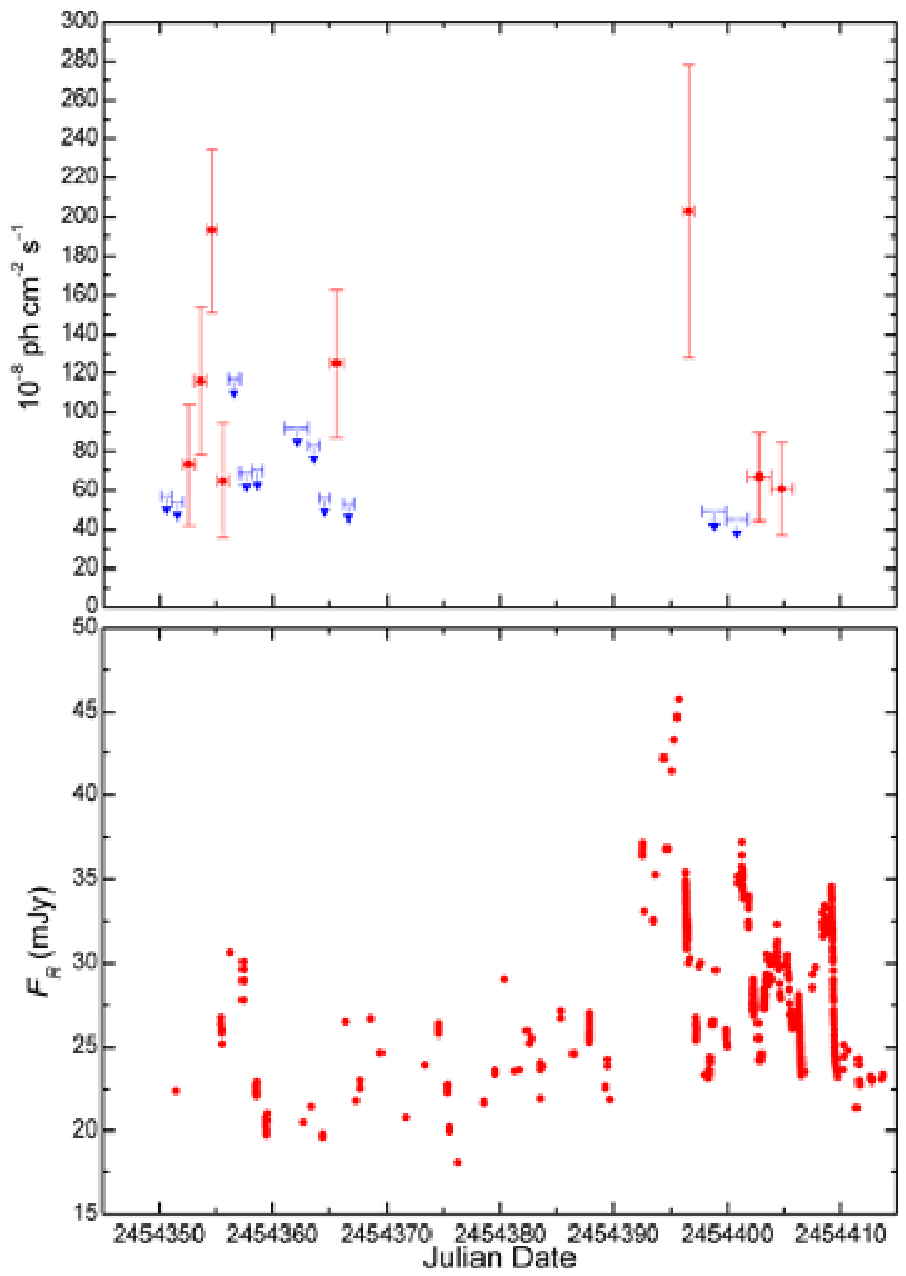}
   \caption{Light curves of S5 0716$+$714 during September-October 2007 in
   $\gamma$-ray (upper panel) and optical band (lower panel) with AGILE and
   GASP-WEBT data, respectively.}
 \end{figure}

\subsection{S5 0716+714}

The intermediate BL Lac (IBL) object S5 0716+714 was observed by AGILE during two
different periods: 4--23 September and 23 October -- 1 November 2007, as
discussed in \cite{Chen}. Between 7 and 12 September 2007 the source showed a high $\gamma$-ray activity with an average
flux of F$_{E>100 MeV}$ = (97 $\pm$ 15)$\times$10$^{-8}$ ph cm$^{-2}$ s$^{-1}$
and a peak level of F$_{E>100 MeV}$ =
(193 $\pm$ 42)$\times$10$^{-8}$ ph cm$^{-2}$ s$^{-1}$, with an increase of flux by
a factor of four in three days (see Fig. 4, upper panel). This source was
detected by EGRET in low/intermediate $\gamma$-ray levels, F$_{E>100 MeV}$ $\simeq$
(20--40)$\times$10$^{-8}$ ph cm$^{-2}$ s$^{-1}$. The flux detected by AGILE
is the highest ever detected from this object and one of the most high flux observed from
a BL Lac object. A simultaneous GASP-WEBT optical campaign started
after the AGILE detection (see Fig. 4, lower panel) and the resulting SED is consistent with a
two-components SSC model (see Fig. 5). Recently \cite{Ni} estimated the
redshift of the source (z = 0.31 $\pm$ 0.08) and this allowed us to calculate the total power
transported in the jet, which results extremely high approaching the maximum
power generated by a spinning black hole of 10$^{9}$ M$_\odot$ through the
pure Blandford and Znajek mechanism \cite {BZ} (see {\cite {Vit09} for a
  detailed discussion). 

\begin{figure}[th!]
   \centering
   \includegraphics[width=8cm]{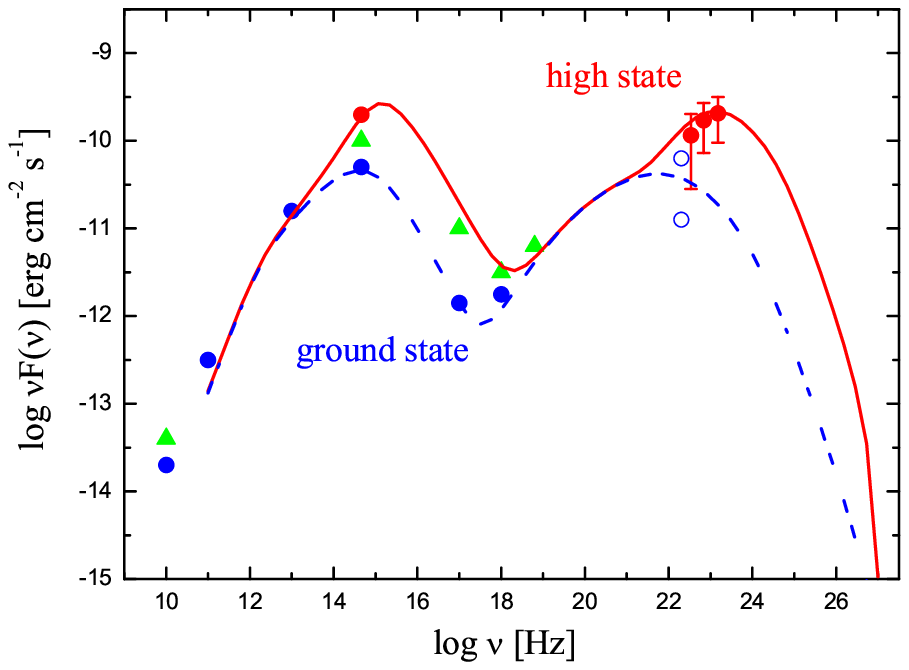}
   \caption{SED of S5 0716$+$714 during mid September
   2007 including optical GASP-WEBT and $\gamma$-ray AGILE data (red dots). Historical
   data relative to a ground state and EGRET data are represented with blue
   dots. Green triangles represent historical data during a high X-ray state.}
 \end{figure}

During October 2007, AGILE detected the source at a flux about a factor of 2
lower than the
September one with no significant variability; simultaneously, $Swift$
observed strong variability in soft X-rays, moderate variability at optical/UV
and approximately constant hard X-ray flux. Also this behaviour is compatible
with the presence of 2 different SSC components in the SED \cite{Gio}.

\subsection{W Comae and Mrk 421}

On 8 June 2008, VERITAS announced the detection of a TeV flare from the IBL
object W Comae
\cite{Sw}. About 24 hours later, AGILE repointed towards the source and
detected it \cite{Ver}. The results of a multiwavelength campaign involving
VERITAS, $Swift$, XMM-$Newton$ and AGILE will appear in a forthcoming paper \cite{Ac09}. 

During the ToO towards W Comae, AGILE
detected also the HBL object Mrk 421. SuperAGILE detected a fast increase of flux
from Mrk 421 up to 40 mCrab in the 15--50 energy band, about a factor of 10 higher
than its typical flux in quiescence. The $\gamma$-ray flux detected by GRID, F$_{E>100 MeV}$ =
(42 $\pm$ 13)$\times$10$^{-8}$ ph cm$^{-2}$ s$^{-1}$, is about a factor
3 higher than the average EGRET value, even if consistent with its maximum. 

An extensive multiwavelength campaign from optical to TeV energy bands
was organized with the participation of WEBT, $Swift$, RXTE, AGILE, MAGIC and
VERITAS. The comparison of the light curves show a possible correlated variability between
optical, X-rays and the high-energy part of the spectrum. The SED can be interpreted
within the framework of the SSC model in terms of a rapid acceleration of
leptons in the jet. 
An alternative more complex scenario is that optical and X-ray emissions come from different
regions of the jet, with the inner jet region that produces X-rays and is partially
transparent to the optical radiation, whereas the outer region produces only
the lower-frequency emission.

\begin{acknowledgments}
FD would like to thank the organizers for a stimulating meeting.
The AGILE Mission is funded by the ASI with scientific and progammatic partecipation by the
Italian Institute of Astrophysics (INAF) and the Italian Institute of Nuclear
Physics (INFN).
\end{acknowledgments}


\newpage

\lhead[\fancyplain{}{}]{\itshape \small{The power from BL Lacs }}
\rhead[\itshape \large{1$^\mathrm{st}$ RYRM Proceedings}]{\fancyplain{}{A. Paggi}}
%
%
%
%
%
%
%




\begin{center}


\setcounter{section}{4}
\section{\large{The Power form BL Lacs}}
\label{paggi}

\normalsize{Alessandro Paggi$^*$}

\emph{\small{Dipartimento di Fisica, Universit\`{a} di Roma Tor Vergata\\ Via della Ricerca scientifica 1, I-00133 Roma, Italy}}




\vspace{.5cm}

\begin{minipage}{.8\textwidth}
\small{Blazars are among the brightest Active Galactic Nuclei (AGNs) observed, with inferred (isotropic) luminosities up to \(L_{iso} \sim {10}^{47} \mbox{ erg}
 \mbox{ s}^{-1}\). Emission from these sources is widely held to originate from a relativistic jet closely aligned to the observer line of sight, with bulk Lorentz factor \(\Gamma \sim 10 \div 20\). Taking into account related relativistic effects the intrinsic jet powers are actually much lower, \(L \sim \Gamma^{-2} L_{iso}\). A physical reference is provided by the power extractable from a central rotating Kerr Hole spun up by past accretion via the electromagnetic interaction with the magnetic field of the accretion disc, the Blandford-Znajek (BZ) mechanism. The limiting power reads \(L_{BZ} \sim 2\times {10}^{45} \left({M_{BH}/{10}^9\,M_{\astrosun}}\right) \mbox{ erg}\mbox{ s}^{-1}\) in an equilibrium magnetic field \(B \sim {10}^4 \mbox{ G}\).

We study in particular gas-poor BL Lacs, which show no evidence of current accretion or well developed accretion disk, and can therefore provide a simple benchmark for this model. A study of several such BL Lacs across the entire electromagnetic spectrum allows robust estimates of the source parameters and energetics, showing that these sources comply with the BZ limit or just exceed it, within the uncertainties on black hole (BH) masses estimation.

In particular, powerful BL Lacs like 0716+714 close to the BZ limit appear to be constrained in the evolution of their flaring activity. It is therefore exciting to look for sources sharply exceeding the BZ limit. If any, they could involve orbits plunging into the BH horizon in a region with very intense magnetic field \(B > {10}^4 \mbox{ G}\) and dominated by strong gravity effects.}

\vspace{.25cm}
$*$ email: \href{mailto:alessandro.paggi@roma2.infn.it}{alessandro.paggi@roma2.infn.it}
\end{minipage}


\end{center}

\setcounter {section} {0}
\setcounter {subsection} {0}
\setcounter {figure} {0}
\setcounter{equation}{0}

\section{Introduction}

Blazars rank among the brightest Active Galactic Nuclei (AGNs) on the basis of their luminosities up to \(L_{iso}\sim {10}^{48}\mbox{ erg}\mbox{ s}^{-1}\) inferred from isotropic flux distribution. Actually, these sources are widely held to be radiating from a relativistic jet closely aligned to the observer line of sight; this emits highly beamed non-thermal radiations, with observed fluxes enhanced by relativistic effects so as to match or overwhelm other, thermal or reprocessed emissions into wider solid angles. Thus, due to special relativity effects in the emitting jets, the luminosities \(L_{iso}\) are far larger than the intrinsic radiated powers \(L_r\); for small viewing angles one easily has \(L_r\approx {10}^{-3} L_{iso}\).

Among Blazars, BL Lacs are in particular those showing no or just weak emission lines. Their spectra may be represented as a continuous spectral energy distribution (SED) \(S_\nu=\nu F_\nu\) marked by two peaks: the one at lower frequency is widely interpreted as synchrotron emission by highly relativistic electrons in the jet; the higher frequency counterpart is believed to be inverse Compton (IC) upscattering by the same electron population of the seed photons either emitted by source external to the jet (external Compton, EC \cite{sikora}), or provided by the synchrotron emission itself (synchrotron-self Compton, SSC \cite{marscher}, \cite{maraschi}). BL Lac Objects also show rapid variability on timescales of days or shorter, during which they undergo strong flux variations often named ``flares'', with intrinsic timescales longer than the apparent ones again by relativistic effects.

BL Lacs are conveniently classified in terms of the frequency of their synchrotron peak \cite{padovani}: for the low frequency BL Lacs (LBLs) the peak lies in the infrared-optical bands, while the IC component peaks at MeV photon energies; instead, high frequency BL Lacs (HBLs) feature a first peak in the X-ray band and the second one at energies of about \({10}^2\mbox{ GeV}\) or beyond.

Here we focus on ``dry" BL Lacs, that is, sources with no (evidence of) surrounding gas and related current accretion. They provide an appropriate benchmark for comparing their intrinsic luminosity with the top power extractable from a maximally rotating supermassive black hole (SMBH) via the electrodynamic Blandford-Znajek (BZ) mechanism \cite{bz}, \cite{livio}, \cite{ghosh}, \cite{cavaliere}; this can emit up to \(L_{BZ} \leq 2\times{10}^{45}\left({{M_{\newmoon}}/{{10}^9\,M_{\astrosun}}}\right)\mbox{ erg}\mbox{ s}^{-1}\), given of a SMBH mass \(M_{\newmoon}\) and a magnetic field  \(B\sim{10}^4\mbox{ G}\) threading the hole horizon and held by the kinetic or radiation pressure in the disk.

\section{SSC model}

To account for the BL Lac SEDs we adopt the simple homogeneous SSC model, assuming the radiations to be produced in
a region containing relativistic electrons in a magnetic field, moving toward the observer with bulk Lorentz
factor \(\Gamma\); they emit primary synchrotron photons, and radiate secondary ones upon IC scattering of the former.

We begin with representing the source by a spherical homogeneous topology of radius \(R\), containing non-relativistic protons and highly relativistic electrons at uniform density \(n\), sharing a bulk Lorentz factor \(\Gamma\sim 10\div 20\); the electrons, however, are given random Lorentz factors \(\gamma\) up to  \({10}^6\).

To match the observed BL Lac spectra we make use of log-parabolic electron energy distributions; such distributions are obtained in the presence of stochastic adding to systematic acceleration processes as shown by \cite{kardashev} and computed in detail by \cite{paggi}. So we write for the particle differential energy distribution
\begin{equation}
N(\gamma)=N_0\,
{\left({\frac{\gamma}{\gamma_0}}\right)}^{-s-r\log{\left({\frac{\gamma}{\gamma_0}}\right)}}\, ,
\end{equation}
where \(s\) is the constant contribution to the slope, \(r\) is the distribution curvature and \(\gamma_0\) is the initial injection energy.
The energetic content of the electron population can be expressed in terms of the root mean square (hereafter rms) Lorentz factor
\begin{equation}
\gamma_p=\sqrt{\frac{\int{\gamma^2 N{\left({\gamma}\right)}\, d\gamma}}{\int{N{\left({\gamma}\right)}\, d\gamma}}}\, ,
\end{equation}
while \(\left\langle{\gamma}\right\rangle=\gamma_p\,{10}^{-1/4r}\) is the distribution mean energy.

The emitted synchrotron SED will correspondingly read \cite{massaro2}
\begin{equation}
S_\nu=S_0
\,{\left({\frac{\nu}{\nu_0}}\right)}^{-a-b\log{\left({\frac{\nu}{\nu_0}}\right)}}\, ,
\end{equation}
with a constant contribution \(a=(s-3)/2\) to the spectral index, a spectral curvature \(b\approx r/5\), and a peak frequency \(\xi\propto B\gamma_p^2\). For the IC an analogous SED holds: in the Thomson regime one has \(a=(s-3)/2\), \(b\approx r/10\) and a peak frequency \(\epsilon\propto B\gamma_p^4\); in the Klein-Nishina (KN) regime one has \(a=s\), \(b\approx r\) and a peak frequency \(\epsilon\propto \gamma_p\) \cite{paggi}. Proton radiations will be quite lower due to the much smaller cross section and lower kinetic energies, and therefore will be neglected here.

Note that observed (primed) frequencies and fluxes are related to rest frame (unprimed) quantities via  \(\nu'=\nu\,\delta\) and \(F'= F\, \delta^4\) \cite{bbr}, where \(\delta={\left[{\Gamma\left({1-\beta\cos{\theta}}\right)}\right]}^{-1}\) is the beaming factor, \(\theta\) being the angle between the jet and the line of sight. For small viewing angles \(\theta\sim 1/\Gamma\), one obtains \(\delta\approx 2 \Gamma\).

We focus on the flaring activity of dry BL Lacs for which sufficient, multi-wavelength data are published, namely 0716+714 in 2009 \cite{vittorini}, Mrk501 in 1997 \cite{massaro}, Mrk421 in 2000 \cite{konopleko} and 2008 \cite{donnarumma} (see Fig. \ref{seds}).

\begin{figure}
\includegraphics[scale=0.3]{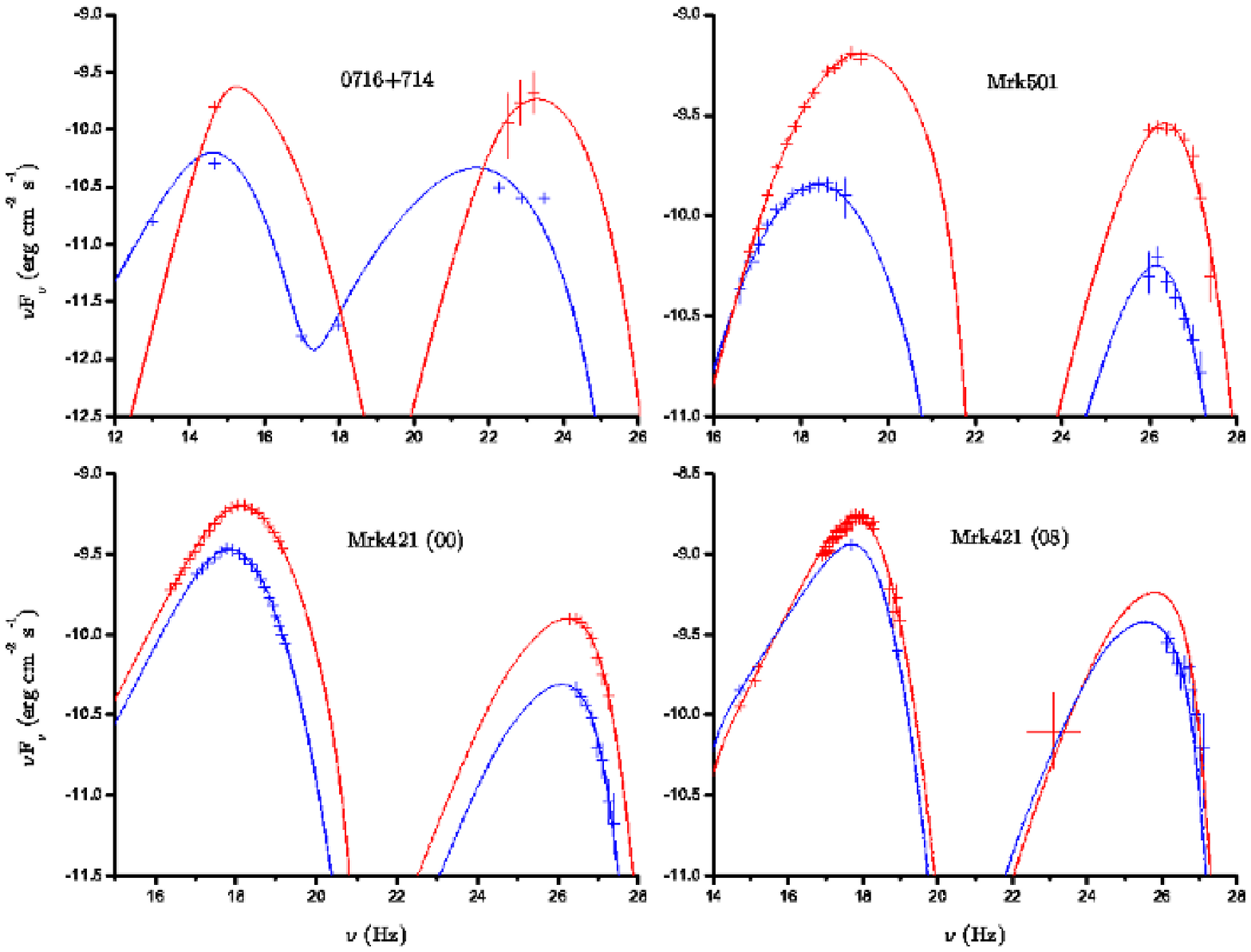}
  \caption{SEDs of dry BL Lacs described in the text: 0716+174 (upper-left frame), Mrk501 (upper-right frame) and Mrk421 (lower-left and lower-right frames), each in a low (blue line) and high (red line) state.}
\label{seds}
\end{figure}

As showed by \cite{paggi}, an homogeneous single zone SSC model can be fully constrained by five key observational quantities: synchrotron peak frequency \(\xi\) and flux \(S\), IC peak frequency \(\epsilon\) and flux \(C\), and the observed variation time \(\Delta t\). Then one obtains in the Thomson regime
\begin{equation}\label{prima}
{\gamma_p}\propto
{\epsilon}^{\frac{1}{2}}{\xi}^{-\frac{1}{2}}
\end{equation}
\begin{equation}
{R}\propto
{\epsilon}^{\frac{1}{2}}\,{S}^{\frac{1}{2}}\,{\Delta t}^{\frac{1}{2}}\,{\xi}^{-1}\,{C}^{-\frac{1}{4}}
\end{equation}
\begin{equation}
{B}\propto
{\xi}^{3}\,{C}^{\frac{1}{4}}\,{\Delta t}^{\frac{1}{2}}\,{\epsilon}^{-\frac{3}{2}}\,{S}^{-\frac{1}{2}}
\end{equation}
\begin{equation}
{\delta}\propto
{\epsilon}^{\frac{1}{2}}\,{S}^{\frac{1}{2}}{\xi}^{-1}\,{C}^{-\frac{1}{4}}\,{\Delta t}^{-\frac{1}{2}}
\end{equation}
\begin{equation}
{n}\propto
{\xi}^{2}\,{C}^{\frac{5}{4}}\, {\epsilon}^{-\frac{3}{2}}\,{S}^{-\frac{3}{2}}\,{\Delta t}^{-\frac{1}{2}}\, ,
\end{equation}
while for the extreme KN regime we have
\begin{equation}
{\gamma_p}\propto
{\xi}^{\frac{3}{5}}\,{\epsilon}^{\frac{3}{5}}\,{C}^{\frac{1}{5}}\,{\Delta t}^{\frac{2}{5}}{S}^{-\frac{2}{5}}
\end{equation}
\begin{equation}
{R}\propto
{\epsilon}^{\frac{2}{5}}\,{S}^{\frac{2}{5}}\,{\Delta t}^{\frac{3}{5}}\,{\xi}^{-\frac{3}{5}}\,{C}^{-\frac{1}{5}}
\end{equation}
\begin{equation}
{B}\propto
{\xi}^{\frac{2}{5}}\,{S}^{\frac{2}{5}}{\epsilon}^{-\frac{8}{5}}\,{C}^{-\frac{1}{5}}\,{\Delta t}^{-\frac{2}{5}}
\end{equation}
\begin{equation}
{\delta}\propto
{\epsilon}^{\frac{2}{5}}\,{S}^{\frac{2}{5}}\,{\xi}^{-\frac{3}{5}}\,{C}^{-\frac{1}{5}}\,{\Delta t}^{-\frac{2}{5}}
\end{equation}
\begin{equation}\label{ultima}
{n}\propto
{\xi}^{\frac{11}{5}}\,{C}^{\frac{7}{5}}\,{\epsilon}^{-\frac{4}{5}}\,{S}^{-\frac{9}{5}}\,{\Delta t}^{-\frac{1}{5}}\,.
\end{equation}

It is possible to evaluate a threshold for the synchrotron peak frequency when the transition between the two IC regimes occurs, that is, for example,
\begin{equation}
\xi_T\approx 7.15\times{10}^{15}\,{\left({\frac{B}{0.1\mbox{ G}}}\right)}^{\frac{1}{3}}\,
{\left({\frac{\delta}{10}}\right)}\,(1+z)^{-1}\mbox{ Hz}\, .
\end{equation}
So, in LBL sources the majority of photons of IC peak are upscattered in the Thomson regime, while in HBL sources the majority of photons of IC peak are upscattered in the KN regime, as also supported by the more curved IC spectra with respect to the synchrotron ones observed in HBL sources (see Fig. \ref{seds}).

In \cite{paggi} it is also shown that spectra from a single electron population are expected to irreversibly broaden under the action of stochastic acceleration process, after \(b\propto 1/t\); so, a sudden increase in the spectral curvature clearly marks the emergence of a second electron population, as may be the case for 0716+714 \cite{vittorini}.

\section{Jet energetics}

We denote by \(L_{iso}= 4\pi\, D_L^2\, F_{obs}\) the luminosity inferred from assuming an isotropic distribution of the flux \(F_{obs}\) observed at the luminosity distance \(D_L\).

We are instead interested in the intrinsic luminosities in the jet frame; assuming one cold proton per electron (satisfying \(\left\langle{\gamma}\right\rangle \leq m_p / m_e\)), we follow \cite{celotti} in writing for the radiative luminosity and for the related powers transported by the jet
\begin{eqnarray}
\label{lumprima}L_r &=& L_{iso}\,\frac{\Gamma^2}{\delta^4}\,,\\
L_e &=& \frac{4}{3}\pi\,R^2\,c\,n\,m_e\,c^2\left\langle{\gamma}\right\rangle\Gamma^2\,,\\
L_p &=& \frac{4}{3}\pi\,R^2\,c\,n\,m_p\,c^2\,\Gamma^2\,,\\
\label{lumultima}L_B &=& \frac{1}{6}\,R^2\,\,c\,B^2\,\Gamma^2\,.
\end{eqnarray}
Here \(L_r\) is the radiated luminosity contributed by synchrotron and IC radiation, \(n\) is the particle number density, \(m_e\) and \(m_p\) are electron and proton masses, respectively. The total jet luminosity will therefore be \(L_{tot}=L_r+L_e+L_p+L_B\), but is dominated by \(L_r\) and next by \(L_e\), with \(L_B < L_p \leq L_e\).

So, in order to evaluate luminosities robust estimates of source parameters are required. These can be achieved with simultaneous, multi-wavelength observations that allow solid spectral reconstructions and evaluation of five the key parameter (beside spectral curvature \(b\)), namely synchrotron peak frequency and flux, IC peak frequency and flux, and variation time; these five parameters can be used in Eqs. \ref{prima} - \ref{ultima} to obtain the five source parameters \(n\), \(R\), \(B\), \(\gamma_p\) (and therefore \(\left\langle{\gamma}\right\rangle={10}^{-1/4r}\), \(r\approx 5b\)) and \(\delta\) (and therefore \(\Gamma\) on assuming a viewing angle of a few degrees).

\section{The Blandford-Znajek powerhouse}

Intrinsic luminosities evaluated via Eqs. \ref{lumprima} - \ref{lumultima} for our sources turns out to top some \({10}^{45}\mbox{ erg}\mbox{ s}^{-1}\); moreover the absence of gas evidences for these sources points to very small current accretion rates \(\dot{m}<{10}^{-2}\) in Eddington units.

Thus an enticing benchmark for these luminosities is provided by the BZ process for extracting energy electrodynamically from a rotating Kerr hole. This yields
\begin{equation}\label{BZ}
L_{BZ} = 2\times{10}^{45}\left({\frac{M_{\newmoon}}{{10}^9\,M_{\astrosun}}}\right)\mbox{ erg}\mbox{ s}^{-1}
\end{equation}
for a hole spun up to maximal rotation by past accretion episodes, with horizon threaded by a poloidal magnetic field produced in the inner disk, and held after \(B^2 /4\pi \leq p_{disk}\) by kinetic or radiation pressure in the disk to yield \(B\sim {10}^4\mbox{ G}\).

The hole mass constitutes a key parameter in Eq. \ref{BZ}; to estimate it we make use of its correlation with the red luminosity of the host galactic bulge \cite{ferrarese}, \cite{gebhardt}, \cite{falomo}, that reads
\begin{equation}\label{falomo}
\log{\left({\frac{M_{\newmoon}}{M_{\astrosun}}}\right)} = -0.5\,M_R -3.0\, ,
\end{equation}
with a scatter of \(\pm 0.5\mbox{ dex}\). For the host galaxy of 0716+714 measurements of the bulge magnitude at \(M_R=18.3\pm 0.5\)  have been recently reported by Nilsson et al. \cite{nilsson}; these yield a mass \(M_{\newmoon}\simeq {5.5 ^{+9.9}_{-3.5}}\times{10}^8\,M_{\astrosun}\); such values are also supported by micro-variability of the optical flux \cite{sasada} \cite{gupta}.

\section{Flaring patterns}

Flaring activity of our BL Lacs is reported in Fig. \ref{seq}. In flares, all sources increase their electron r.m.s. energy (and so their peak frequency) as well as their luminosity, but the powerful source 0716+714 is apparently constrained to move sideways, so as to skim the BZ limit.

\begin{figure}
\includegraphics[scale=0.3]{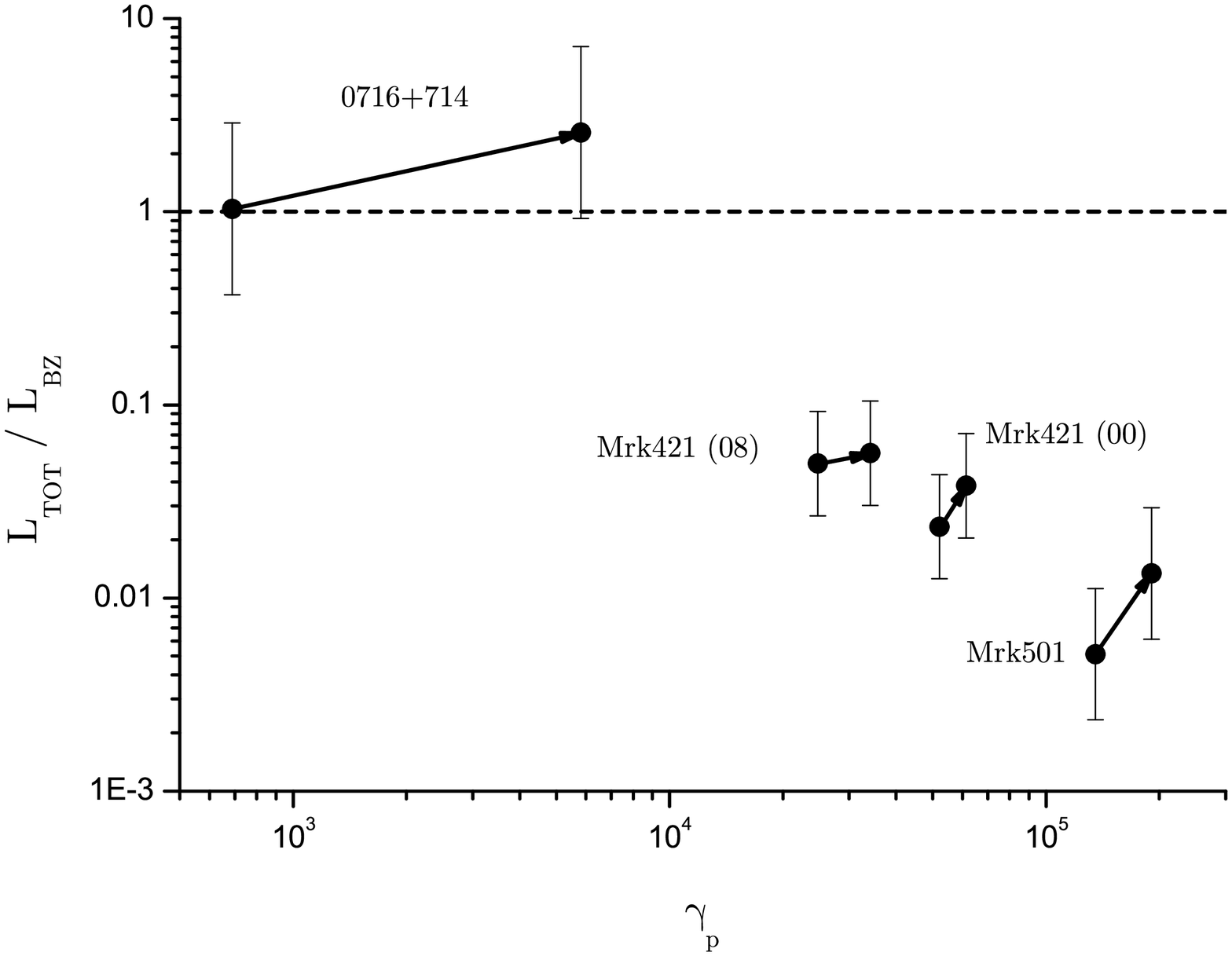}
  \caption{For the named sources total jet luminosities (normalized to BZ luminosity) are plotted against the electron rms energy. Bars represent hole mass uncertainties reflecting those in bulge luminosities of the host galaxies and the scatter in the relation Eq. \ref{falomo}.}\label{seq}
\end{figure}

We note that during flare activity the sources move in the \(L_{tot}\) - \(\gamma_p\) plane to higher \(\gamma_p\) and \(L_{tot}\) almost perpendicularly to the locus of the bright BL Lacs (see Fig. \ref{bs}); this indicates that the flares are not due to disk activity, but rather to emitting electrons acceleration.

\begin{figure}
\includegraphics[scale=0.3]{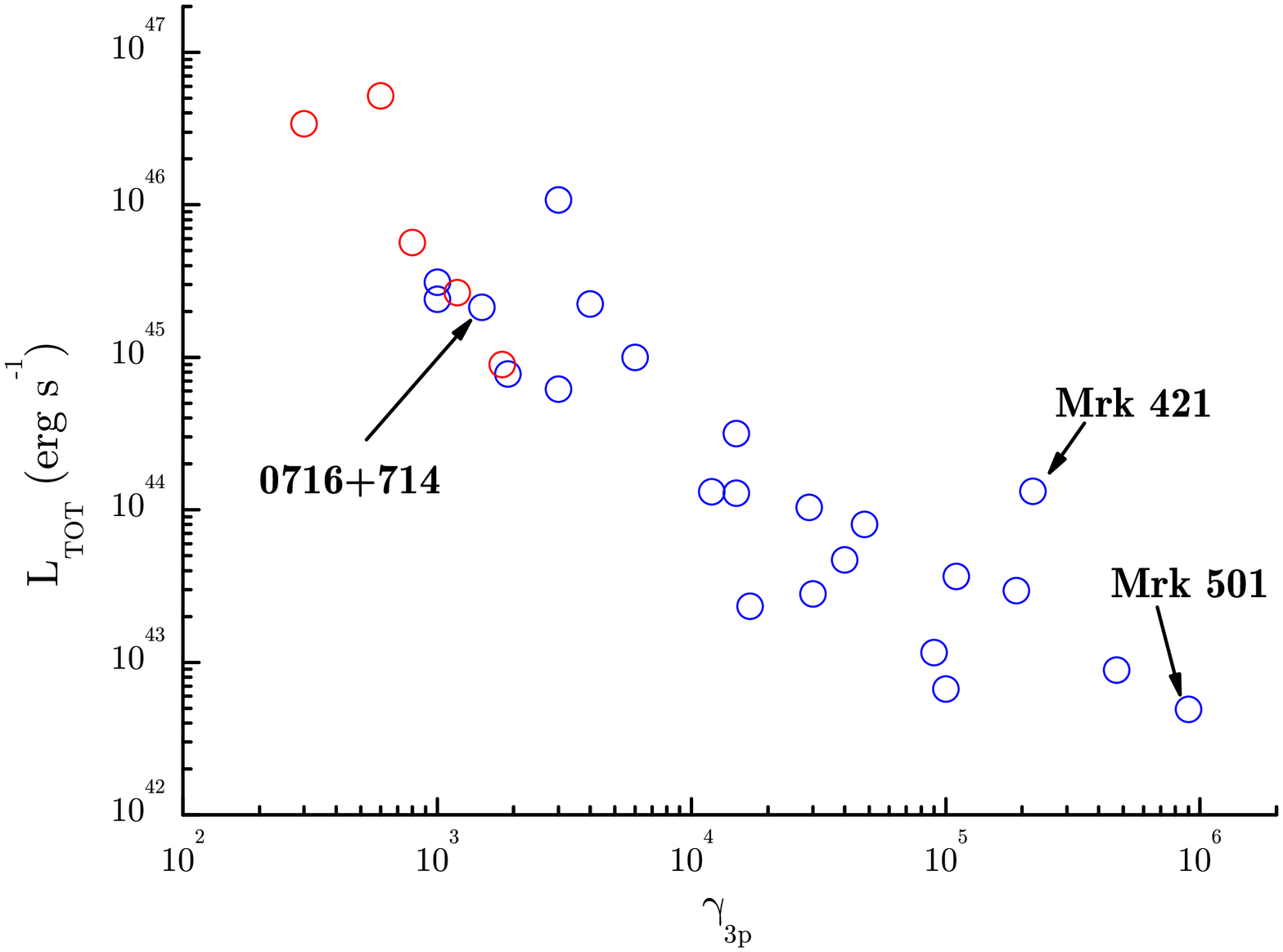}
  \caption{Bright Blazars (data from \cite{celotti}). \(\gamma_{3p}\propto\gamma_p\) is the Lorentz factor of the electrons responsible for synchrotron peak emission; blue circles indicates ``dry" BL Lacs (with no evidence of surrounding gas), while red circles indicates ``wet" BL Lacs (showing gas evidence).}
\label{bs}
\end{figure}

Moreover, during their flaring activity, sources move into the region of faster radiative cooling; this implies short-lived flares with timescales \(\sim 1\mbox{ day}\), or requires more structured jets than the homogeneous SSC model, e.g., decelerating jets \cite{gk}, spine-sheath layer jet \cite{tavecchio2008} or jets in a jet \cite{giannios}.

\section{Conclusions}

The SSC radiation model provides a robust evaluation of the jet luminosity for 0716+714 and for other ``dry" BL Lacs (\(\dot{m}<{10}^{-2}\)). During the observed flare activity this appears to be effectively constrained by the power extractable from the central rotating black hole after BZ process.

We add that higher luminosities may be provided in ``wet" Blazars by the disk contribution enhanced in the Blandford-Payne regime (\(\dot{m}\sim{10}^{-1}\)) \cite{bp}; this, however, involves substantial current accretion and would imply related gas evidence, that is missing in 0716+714. The next step in this sequence, as extrapolated from Fig. \ref{bs}, is represented by FSRQs with Eddington-accreting disk with \(\dot m \sim 1\).

If the benchmark will be found to be substantially exceeded, and intrinsic luminosities \(L> {10}^{46}\mbox{ erg}\mbox{ s}^{-1}\) detected in dry BL Lacs (especially at higher redshifts), with \(M_{BH}< 10^9 M_{\astrosun}\) in Eq. \ref{BZ}, they will require \(B > {10}^4\mbox{ G}\) at the BH horizon; these fields may relate to the larger dynamical stresses up to \(B^2/4 \pi  \leq\rho c^2\) associated with orbits plunging from the disk toward the BH horizon \cite{meier} into a region dominated by strong gravity effects; these conditions will provide a powerful test for GR at work. All that will represent exciting test ground for available \emph{FERMI} data.


%

\newpage

\lhead[\fancyplain{}{}]{\itshape \small{Studying the X-ray/UV Variability of Active Galactic Nuclei with data from Swift and XMM archives}}
\rhead[\itshape \large{1$^\mathrm{st}$ RYRM Proceedings}]{\fancyplain{}{S. Turriziani}}


\begin{center}


\setcounter{section}{5}
\section{\large{Studying the X-ray/UV Variability of Active Galactic Nuclei with data from Swift and XMM archives}}
\label{turriziani}

\normalsize{Sara Turriziani$^*$}
 
 \emph{\small{Dipartimento di Fisica, Universit\`a di Roma Tor Vergata\\
Via della Ricerca Scientifica n. 1, I-00133 Roma, Italy \\
ASI Science Data Center, ASDC c/o ESA-ESRIN, Frascati}}
 

\vspace{.5cm}

\begin{minipage}{.8\textwidth}
\small{Many efforts have been made in understanding the underlying origin of variability in Active Galactic Nuclei (AGN), but at present they could give still no conclusive answers. Since a deeper knowledge of variability will enable to understand better the accretion process onto supermassive black holes, here we present preliminary results of the first \textit{ensemble} struction function analysis of the X-ray variability of samples of quasars with data from Swift and XMM-Newton archives. Moreover, it is known that UV and X-ray luminosities of quasars are correlated and recent studies quantified this relation across 5 orders of magnitude. In this context, we present here some preliminary results on the X-ray/UV ratio from simultaneous observations in UV and X-ray bands of a sample of quasars with data from XMM-Newton archive.}

\vspace{.25cm}
$*$ e-mail: \href{mailto:Sara.Turriziani@roma2.infn.it}{Sara.Turriziani@roma2.infn.it}

\end{minipage}

\end{center}

\setcounter {section} {0}
\setcounter {subsection} {0}
\setcounter {figure} {0}
\setcounter{equation}{0}

\section{Introduction}

Active Galactic Nuclei (AGNs) show flux variations over the entire electromagnetic spectrum. Indeed, variability was one of the first recognized properties of quasars. The variations appear to be aperiodic and have variable amplitude. Although variability plays a key role in constraining the size of the central engine of AGNs, its physical origin remains substantially unknown.  Many mechanisms have been proposed to explain optical observations, such as supernova explosions, star capture, gravitational microlensing or disk instabilities. Some indication on the nature of variability can be obtained from the analysis of the power spectrum, or the structure function, of single-band lightcurves. Besides the study of individual lightcurves, \textit{ensemble} properties of statistical quasar (QSO) samples can provide further constrains on the origin of variability. 

Rapid X-ray variability is a hallmark of AGNs. X-ray short time scale ($10^3-10^5$ s) variability provides 
evidence that the emission comes from a compact region around the central supermassive black hole. 
Whereas theoretical studies \cite{Shak73} provide an explanation of the optical-UV radiation from a steady, optically thick, accretion disk, they cannot explain the X-ray emission. In 
recent years consensus has grown on a standard scenario where UV thermal photons from the disk 
are Comptonized by hot (T $ \sim $ 100-200 keV) electrons in an optically thin corona. This scenario 
accounts, to a first approximation, for the observed power-law spectrum with a high energy cut-off, and it can also produce the so-called Compton reflection component, including emission lines, with different assumptions about the geometry of the X-ray emitting corona. However, a self-consistent theory of X-ray emission is still missing, and even the global distribution of the radiated 
energy between X-rays and optical bands, as a function of the total emitted power,  is still a subject of debate. This is because the  origin of the hot corona and its geometry is far from clear. Magnetic 
flares, clumpy disks and aborted jets are among the suggested mechanisms to heat the corona (see e.~g.~\cite{Matt2007}. All of these mechanisms are associated with variable and clumpy structures, thus  variability itself may provide clues to identify the most appropriate model. 

 In the optical bands the \textit{ensemble} analysis on large optical samples (25,000 objects) was made possible by the Sloan Digital Sky Survey (SDSS), and provided a characterization of the dependence of optical variability on luminosity, redshift, wavelength and time delay  \cite{VB04}, \cite{devries2005}. A similar analysis has not yet been performed in the X-ray, and now becomes possible thanks to the relatively wide field-of-view of typical X-ray instrumentation, such as those on-board XMM-Newton and Swift satellites, to retrieve field data from individual pointed observations. Two available databases are suitable for this analysis, i.e. the Second XMM-Newton serendipitous source catalogue (XMMSSC) \cite{Watson2009}, and the Swift XRT Serendipitous Source Catalog \cite{Puccetti2009}.  The former is limited, by orbital constraints, to long time scale (several months) variations and suffers for rather sparse sampling.  On the contrary the Swift database provides a sampling at intermediate time-scales (hours to  a few months) for a number of objects sufficient to calculate an \textit{ensemble} SF. To build an {\textit{ensemble} SF of the AGNs, it is necessary to ascertain the AGN nature of the X-ray serendipitous sources, excluding possible X-ray emitting stars or galaxies, and to know the redshifts of the sources in order to group all the individual flux variations of different objects in bins of rest-frame time lag. Here we present preliminary results of an X-ray variability analysis of two samples of quasars with optical spectra in the SDSS: we used archival data from the Swift and XMM serendipitous source catalogs to perform a study of the \textit{ensemble} X-ray variability. We used a structure function (SF) analysis to express a curve of growth of variability with time lag. The index of the power law portion of the SF contains important information on the variability mechanism and could be used to put constraints on emission models. We also merged the data from the two dataset to obtain a combined Swift-XMM SF \cite[see][for details]{Vagnetti09b}.

Moreover, it is known that UV and X-ray luminosities of quasars are correlated 
and recent studies quantified this relation across 5 orders of magnitude (e.~g.~\cite{stra2005} and \cite{Gibson2008}).
Such studies inform ongoing efforts to understand the structure and the physics
of quasars nuclear regions, providing constraints on models of physical associations 
between UV and X-ray emissions. Because UV photons are generally
thought to be radiated from the accretion disk whereas X-rays are produced in the
disk corona, the UV/X-ray luminosity relation is an indication of the balance
between accretion disks and their coronae. In this context, we present here some
preliminary results on the X-ray/UV relation from simultaneous observations
in UV and X-ray bands of a sample of quasars with data from XMM-Newton archive \cite[see][for details]{Vagnetti09a}.

\section{X-ray variability}

\begin{figure}
\includegraphics[width=9.0cm, angle=0]{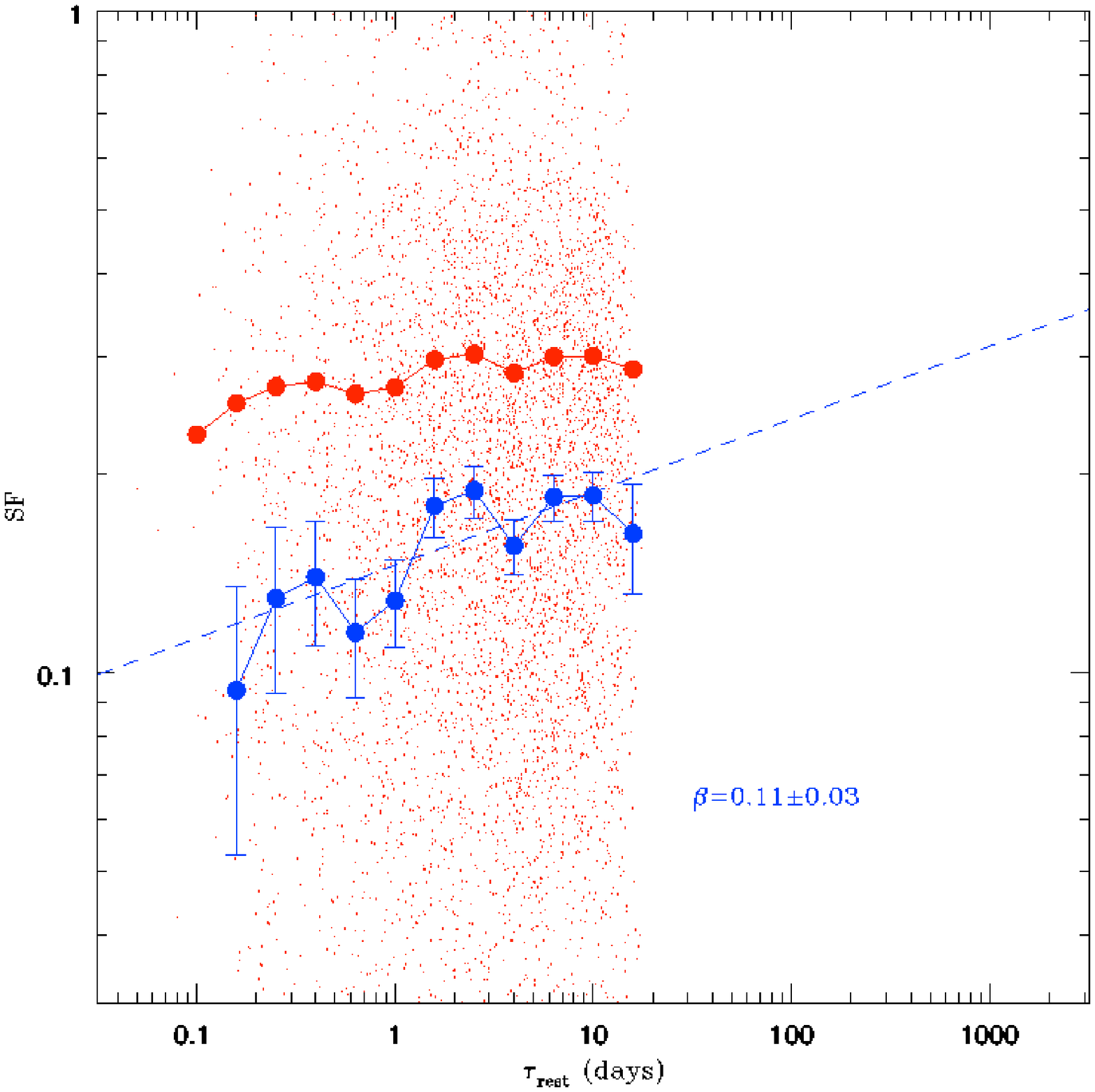}
\caption{\label{fig:sfswift}  \textit{Ensemble} Structure Function (SF), in the rest-frame, for the Swift Sample.  The Figure shows both the noise subtracted SF (blue line) and noise unsubtracted (red line) SF. The photometric noise dominates the first bin and was subtracted in quadrature \cite{diclem96} to obtain the noise subtracted (blue line) SF. The red dots are the contributions to the SF from the individual source lightcurves. The dashed line is the linear fit of the SF.}
\end{figure}

\begin{figure}
\includegraphics[width=9.0cm, angle=0]{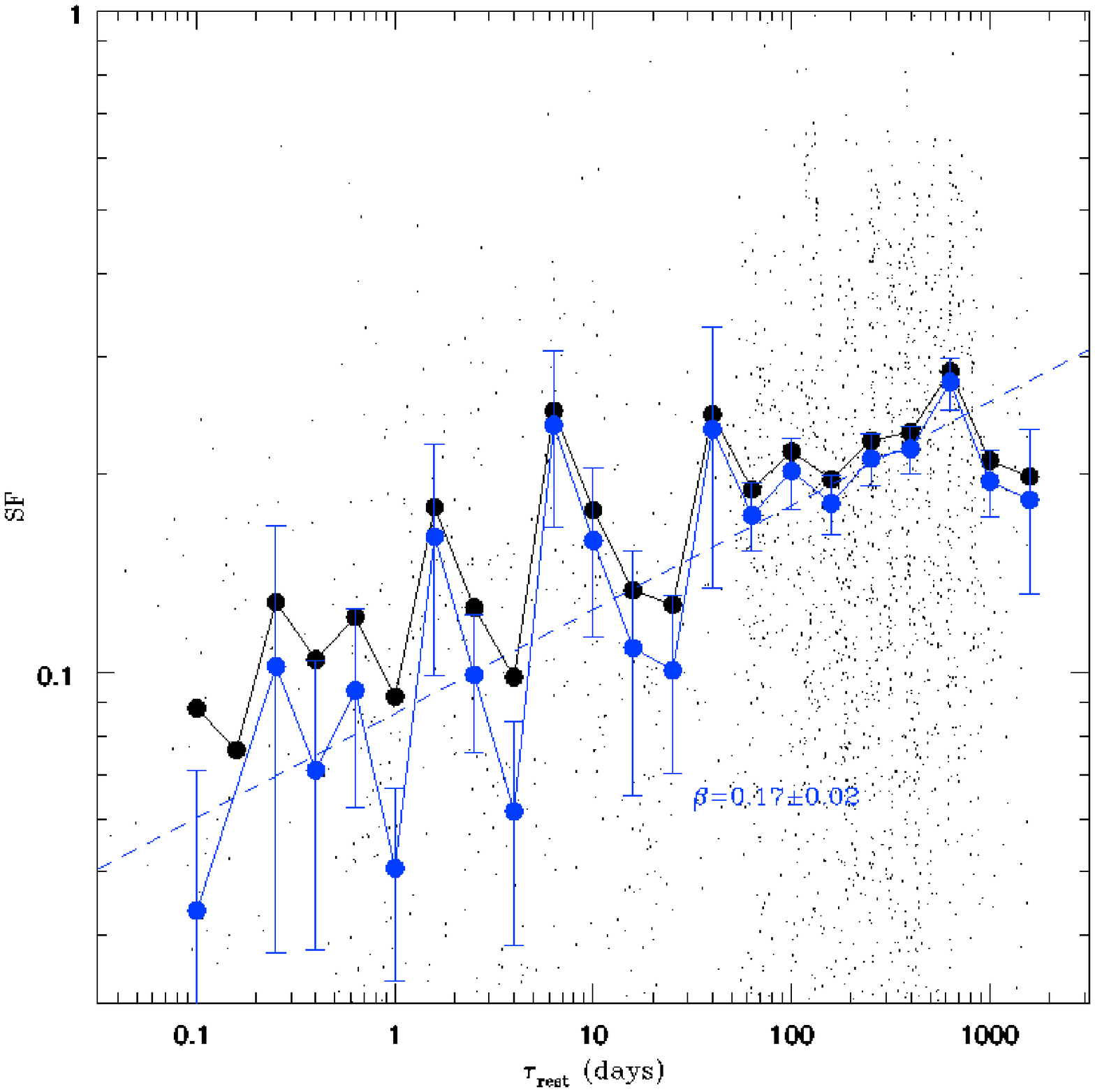}
\caption{\label{fig:sfxmm} \textit{Ensemble} Structure Function (SF), in the rest-frame, for the XMM sample. The Figure shows both the noise subtracted SF (blue line) and noise unsubtracted SF (black line). The photometric noise was evaluated in the first bin and subtracted in quadrature\cite{diclem96} to obtain the noise subtracted (blue line) SF. The black dots are the contributions to the SF from the individual source lightcurves. The dashed line is the linear fit of the SF.}
\end{figure}

\begin{figure}
\includegraphics[width=9.0cm, angle=0]{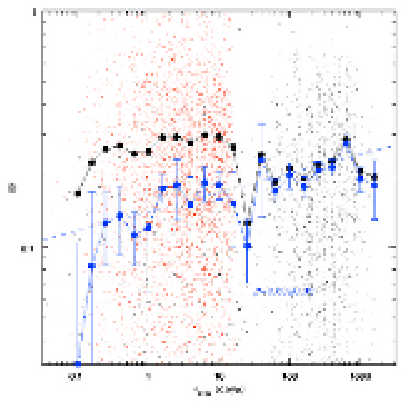}
\caption{\label{fig:sfcomb} \textit{Ensemble} Structure Function (SF), in the rest-frame, obtained combining the data of the Swift sample and the XMM sample. The Figure shows both the noise subtracted (blue line) and noise unsubtracted (black line). The photometric noise was evaluated in the first bin and subtracted in quadrature \cite{diclem96} to obtain the noise subtracted (blue line) SF. The red and black dots are the contributions to the SF from the Swift and XMM individual source lightcurves, respectively. The dashed line is the linear fit of the SF.}
\end{figure}

\subsection{Struction Function (SF)}

 The \textit{ensemble} statistic on a large sample of objects together with the spread in rest-frame time, caused by the redshift distribution of the sample, allows an accurate characterization of variability through the study of the dependence of the \textit{ensemble} structure function (SF):
\begin{equation}
\label{eq:sf}
 SF(\tau)= \sqrt{\pi/2\langle|logF(t+\tau) - logF(t)|\rangle^2 -\sigma_n^2}
\end{equation} 
on  luminosity, redshift, time lag $\tau $ and wavelength, where $\sigma_n$ is the contribution of the noise to the r.m.s. logarithmic flux changes and the angular brackets indicate the ensemble average over appropriate bins of time lag and $F$ is the flux\cite{diclem96,VB04}. 

\subsection{Data}

The Swift mission was specifically designed to sample the X-ray afterglow of gamma-ray bursts (GRBs) on time scales from hours to months. A catalog of serendipitous X-ray sources in the deep fields centered on GRBs (The Swift-XRT GRB Deep Field Serendipitous Survey)
is being completed at ASDC \cite{Puccetti2009}. These observations make up an unbiased X-ray survey since GRBs explode at random
positions in the sky. The catalog contains $\sim$ 7000 serendipitous sources with fluxes down to $\sim 10^{-15} erg~ cm^{-2} s^{-1}$ in the 0.5-10 keV band and $\sim$ 800 down to $5 \times 10^{-14} erg~ cm^{-2} s^{-1}$. We cross-correlated this catalog with the spectroscopic catalog provided by the SDSS, Data Release 7 \cite{dr7}, in order to built a sample of confirmed quasars. We found 27 confirmed quasars with enough sampling (at least, 100 photons in the lightcurve) to be used in the following SF analysis. Hereafter, we will refer to this sample as Swift sample.

To have a second confirmed quasars sample, we cross-correlated the repeated X-ray observations in the updated incremental version 2XMMi of the Second XMM-Newton Serendipitous catalogue (XMMSSC) with the SDSS Quasar Catalog, Fifth Data Release \cite{qsodr5}. This results in 272 quasars.   Hereafter, we will refer to this sample as XMM sample. The 2XMMi catalogue contains information for 192,000 serendipitous XMM sources, 27,000 of which possess lightcurves and time-dependent spectra. 

\subsection{Data Analysis and Results}

Figure \ref{fig:sfswift} shows the results from the SF analysis on the Swift sample, whereas Figure \ref{fig:sfxmm} shows the results from the SF analysis on the XMM sample. The figures show both the noise subtracted SF (blue line in both figures) and noise unsubtracted SF (red line in Fig. \ref{fig:sfswift}, black line in Fig. \ref{fig:sfxmm} and \ref{fig:sfcomb}). The photometric noise dominates the first bin and was subtracted in quadrature \cite{diclem96} to obtain the noise subtracted (blue line) SFs. The red and black dots are the contributions to the SF from the Swift and XMM individual source lightcurves, respectively. The linear fits of the SFs give the slopes $\beta=0.11\pm0.03$ (Swift sample) and $\beta=0.17\pm0.02$ (XMM sample).
 
 Figure \ref{fig:sfcomb} shows the results  from the SF analysis on the combined Swift-XMM datasets. The linear fit of this combined SF gives a slightly flatter slope ($\beta=0.08\pm0.02$) than in the two separated samples. The slope of the SFs can be related to the slope $\alpha$ of the power spectrum (PDS) \cite[see for details][$\alpha=1+2\beta$]{Kaw98}. We found that the slopes of the SFs are consistent with the PDS slopes computed on single object lightcurves ($1\alt \alpha \alt 2$), although smaller than the average PDS slope ($\alpha_m=1.55$) \cite{LP1993}.

\section{X-ray/UV ratio}

\begin{figure}
\includegraphics[width=9.0cm, angle=0]{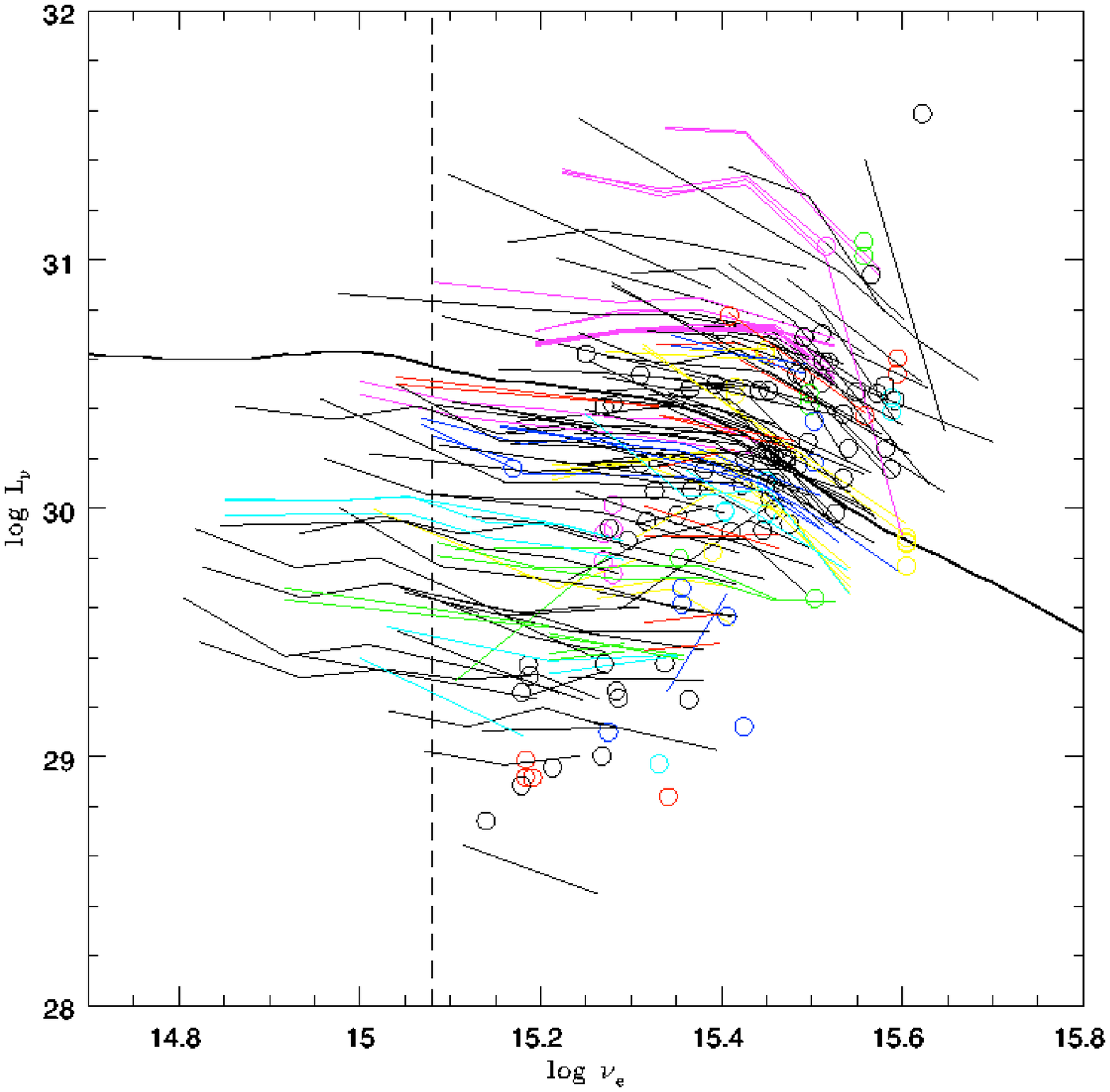}
\caption{\label{fig:sed} Optical-UV spectral energy distributions (SEDs) for each source with data from the XMM-Newton Optical Monitor (OM). SEDs with 2-6 frequency points are shown as lines, whereas small circles represents sources with only 1 frequency-point. Black lines and circles refer to sources with single-epoch data, whereas colored data refer to sources with multiepoch observations. The continuous curve spanning all the plot range is the average SED for Type 1 Quasars from the SDSS \cite{sed2006R}. A vertical dashed line is shown at $\log \nu_{e} = 15.08$ ($\lambda_{e} = 2500 \text{\AA}$).}
\end{figure}

\begin{figure}
\includegraphics[width=9.0cm, angle=0]{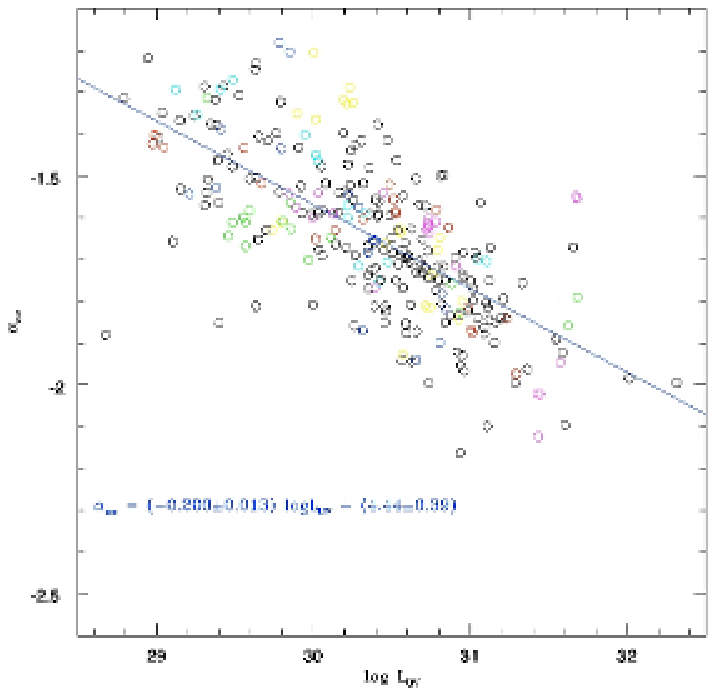}
\caption{\label{fig:alox} The Figure shows the $\alpha_{ox} - L_{UV}$ anticorrelation. Black open circles refer to sources with single-epoch data, whereas colored data refer to sources with multiepoch observations. The blue line represents the linear fit of the data.}
\end{figure}

\subsection{The dataset}

We used TOPCAT to built a sample of objects with simultaneous UV-X-ray obervations, matching the Second XMM-Newton Serendipitous catalogue XMMSSC with the XMM-OM Serendipitous Ultra-Violet Source Survey (SUSS), a catalog of UV sources serendipitously detected by the Optical Monitor on board XMM-Newton. Among them we selected a sample of confimed quasars in the SDSS Quasar Catalog, Fifth Data Release\cite{qsodr5}. The sample consists of 209 radio-quiet quasars with no Broad Absorption Line (BAL). Among them 44 quasars have repeated observations (up to 9).

\subsection{Data analysis and preliminary results}

The relationship between the X-ray and optical/UV luminosity of AGN is usually described by the slope of an hypothetical power law between $2500 \text{\AA}$ and 2 keV rest-frame, $\alpha_{ox}=0.3838\log(L_X/L_{UV})$, where $L_{UV}$ and  $L_{X}$ are the  specific luminosities at $2500 \text{\AA}$ and 2 keV, respectively. The XMM-OM SUSS catalog provides for each source one or more specific fluxes (up to 6). We computed an Optical-UV SED for each source. The results are plotted in Fig. \ref{fig:sed}.  We used these SEDs to evaluate $L_{UV}$, whereas we estimated $L_{X}$ from the integrated flux in the band 4 (2-4.5 KeV), available in the XMMSSC catalog. Details on the calculation of the specific luminosities will be found in \cite{Vagnetti09a}. Figure \ref{fig:alox} shows the $\alpha_{ox} - L_{UV}$ anticorrelation in our sample.  The dashed line represents the linear fit of the data, giving $\alpha_{ox} = (-0.200 \pm 0.013)\log L_{UV} + (4.440 \pm 0.390)$. The slope is steeper than those provided by \cite{stra2005,just2007}, but it is in good agreement with the recent results from \cite{Gibson2008}. We found the scatter in the data to have the same order of magnitude reported by \cite{Gibson2008}.

\section{Conclusions}

We produced the first \textit{ensemble} SFs of AGNs in the X-ray band using data from the Swift and XMM-Newton archives. The slopes of the \textit{ensemble} SFs are consistent with those provided by PDS analyses, althought flatter that the average value found in \cite{LP1993}.

 We presented the first attempt to use simultaneous UV and X-ray observations to study the  $\alpha_{ox} - L_{UV}$ relation using XMM-Newton observations. We found a steeper slope than many precedent works\cite[e.g.][]{stra2005,just2007}, but in agreement with a recent study \cite{Gibson2008}. Since the scatter in the data has the same order of magnitude reported by \cite{Gibson2008}, we could suppose that variations within a same object and from object to object are more important than the simultaneity of the X-ray/UV data.

\begin{acknowledgments}
Part of this work is based on archival data, software or on-line services provided by the ASI Science Data Center (ASDC). This research made use of the XMM-Newton Serenidiptous Source Catalog, which is a collaborative project involving the whole Science Survey Center Consortium, and of the XMM-OM Serendipitous UV Source Survey (SUSS), which has been created at MSSL on behalf of ESA. This work makes use of EURO-VO software, tools or services. The EURO-VO has been funded by the European Commission through contract numbers RI031675 (DCA) and 011892 (VO-TECH) under the 6th Framework Programme and contract number 212104 (AIDA) under the 7th Framework Programme. This work makes use of SDSS data. ST acknowledges financial support through Grant ASI I/088/06/0.
\end{acknowledgments}


%

\newpage

\lhead[\fancyplain{}{}]{\itshape \small{The problem of surgical wound infections: air flow simulation in operating room}}
\rhead[\itshape \large{1$^\mathrm{st}$ RYRM Proceedings}]{\fancyplain{}{P. Abundo}}


\begin{center}

\setcounter{section}{6}
\section{\large{The problem of surgical wound infections: air flow simulation in operating room}}
\label{abundo}

\normalsize{Paolo Abundo$^*$, Luca Armisi}

\emph{\small{Servizio di Ingegneria Medica, Fondazione Policlinico Tor Vergata, Viale Oxford 81 - 00133 Roma}}


\vspace{.25cm}
\normalsize{Fabio Gori}

\emph{\small{Dip.to di Ingegneria Meccanica, Universit\`a degli Studi di Roma Tor Vergata, Via del Politecnico 1 - 00133 Roma}}

\vspace{.25cm}
\normalsize{Nicola Rosato}

\emph{\small{Dip.to di Medicina Sperimentale e Scienze Biochimiche, Via Montpellier 1 - 00133 Roma}}

\vspace{.5cm}

\begin{minipage}{.8\textwidth}
\small{In operating rooms, the air is required to have suitable characteristics to preserve the optimal conditions for well-being of the medical staff and, more important, of the patients, who undergo surgical operations. It is necessary to carefully take into account the Colony Forming Unit floating in the air, which can cause an infection of the surgical wound.
 As regards the causes of the infections in operating room, we built a Fluid-Dynamics simulative model, whose aim is to study the air flows inside the operating room.
 Particularly, our scenario is a real operating room existing in the Policlinico Tor Vergata of Rome.
 With the most common operating room set up, we observe the presence of eddies close to the most critical zone, i.~e. the surgical wound, mainly due to the scialitic lamps.
 In this situation the air stagnation around the surgical wound is accentuated and the potentially infective particles may easily cause an infection. In order to find a possible solution for this serious problem, we simulate a scenario of operating room in which a device able to produce an additional sterile air flow was added. If the target of this additional air flow is the surgical wound, simulations show a decreasing of the turbulence and a smaller air stagnation around the surgical wound, which imply a lower probability of an infective event.
}

\vspace{.25cm}
$*$ e-mail: email: \href{mailto:paolo.abundo@ptvonline.it}{paolo.abundo@ptvonline.it}
\end{minipage}
\end{center}

\setcounter {section} {0}
\setcounter {subsection} {0}
\setcounter {figure} {0}
\setcounter{equation}{0}
\setcounter{figure}{0}

\section{Introduction: surgical wound infection}

The present study deals with the management of the clinical risk, defined as "the probability that a patient undergoes an adverse event, i.e. damage, also unintentional, caused by the medical treatment during the hospitalization that causes a worsening of the health condition of the patient or his death".
\\The clinical risk can be limited in a hospital by the institution of Clinical Risk System and above all, of a Medical Engineering Service, that assure the efficient and economical use of a medical device, producing more tightly procedures also in the starting test of a device. Every year in Italy almost 10\% of the hospital-patients is affected by an infection, with an increase of the costs for the National Health Service. Specifically, surgical infections are 10-30\% of the total nosocomial infections. The main microrganisms causing the surgical infections are Stafilococcus Aureus, Pseudonomas and Escherichia Coli. When these particles join the "right" target they can make a colony, causing an infection of the surgical wound \cite{8}. Beside the state of the patient, it is present also an increase of costs for the National Health System. A typology of surgical activities where the infection can be very serious is the implant of haunch or knee arthroprosthesis \cite{2}, \cite{4}, when an infection can even lead to the mobilization of the prosthesis.

\section{Aim of the study}

Despite the precautions realized in operating room (OR), surgeons and other clinical operators emit physiological particles fluctuating in the air overlying the operating table, thus producing a potential risk for the patient. As far as the causes of the infections in operating room is concerned, a Fluid-Dynamics model has been used with the aim of studying air flows inside the operating room and evaluating the probability of an occurring infection of surgical wound. The geometrical model of a real operating room has been built using the software Gambit, and the Fluid-Dynamics simulations have been carried out with Fluent. The reviewing of the numerical simulations allowed to make proposals in order to minimize the contamination of the air near the patient surgical wound.

\section{Operating room}
The official definition of the operating room is a medical local of Group 2, i.e. following the norm CEI 64-8/7, "a medical local where the parts applied to the patient are finalized to be used in cardiac operations or where patient undergoes vital treatment and loss of energy can cause a life danger".
\begin{figure}[h!]
   \centering
   \includegraphics[width=8cm]{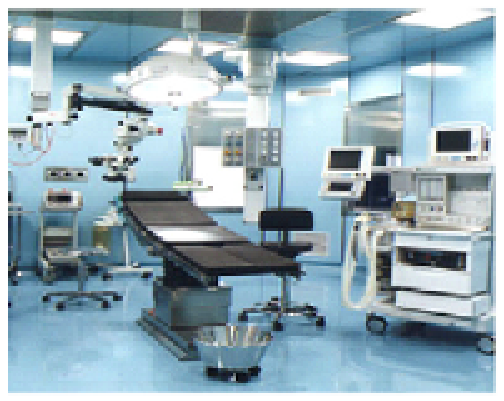}
   \caption{A standard Operating Room.}
\end{figure}

A standard operating room is characterized by an extreme technological complexity, where are present many electromedicals, as operating table, scialitic lamps, anesthesia devices, electrosurgical devices, defibrillators, etc. In operating rooms air is required to have suitable characteristics to preserve optimal conditions for well-being of the medical staff and of the patients which undergo surgical operations. It is then necessary to be careful about Colony Forming Unit (CFU) floating in air.
\\If these particles join the "right" target, they can make a colony causing an infection of the surgical wound. Once this control is realized, climatization system is defined a "Controlled Air - Contamination System" \cite{5},\cite{6},\cite{3},\cite{7}.

\section{Operating Room Modeling}
The core of the present study is to use a Fluid-Dynamics model with the aim of studying air flow inside the operating room. Computational Fluid-Dynamics can simulate the velocity field of a fluid crossing a surface of every shape, under esteblished boundary conditions.
\\The first part of the work is modelling an Operating Room using the software Gambit. The real operating room is that existing at the Policlinico Tor Vergata of Rome (N.5 of the next opening Operating Zone), with a laminar diffusion of air from the top of the room. The real dimensions of the OR are used, integrated with data obtained by in situ inspection. The following step is the creation of the mesh which can discretize the built volumes.
\begin{figure}[h!]
   \centering
   \includegraphics[height=3.5cm]{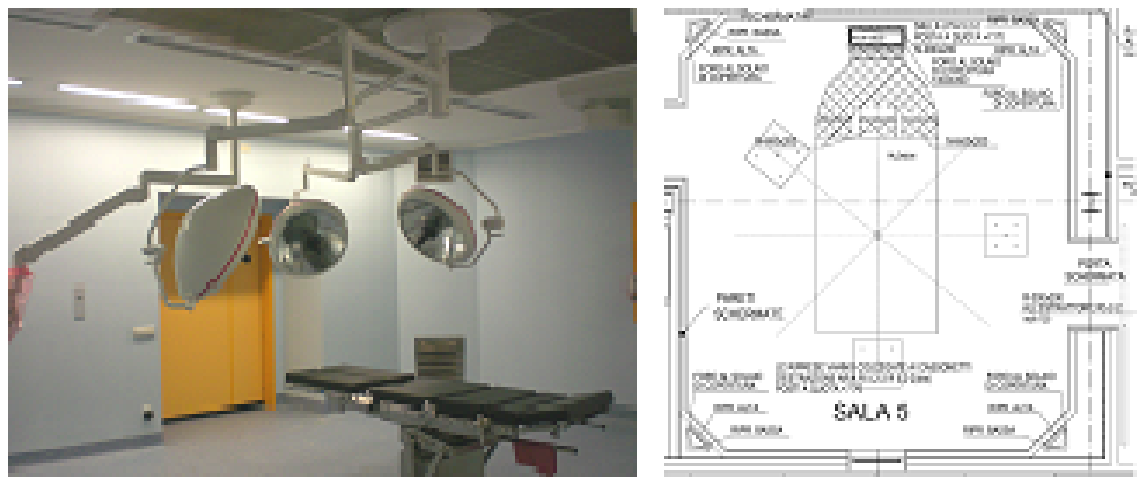}
   \includegraphics[height=3.5cm]{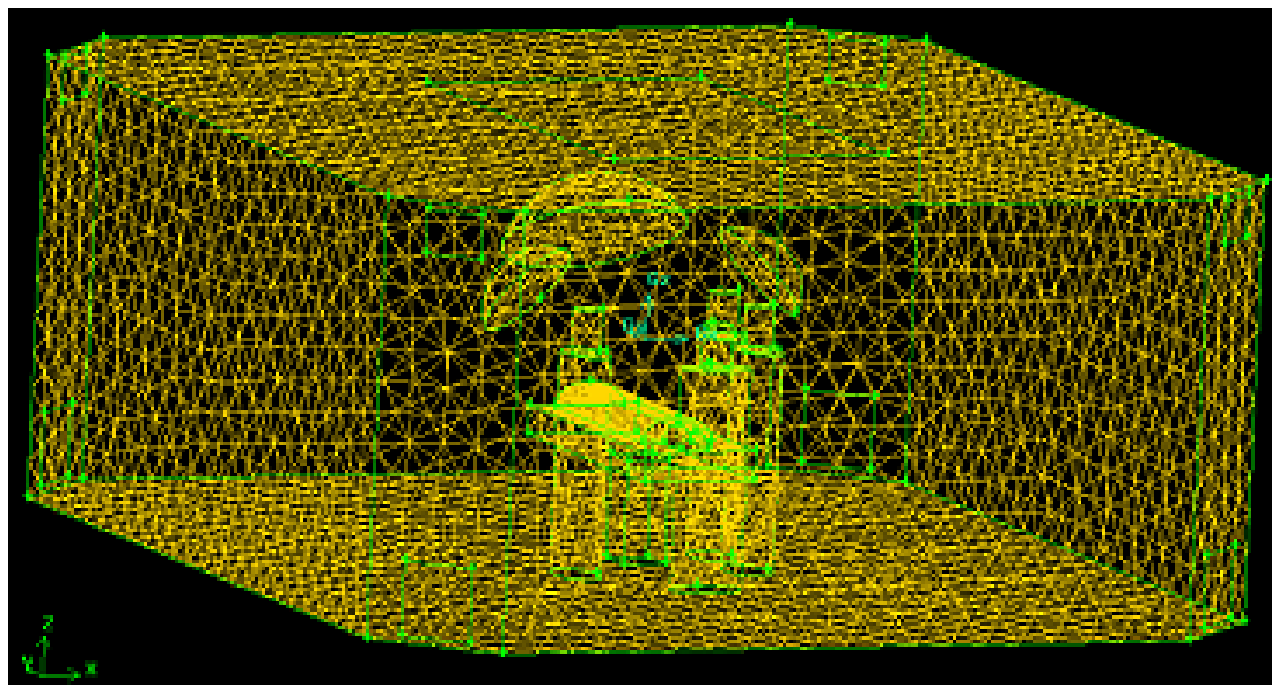}
   \caption{The real OR (left panel); its scheme "as built" (middle panel) and the mesh adopted (right panel).}
\end{figure}


In the analysis of the chosen OR the grids of air recirculation are considered and the air-laminar diffusion from the ceiling of the room. The main purpose of this study is to determine the air velocity inside the OR during a normal activity, i.e. when several equipments are influencing the flow. It is then necessary to investigate every part of OR, considered important for the Fluid-Dynamics simulation. The two main important devices to investigate are the real Maquet operating table and the real Martin Scialitic lamps. In order to obtain important information about the problem of sterility in OR, the complete OR with the operating table, the patient, the scialitic lamps and the health operators are investigated in different configurations.
\begin{figure}[hb!]
   \centering
   \includegraphics[width=.95\textwidth]{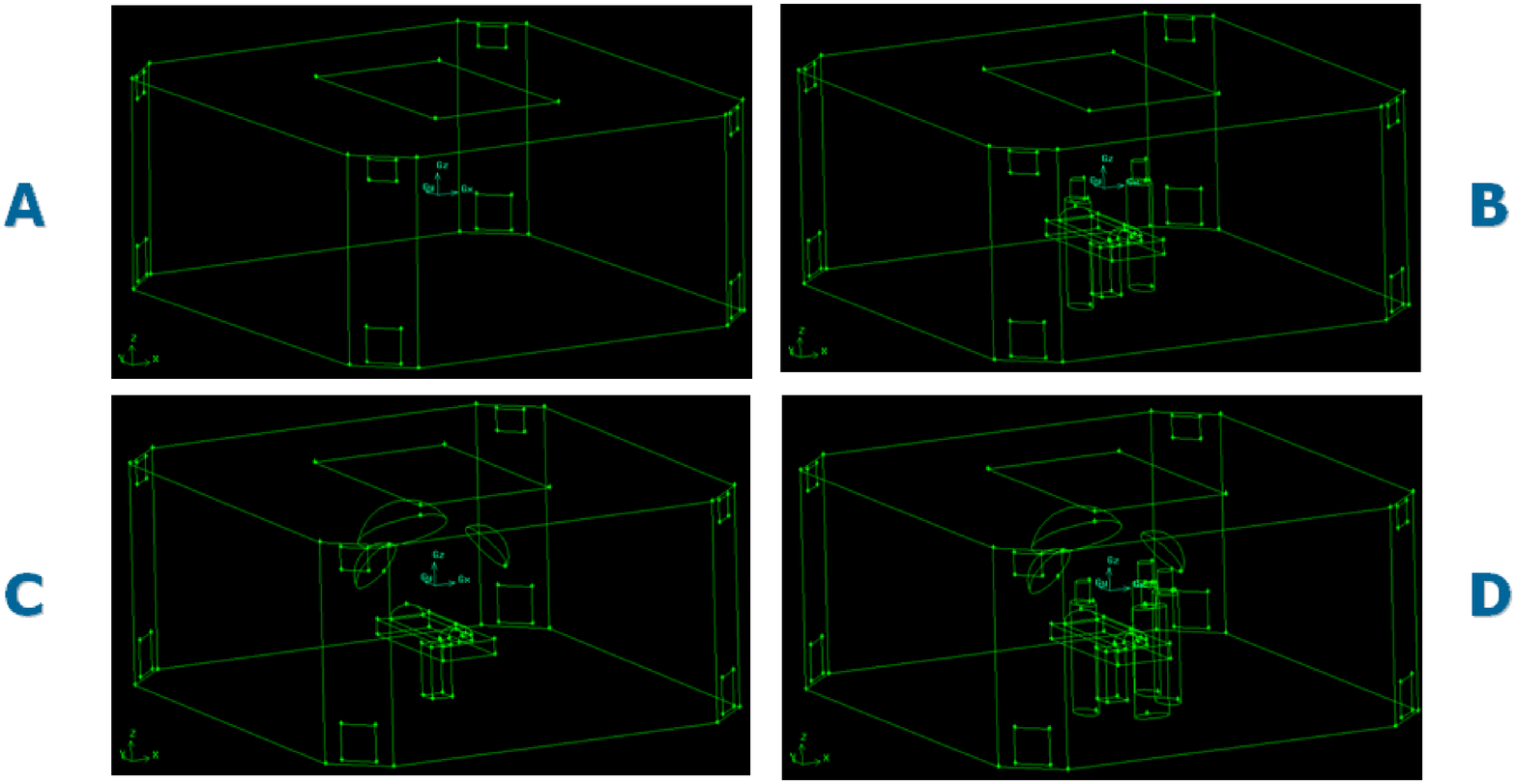}
   \caption{Simulation configurations.}
\end{figure}

In the first configuration the OR without any items inside is analyzed, while in the B configuration the operating table with the patient and 2 surgeons are considered. Because the target is to study in detail the trend of the sterile air flow near the surgical wound, which can potentially affect with an infection, a denser mesh is put in the interested zones. Specifically, a mesh with a step of 20 mm has been put along the surface delimiting the patient body.

In the configuration C, the operating table with the patient and the scialitic lamps are present, while in the last D configuration the complete OR is considered, i.e. with the patient, the operating table, the scialitic lamps and the surgeons.

The boundary conditions for the single faces of the total volume are necessary. We report them according to the different configurations in Tab.~\ref{tab:confs}.
\begin{table}[h]
\begin{tabular}{|c|c|c|c|c|}
\hline
{\bf Type} & {\bf Conf.~A} & {\bf Conf.~B} & {\bf Conf.~C} & {\bf Conf.~D} \\
\hline
Laminar-air diffusion ceiling & INLET\_VENT & VELOCITY\_INLET & VELOCITY\_INLET & VELOCITY\_INLET \\
Recirculation air grid & PRESSURE\_OUTLET & PRESSURE\_OUTLET & PRESSURE\_OUTLET & PRESSURE\_OUTLET \\
OR walls & WALL & WALL & WALL & WALL \\
Operating table & - & WALL & WALL & WALL \\
Patient & - & WALL & WALL & WALL\\
Scialitic lamps & - & - & WALL & WALL \\
Health operators & - & WALL & -  & WALL \\
\hline
\end{tabular}
\caption{OR mesh: boundary conditions for each configuration analyzed.}
\label{tab:confs}
\end{table}


The first boundary condition (BC) is relative to the air velocity coming from the top of the OR (0.45 m/s) along the surface of the OR air diffusion ceiling. The second BC is relative to the pressure gradient between inside and outside the OR, which is assumed as equal to 10 Pa, on the surface of air exit, e.~g.~ on the surfaces of recirculation air grids. The other boundary conditions describe the OR wall and the relative items surfaces inside, which are  assumed just as "wall".
 
\section{Simulations}
The simulation can be carried out with the software Fluent with the meshing volumes made by Gambit, after optimization of the shapes of the tetrahedrons of the mesh, confirmation of the BC and the verification of the convergence criteria.
Next figures show several simulations made with different boundary conditions: Figure ~\ref{fig6} shows the simulations in the OR without patient and surgeons (configuration A).
\begin{figure}[h!]
   \centering
   \includegraphics[width=.45\textwidth]{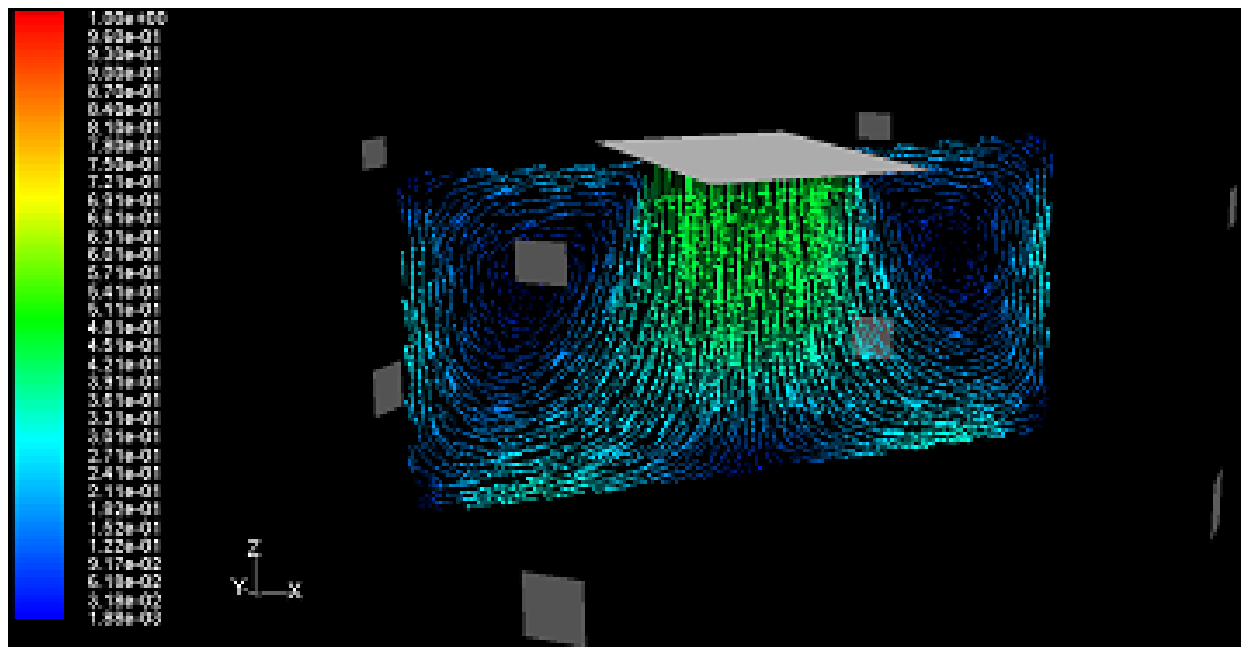}
   \includegraphics[width=.45\textwidth]{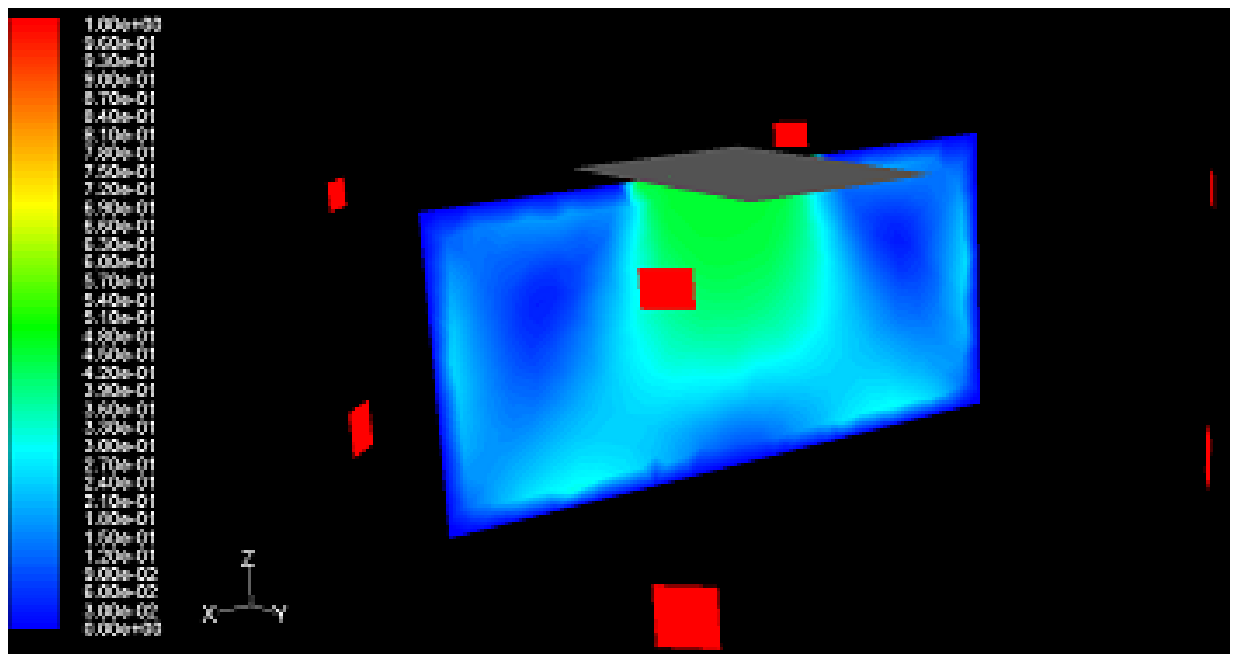}
   \caption{Configuration A: air velocity vectors (m/s) on a single plane (left panel); air velocity profile representation (m/s) (right panel).} \label{fig6}
\end{figure}

Figure ~\ref{fig9} is relative to the simulations in the OR with the patient on the operating table and two surgeons (configuration B). Among the most serious surgery with danger of infection is the knee arthroprosthesis. Left and middle panels show the air velocity field on two special planes respectively: plane 1 is a XZ plane passing on the knee of the patient, while plane 2 is a YZ plane passing through the patient.
The right panel in Figure ~\ref{fig9} presents the turbulence viscosity of air inside the OR.
\begin{figure}[h!]
   \centering
   \includegraphics[width=.3\textwidth]{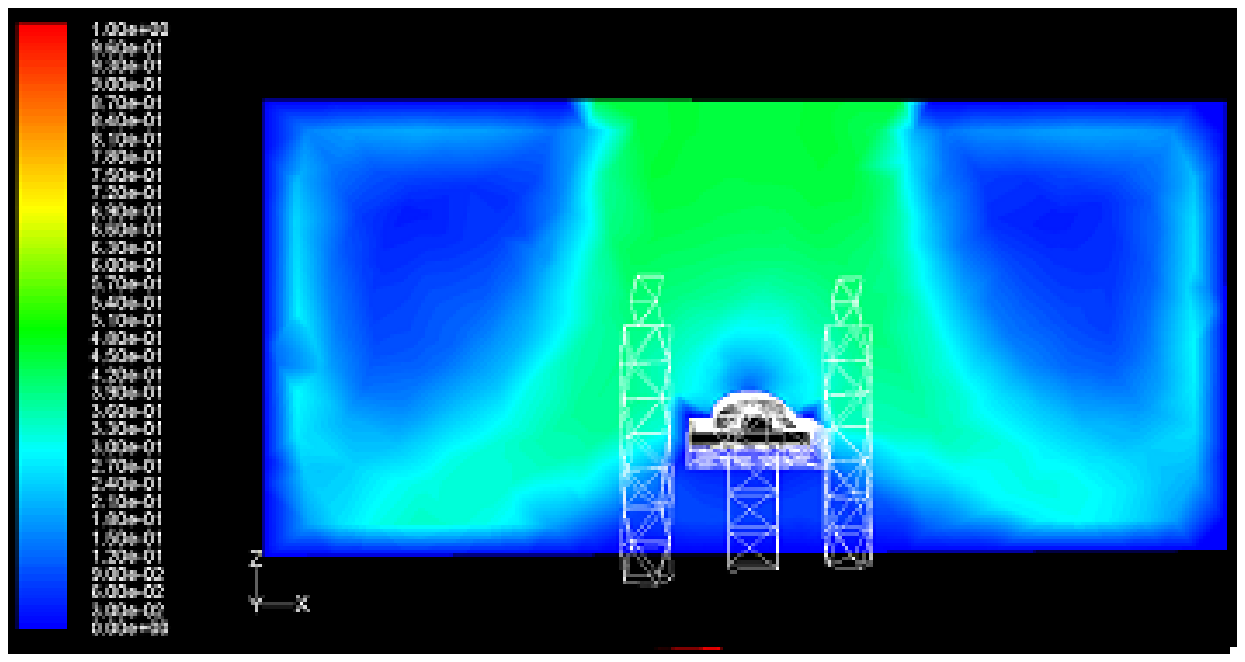}
   \includegraphics[width=.3\textwidth]{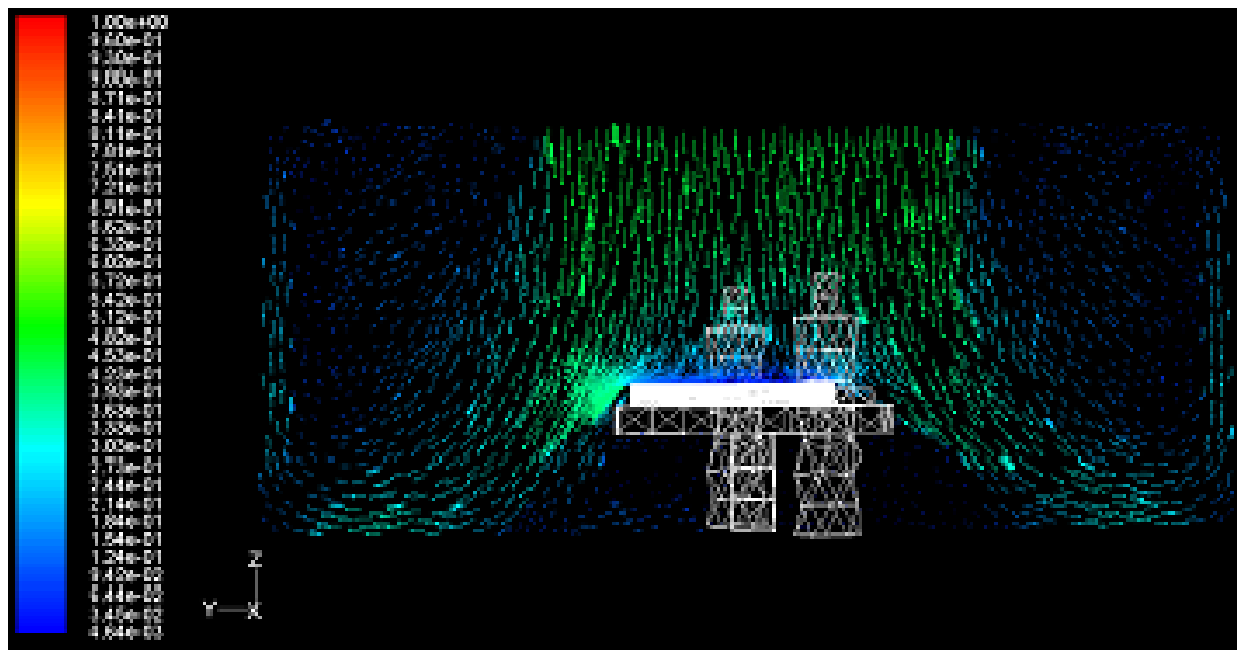}
   \includegraphics[width=.3\textwidth]{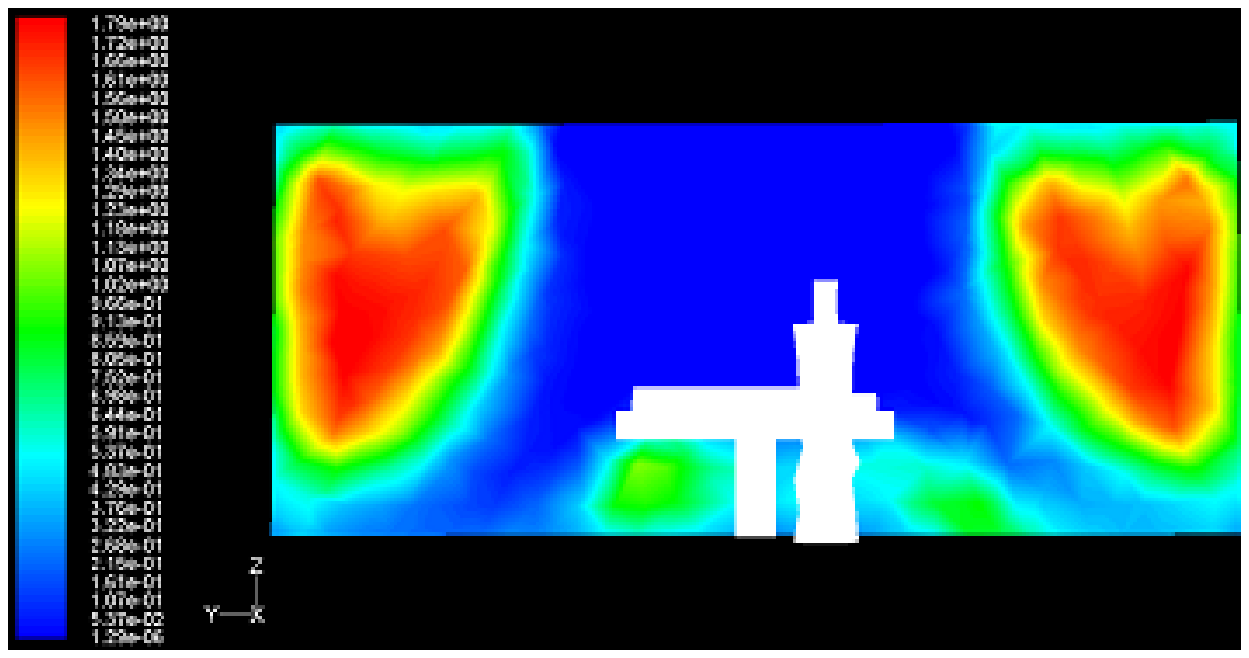}
   \caption{Configuration B: air velocity vectors (m/s) on plane 1 (left panel) and 2 (middle panel) and air turbulence viscosity (Kg/m-s) on plane 2 (right panel).} \label{fig9}
\end{figure}


Configurations A and B show a good laminar diffusion of air between the ceiling of the room and the operating table zone, also without air turbulence near the patient. The main reason is the absence of obstacle in this area, namely the scialitic lamps, as it is summarized in Tab.~\ref{tab:confs}.
 This is the ideal case for the patient, because the sterile air flow is close to the surgical wound, and the probability of an infection is very small.

The most realistic configuration D, representing the complete OR with the operating table, the patient, the lamps and 4 surgeons, shows different results as it is plotted in Figure ~\ref{fig12}.
\begin{figure}[h!]
   \centering
   \includegraphics[width=.3\textwidth]{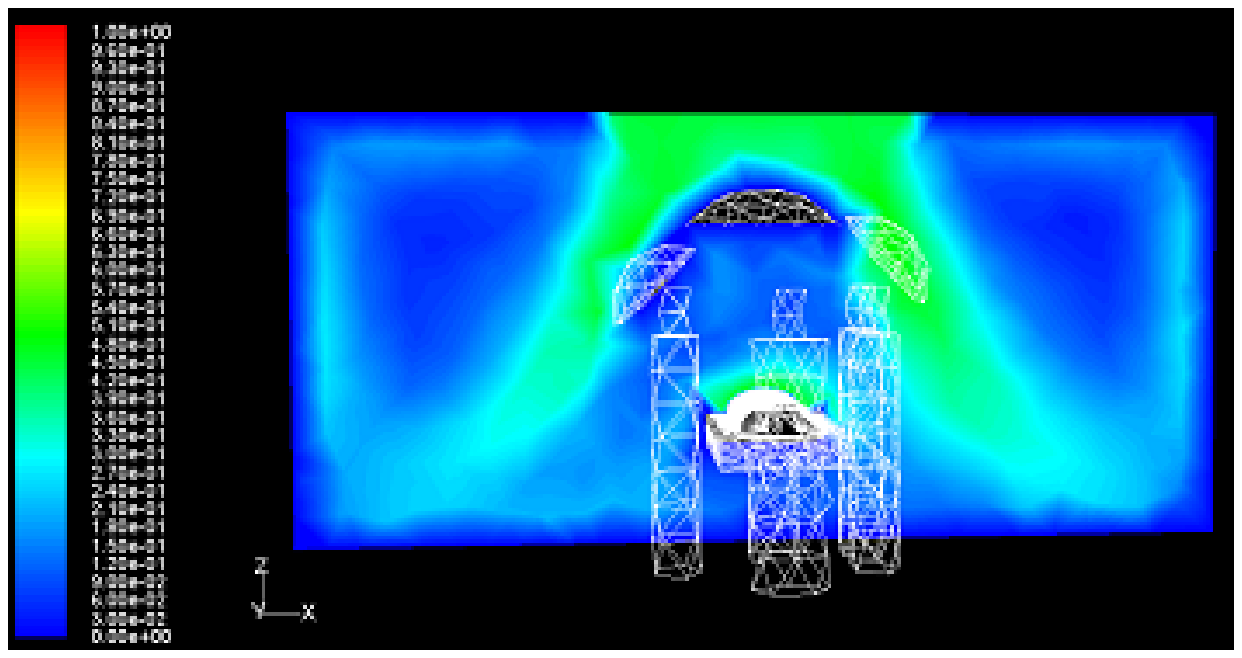}
   \includegraphics[width=.3\textwidth]{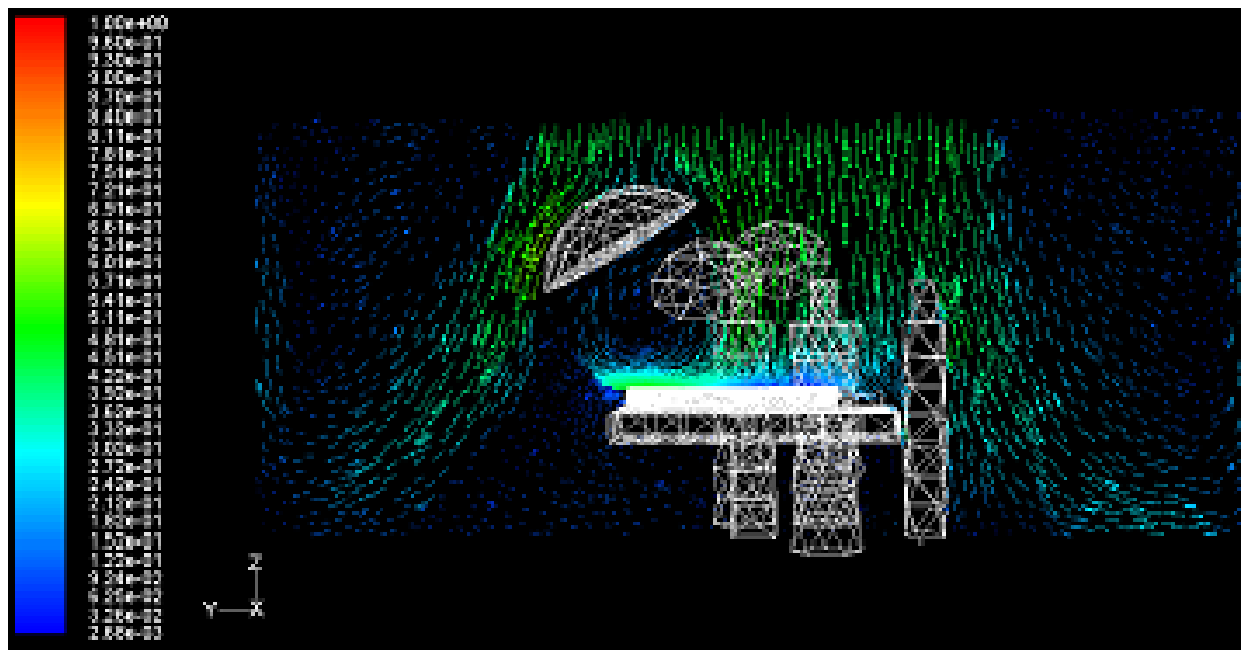}
   \includegraphics[width=.3\textwidth]{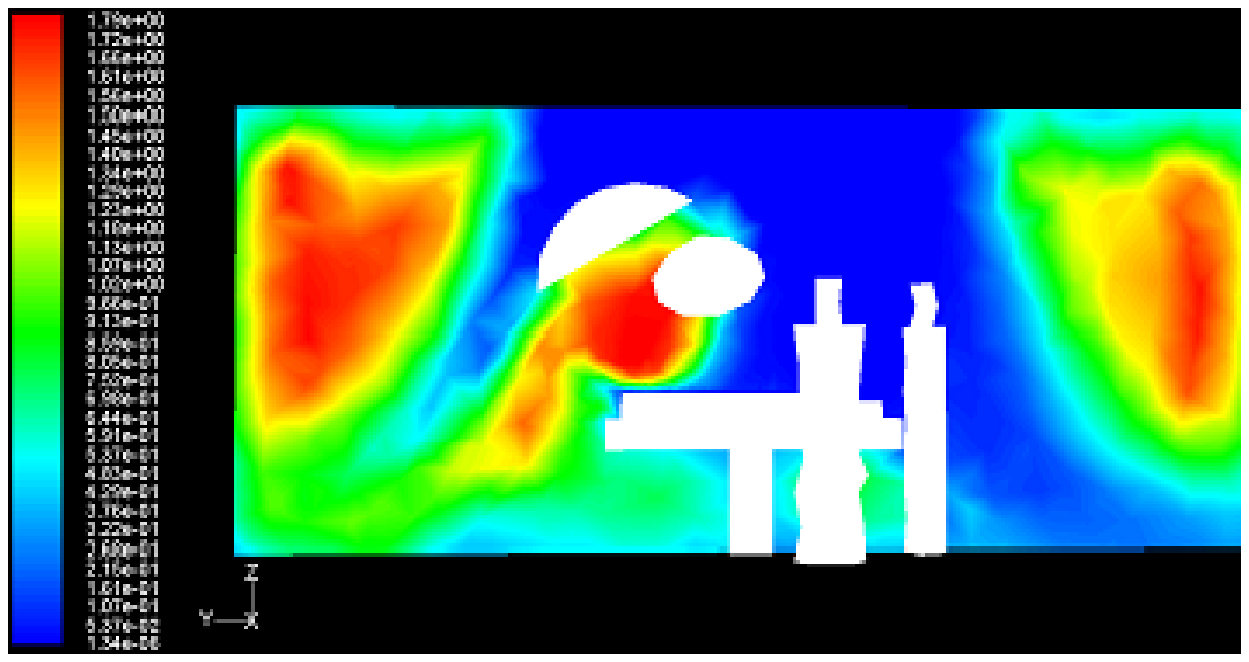}
   \caption{Configuration D: air velocity vectors (m/s) on plane 1 (left panel) and 2 (middle panel) and air turbulence viscosity (Kg/m-s) on plane 2.} \label{fig12}
\end{figure}

Air velocity profile on plane 1 and 2 and turbulence viscosity on plane 2 are very complex and irregular due to the many items present in this configuration. Air flow near the patient is not laminar. In this configuration are also present surgeons who contribute to the non laminar behaviour of air flow near the operating table. Figure ~\ref{fig12} shows high values of air turbulence below the lamps, which are very important in the present simulation. Eddies of air are maximum just close to the surgical site, which is the worst case for the infection probability. In this case air is stagnant and laminar sterile air flow does not reach the wound, with a greater probability that a potentially infective particle could reach the surgical wound.

A new mesh with a step of 100 mm for the OR space and 20 mm for the patient surface and the recirculation air grid has been used and the results are shown in Fig.~\ref{fig13}.
\begin{figure}[h!]
   \centering
   \includegraphics[width=.8\textwidth]{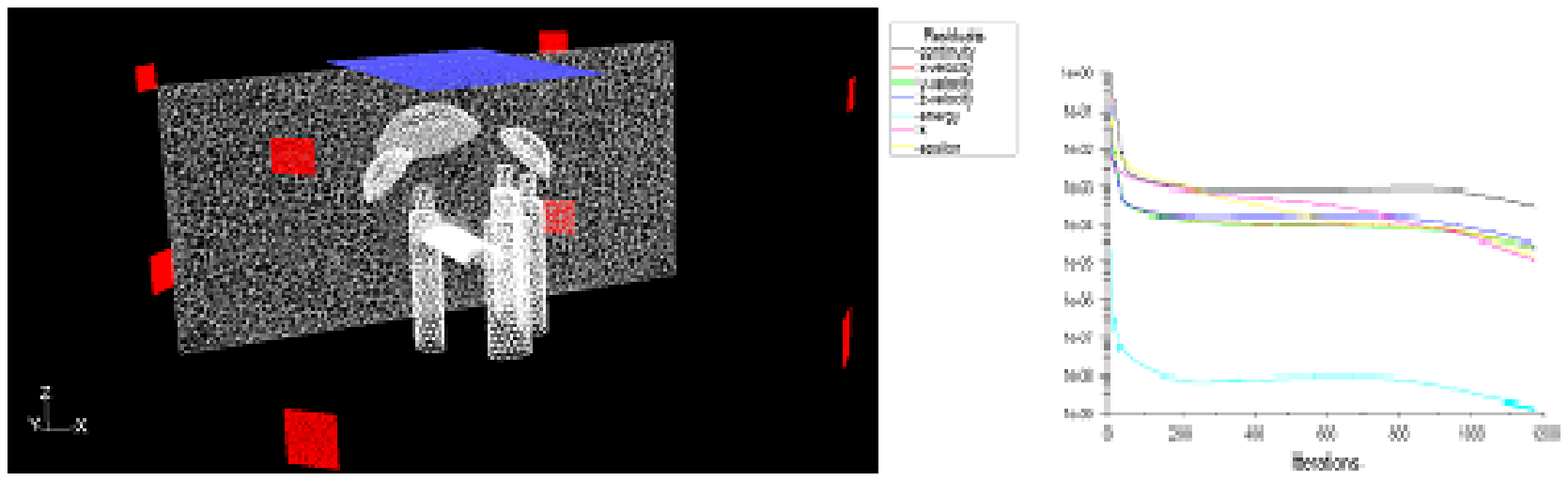}
   \caption{Grid display on a plane and residual trend.} \label{fig13}
\end{figure}
The residuals of the several computed parameters show a regular trend.
Figure~\ref{fig15} shows air velocity vectors and turbulence viscosity on plane 2 with an evident higher level of spatial resolution of the simulations.
\begin{figure}[h!]
   \centering
   \includegraphics[width=.45\textwidth]{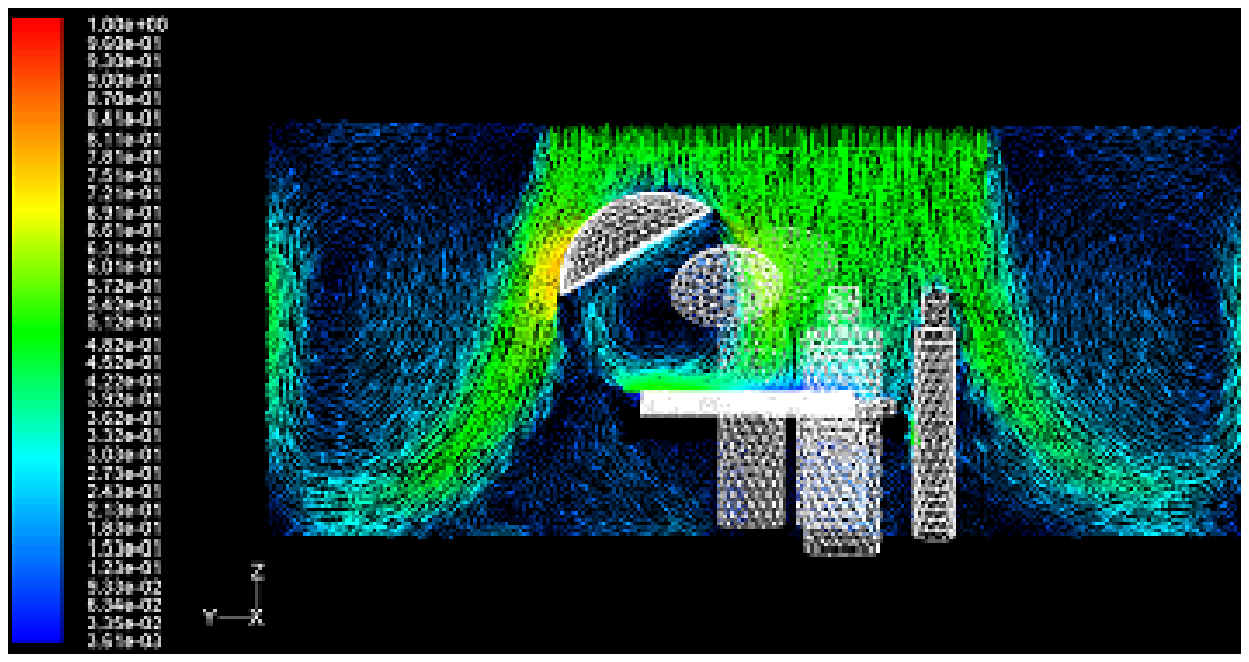}
   \includegraphics[width=.45\textwidth]{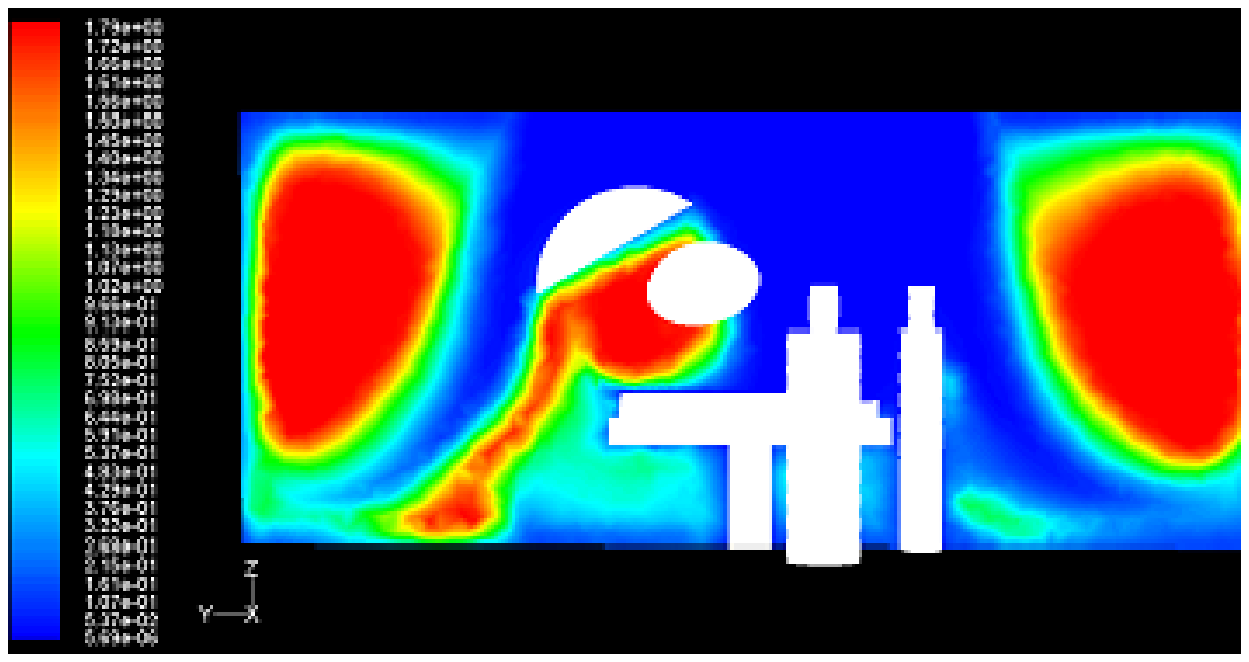}
   \caption{Configuration D: air velocity vectors (left panel) and air turbulence viscosity (Kg/m-s) (right panel) on plane 2 by using a thicker simulation mesh.}
 \label{fig15}
\end{figure}

Compared to Figure ~\ref{fig12}, Figure ~\ref{fig15} presents some eddies invisible with the previous mesh.

\section{Assessment of using an additional sterile laminar air flow}
It is here proposed to use a solution to optimize the laminar flow in the zone near the surgical wound, by introducing inside the OR one more device, i.e. Toul Maquet, which is already existing in the market. The proposed device can produce another laminar flow moving toward the patient with a velocity of 0.4 m/s, after a HEPA filter, \cite{1}. This filter can make cleaner the air near the wound, inducing a lower probability of infection.

The final part of the present paper presents simulations carried out in presence of the HEPA filter according to the configuration presented in Fig. ~\ref{fig16}. Simulations are done similarly to the previous approach for all the OR configurations. The boundary condition used with this filter is relative to the surface in front of the surgical wound of the patient, i.e. knee in the present paper. From this surface is coming a sterile air with velocity 0.4 m/s and 295$^{\circ}$K temperature. The other boundary conditions of the OR are the same.
\begin{figure}[h!]
   \centering
   \includegraphics[width=8cm]{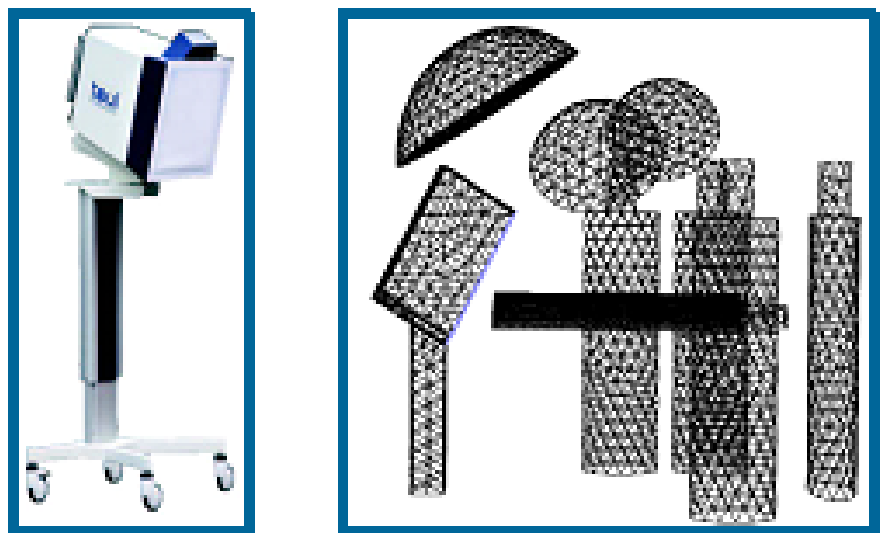}
   \caption{Additional laminar air flow producer and its modeling.} \label{fig16}
\end{figure}

Results of the simulations are presented in Figures ~\ref{fig17} and ~\ref{fig19}, showing that air velocity vectors are more parallel than in absence of the new device, which produces additional laminar air flow toward the surgical wound. Air velocity near the wound is close to 0.4 m/s, which implies a lower degree of air stagnation and a lower probability of infective event. Figures ~\ref{fig17} and ~\ref{fig19} show a decrease of turbulence viscosity near the surgical wound in presence of the additional sterile air flow.
\begin{figure}[h!]
   \centering
   \includegraphics[height=3.5cm]{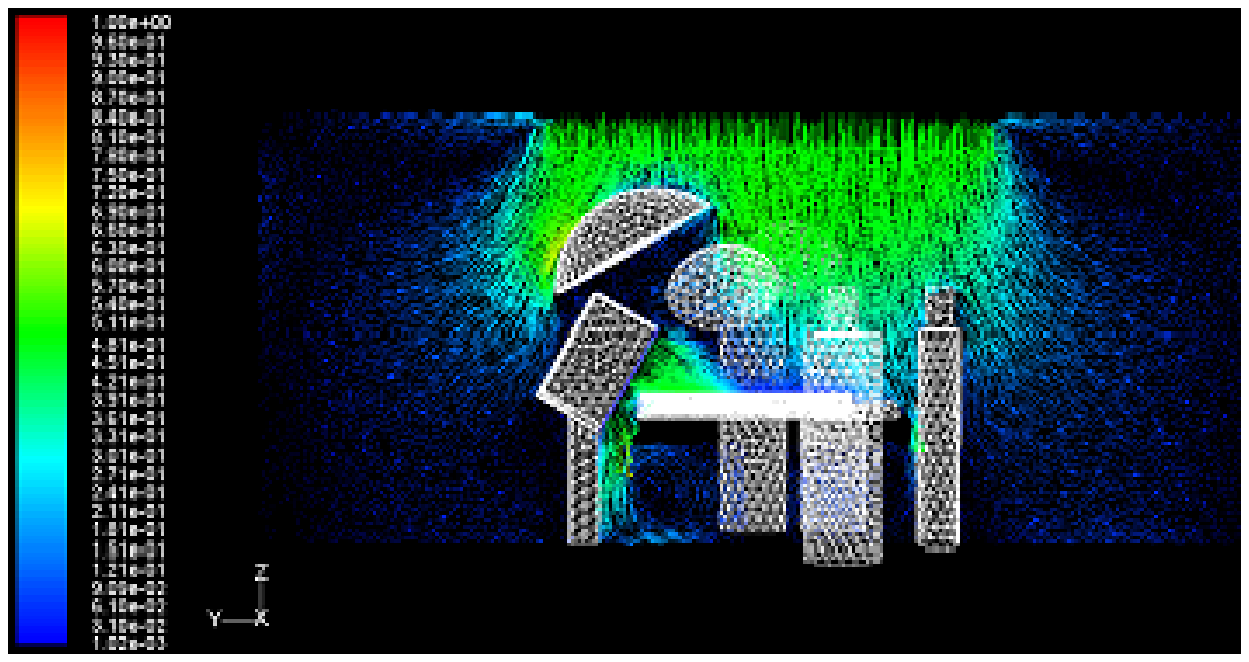}
   \includegraphics[height=3.5cm]{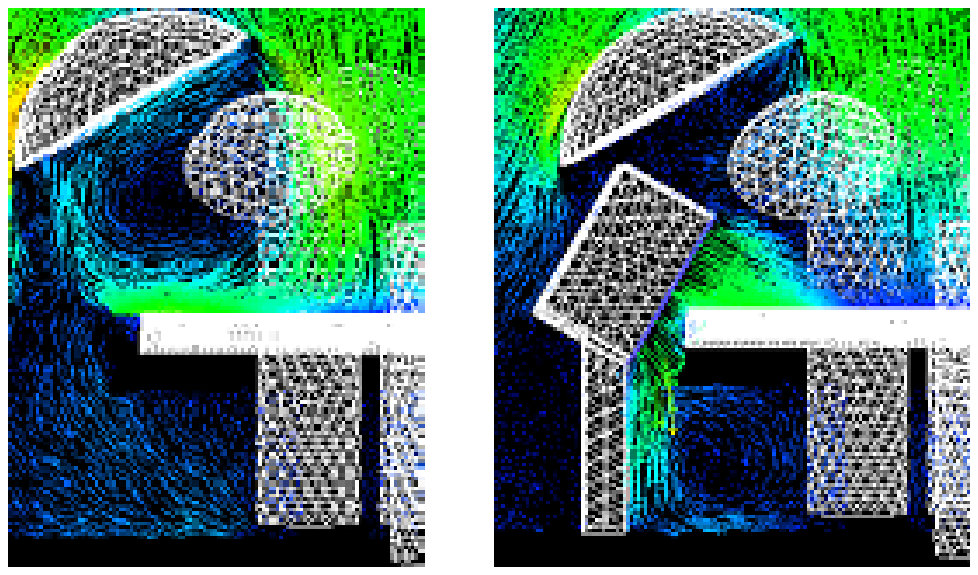}
   \caption{Air velocity vectors on plane 2 (m/s) in a complete OR with an additional sterile laminar air flow.}
    \label{fig17}
\end{figure}

\begin{figure}[h!]
   \centering
   \includegraphics[width=8cm]{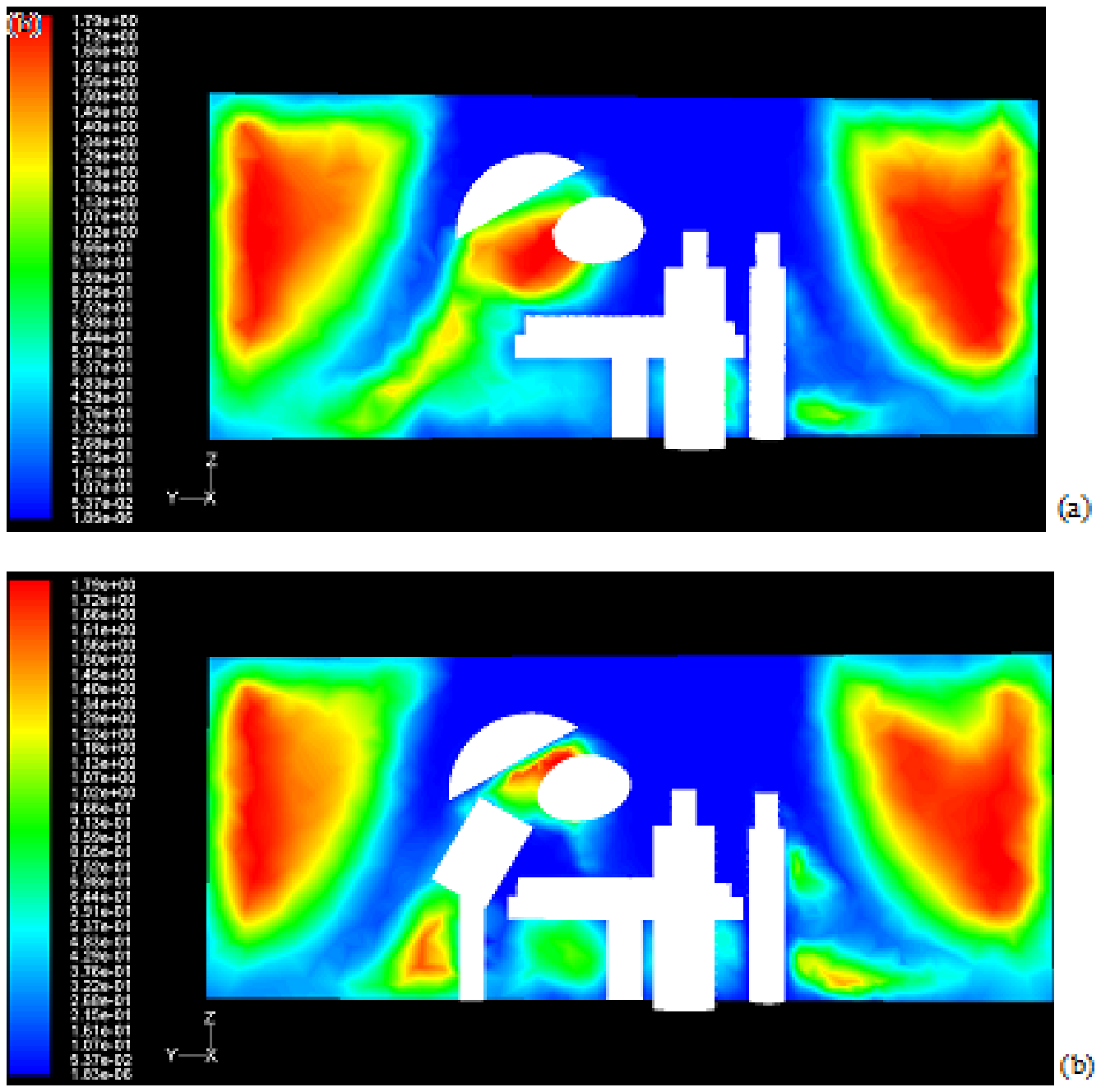}
   \caption{Air turbulence viscosity (Kg/m-s) on plane 2 without (a) and with (b) the additional sterile laminar air flow.}
   \label{fig19}
\end{figure}

\section{Conclusions}
A Fluid-Dynamics model has been used to study air flow inside a real Operating Room of the Policlinico Tor Vergata of Rome. The laminar diffusion of air from the top of the room has been investigated in different configurations, i.e. in presence of a complete operating room, e.g. with the operating table, the patient, the scialitic lamps and the health operators. The simulations on the OR with the operating table, the patient and without the lamps, have shown that the air laminar diffusion on the top of OR produces a privileged space without turbulence, maximizing the air cleanness near the surgical site. The situation changes dramatically when the OR is complete, i.e. including the scialitic lamps overlying the patient. It has been observed the presence of eddies, especially near surgical wound. Similar conclusions can be made analyzing the parameters directly related to the turbulence of the air flows, which show a turbulence activity above the scialitic lamps. Then, air stagnation around the surgical wound is increased and the potentially infective particles may easily cause an infection.
In order to find a solution to the problem of infective events, the complete operating room has been investigated with the presence of a device able to produce additional sterile air flow. The resulting simulations show that air flow vectors near the surgical wound are more regular and less turbulent than without it. A lower air stagnation around the surgical wound has a smaller probability of an infective event.


\newpage

\lhead[\fancyplain{}{}]{\itshape \small{The Resistive Plate Chambers of the ATLAS experiment: performance studies }}
\rhead[\itshape \large{1$^\mathrm{st}$ RYRM Proceedings}]{\fancyplain{}{G. Cattani}}




\begin{center}


\setcounter{section}{7}
\section{\large{The Resistive Plate Chambers of the ATLAS experiment: performance studies}}
\label{cattani}

\normalsize{Giordano Cattani$^*$}

 \emph{\small{
Department of Physics\\ University of Rome ``Tor Vergata'' and INFN Roma 2.\\
Via della Ricerca Scientifica, 1 00133 Rome (Italy)}}


\vspace{.25cm}
\normalsize{On behalf of the RPC group}
             
\vspace{.5cm}

\begin{minipage}{.8\textwidth}
\small{ATLAS (A Toroidal LHC ApparatuS) is one of the four
experiments ready for p-p collisions scheduled in autumn 2009
at the LHC at CERN.
It is a general purpose experiment, with a physics program
that spans from the search of the Higgs Boson to the evidence of
physics Beyond the Standard Model (BSM).
Events with final state muons are signature of the most promising
physics channels, e.g. the Standard Model Higgs Boson decaying into four muons (H $\rightarrow$ 4$\mu$).
For this reason a completely independent in-air Muon Spectrometer
with a toroidal magnetic field for muon track measurement has been realized to trigger and measure
the momentum of high energy muons.
Resistive Plate Chambers (RPCs) provide the first-level muon
trigger and the measurement of the coordinate in the non-bending direction
in the barrel region of the Muon Spectrometer.
The installation in the experimental hall of RPCs has been completed and
extensive tests of their performance have been performed with cosmic-rays muons.
An overview of the ATLAS experiment, focusing on the results of the studies on the RPCs
performance will be presented.}

\vspace{.25cm}
 $*$ e-mail: \href{mailto:giordano.cattani@roma2.infn.it}{giordano.cattani@roma2.infn.it}
\end{minipage}

\end{center}


\setcounter {section} {0}
\setcounter {subsection} {0}
\setcounter {figure} {0}
\setcounter{equation}{0}

\section{\label{sec:lhc}The Large Hadron Collider}

The Large Hadron Collider (LHC) \cite{lhc} is a two-ring-superconducting-hadron accelerator and collider
installed in the existing 26.7 km tunnel that was constructed between 1984 and 1989 for the CERN
LEP machine. It is designed to collide two proton beams with a centre-of-mass energy up to 14 TeV and
an unprecedented luminosity of 10$^{34}$ cm$^{-2}$s$^{-1}$. It can also collide heavy ions (Pb) with an
energy of 2.8 TeV per nucleon and a peak luminosity of 10$^{27}$ cm$^{-2}$s$^{-1}$.\\
Beams can cross in four points, each hosting a different experiment. At the crossing point the angle
between the beams is 200 $\mu$rad.\\
The two high luminosity insertions in point 1 and point 5, diametrically opposed, host
the general purpose experiments ATLAS and CMS respectively. Two more experiments, one aimed at the study of
heavy ions collisions (ALICE) and one designed to perform accurate studies on B physics (LHCb)
are located in point 2 and point 8 respectively.\\
Each LHC proton beam contains 2835 bunches of 10$^{11}$ protons each.
The existing CERN accelerator chain is used as injection system for the LHC. The bunches, with an energy of
26 GeV are formed in the PS, and are characterized by a 25 ns spacing. Three
trains of 81 bunches, corresponding to a total charge of 2.43 10$^{13}$ protons, are
then injected in the SPS on three consecutive PS cycles, thus filling 1/3 of the
SPS circumference. The resulting beam is accelerated to 450 GeV before being
transferred to the LHC. This cycle has to be repeated 12 times in order to fill both
the LHC counter-rotating beams.

\section{\label{sec:atlasms}The ATLAS Muon Spectrometer}

The Muon Spectrometer of the ATLAS \cite{GenPap} experiment has been designed to measure the momentum
of particles in the pseudorapidity range $|\eta|<$ 2.7 and to trigger on these particles in the
region $|\eta|<$ 2.4.\\
The target performance is a stand-alone transverse momentum resolution of approximately 10\% for 1 TeV tracks
(Fig.~\ref{fig:muonspec:momentum}), which translates into a sagitta along the z (beam) axis of about
500 $\mu$m, to be measured with a resolution of $\leq$ 50 $\mu$m.\\
Three air-core toroid magnets, one in the barrel (the central region of the detector, $|\eta|<$ 1.05) and two in the end-caps
(the regions at the edge of the experiment), host the particle detectors for the trigger and for precision
tracking. The bending power, which ranges between 1 and 7.5 Tm (depending on the pseudorapidity) and
the low amount of material crossed by the muons in the spectrometer, allow the precise determination
of the transverse momentum.\\
Resistive Plate Chambers (RPCs) and Thin Gap Chambers (TGCs) provide the information to the muon
trigger in the barrel and in the end-caps respectively, while Monitored Drift Tubes
(MDTs) in the barrel and end-caps regions and Cathode Strip Chambers (CSCs) in the forward region ($|\eta|>$ 2), precisely
measure the position in the bending plane \footnote{The bending plane is parallel to the beam axis and is called the $\eta$ view.
The plane perpendicular to the beam axis is the non-bending plane and is called $\phi$ view}.\\
Chambers in the barrel region are located between and on the eight coils of
the superconducting barrel toroid magnet, while the end-cap chambers are in front and behind the
two end-cap toroid magnets. The $\phi$ symmetry of the toroids is reflected in the symmetric structure
of the muon chamber system, consisting of eight octants. Each octant is subdivided in the azimuthal
direction in two sectors with slightly different lateral extensions, a large (L) and a small (S) sector, leading
to a region of overlap in $\phi$ (Fig.~\ref{fig:muonspec:phiview}).
\begin{figure}
\begin{center}
\begin{minipage}{14pc}
	\includegraphics[width=14pc]{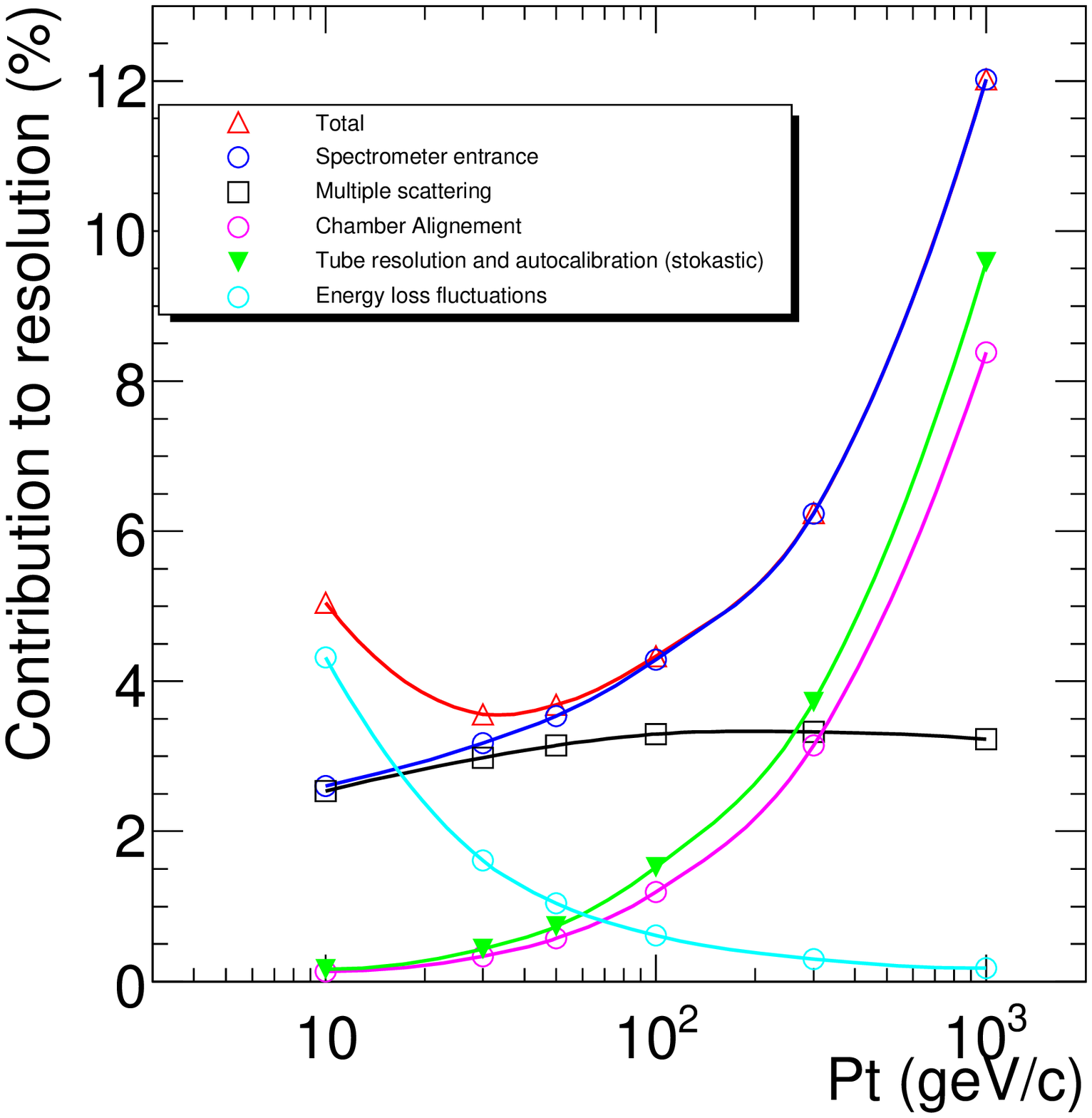}
\end{minipage}
\begin{minipage}{20pc}
	\caption{\label{fig:muonspec:momentum}Contributions to the momentum resolution for muons reconstructed
in the Muon Spectrometer as a function of transverse momentum for $|\eta|<$ 1.5.
The alignment curve is for an uncertainty of 30$\mu$m in the chamber positions.
}
\end{minipage}
\end{center}
\end{figure}
The chambers in the barrel are arranged in three concentric cylindrical shells around the beam
axis at radii of approximately 5 m, 7.5 m and 10 m.
In the two end-cap regions, muon chambers form large wheels, perpendicular to the z-axis and
located at distances of $|z|\sim$ 7.4 m, 10.8 m, 14 m and 21.5 m from the interaction point.
\begin{figure}
\begin{center}
\begin{minipage}{14pc}
	\includegraphics[width=14pc]{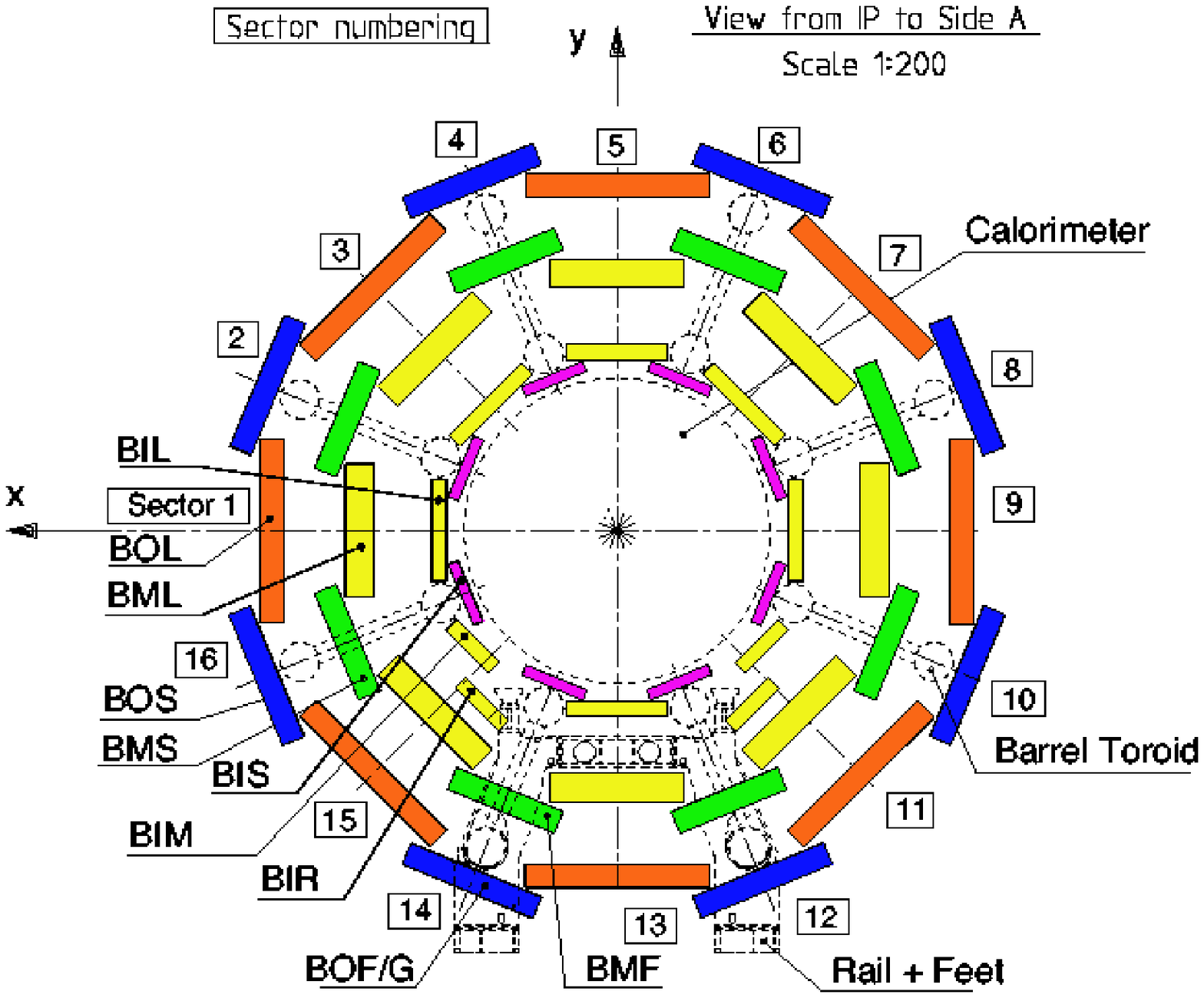}
\end{minipage}
\begin{minipage}{20pc}
	\caption{\label{fig:muonspec:phiview}Cross-section of the barrel muon system perpendicular
to the beam axis (non-bending plane), showing three concentric cylindrical layers of eight large
and eight small chambers. Chambers naming convention depending on their position is shown as well.}
\end{minipage}
\end{center}
\end{figure}

RPCs are assembled together with a MDT of equal dimensions in a common mechanical support structure.
RPCs provide the first-level muon trigger: two RPC chambers are installed together in the barrel middle station (BM)
which provide the low-p$_T$ trigger information.
The high-p$_T$ trigger makes use of the RPC modules installed on the outer barrel chambers (BO) combined
with the trigger result form the low-p$_T$ system. Each chamber consists of two independent detector layers,
each measuring $\eta$ and $\phi$ coordinates via two orthogonal sets of read-out strips.
A muon track that crosses three stations delivers six measurements in each view.
This redundancy in the track measurement allows the use
of a 3-out-of-4 coincidences in both projections for the low-p$_T$ trigger
and a 1-out-of-2 OR for the high-p$_T$ trigger. This coincidence schema increase the trigger robustness, 
rejecting fake tracks form noise and cavern background and ensuring good trigger efficiency with different
detector performance. 
A system of programmable coincidence logic allows concurrent operation with a total of six thresholds,
three associated with the low-p$_T$ trigger (threshold range approximately 6-9 GeV) and three associated
with the high-p$_T$ trigger (threshold range approximately 9-35 GeV).
The RPCs are also used to provide the coordinate along the MDT tubes in the non-bending plane.\\
Similarly, in the end-cap two TGC doublets and one triplet are installed close to the middle station
and provide the low-p$_T$ and high-p$_T$ trigger signals.
The TGCs are also measuring the coordinate of the muons in the direction parallel to the MDT wires.
In addition some TGC chambers are also installed close to the inner MDTs to improve the position measurement
accuracy along this coordinate.

\section{\label{sec:rpcs}The Resistive Plate Chambers of the ATLAS experiment}

The first published work on Resistive Plate Chambers (RPCs) was in 1981 \cite{Sant}. Since then
they had great development and they are now widely used in both high-energy and astro-particle physics
experiments.\\
RPCs are gaseous detectors and their operation is based on the detection of the gas ionization produced by
charged particles when traversing the active area of the detector, under a strong
uniform electric field applied by resistive electrodes. The electrodes are made
of a mixture of phenolic resins (usually called bakelite), which has a volume resistivity
$\rho_V$ between 10$^9$ and 10$^{12}$ $\Omega$cm. The plates are kept spaced by insulating spacers.
The thickness of the plates and the distance at which they are kept (the gas gap) are different for different
experiments. If RPC chambers are used as trigger detectors the gas gap is of the order of 2 mm while
in the case they are used as time of light detectors the gas gap reduces to 200 - 300 $\mu m$.
In the following we refer to the characteristics of RPCs for the ATLAS
experiment (Fig.~\ref{fig:rpc:struct}).
\begin{figure}
\begin{center}
\begin{minipage}{14pc}
	\includegraphics[width=14pc]{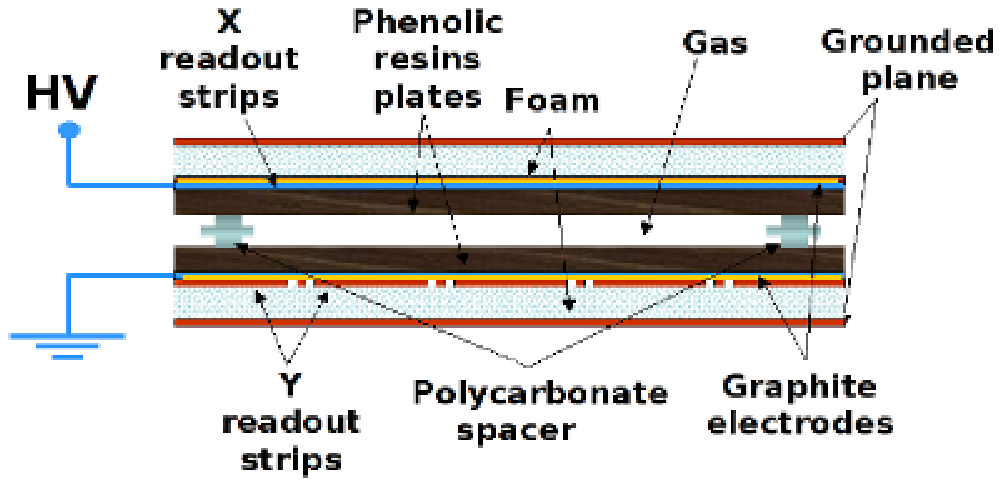}
\end{minipage}
\begin{minipage}{20pc}
	\caption{\label{fig:rpc:struct}Structure of a Resistive Plate Chamber of the ATLAS experiment.}
\end{minipage}
\end{center}
\end{figure}
RPCs used in the ATLAS experiment have 2 mm thick plates, kept at 2 mm one
from each other. The spacers are cylindrical with 12 mm diameter and are placed
one every $\sim$ 10 cm in both directions. The plate external surfaces are coated by thin
layers of graphite painting. The graphite has a surface resistivity of $\sim$ 100 k$\Omega$,
thus allowing uniform distribution of the high voltage along the plates without creating
any Faraday cage that would prevent signal induction outside the plates (as
would happen using, for example, metallic electrodes). Between the two graphite
coatings is applied a high voltage of about 9.8 kV (at room temperature and standard atmospheric pressure),
resulting in a very strong electric field which provides avalanche multiplication
of the primary ionization created by the incident particle. The presence
of such high electric field calls for extreme smoothness of the inner surfaces of the bakelite plates,
which is obtained covering the surfaces with a thin layer of linseed oil. The discharge
electrons drift in the gas and the signal induced on pick-up copper strips is
read out via capacitive coupling, and detected by the front-end electronics.
Read-out strips have a typical width of $\sim$ 30 mm and are grouped in two ($\eta$ and $\phi$)
read-out panels orthogonal to each other.\\
The gas together with the operating voltage determine the detector woking mode: avalanche or streamer.
Avalanche mode is the typical Townsend mechanism. According to this mechanism the avalanche multiplication
is due to the drifting electrons which collide with the gas molecule and produce further ionization.
For extremely high values of the electronic charge (established by Meek in
$\sim$ 10$^8$ electrons for noble gases) the avalanche becomes the precursor of a new
process, called streamer. In this phase, the electrons have low kinetic energy
and electron-ion recombination can occur with photon emission. Then the resulting photons
can further ionize the gas molecules, restarting one more avalanche multiplication,
delayed with respect to the first one. When the number of photons
is large and the electric field strong enough there are several secondary avalanches,
until the local density and electrons and ions distribution are such to connect the
two electrodes: a plasma column between the electrodes has been formed.
An extremely high current flows in the gas ($\sim$ 100 times larger
than the typical avalanche), until all the electrons and ions are collected.

In the ATLAS experiment RPCs operate in avalanche mode and this is achieved choosing an appropriate
high voltage and gas mixture: C$_2$H$_2$F$_4$ 94.7 $\%$ - C$_4$H$_{10}$ 5$\%$ - SF$_6$ 0.3 $\%$.\\
Indeed to work in avalanche mode the main component should be an electronegative gas,
with high enough primary ionization production, but with low free path for electron capture.
The high electronegative attachment coefficient limits the avalanche
electrons number below the Meek limit. A gas showing these characteristics is the
Tetrafluorethane (C$_2$H$_2$F$_4$).

Another component is a polyatomic gas, usually a hydrocarbon (C$_4$H$_{10}$), which has a
high absorption probability for ultra violet photons, produced in electron-ion recombinations.
This component allows to dissipate the photon energy by rotational-vibrational energy levels.
Then the introduction of a quencher (as SF$_6$) furtherly suppresses streamer formation allowing to
work in a pure avalanche mode, \cite{Camarri} avoiding photoionization with related multiplication
and limiting the lateral charge spread.\\
RPCs spatial resolution depends both on the read-out geometry an electronics.
Using an analog readout it is possible to obtain resolution of $\sim$ 1 cm, but with a
digital read-out the resolution is limited by the strip width, typically of the order
of a few centimeters.\\
Concerning time resolution, the high and uniform electric field applied to the gas gap by the
electrodes is the same for all primary clusters, producing at fixed time the same
avalanche growth, limited by the distance of the primary clusters from the anode.
The signal at any time is the sum of simultaneous contributions from all primary clusters
multiplications. The resulting time jitter for detectable signals is always $<$ 2 ns.\\
The excellent time resolution makes the RPCs a very good candidate for trigger
detector.


\section{\label{sec:rpcperf}RPCs performance}

After the installation of the ATLAS detector in its experimental cavern, the main
focus of activity is presently to study in situ its performance.
Using the reconstructed tracks of cosmic muons, the performance of the detectors have been verified
as by design, after the integration of the chambers in the full apparatus.
Results shown in the following sections are from data recorded over the full Muon Spectrometer
( $\sim$ 8000 RPC panels ).

\subsection{\label{sec:efficiency}Efficiency}

In order to determine the RPC efficiency, it has to be taken into account that RPCs are actually providing
the muon trigger thus introducing a bias on the efficiency measurements.\\
Since the trigger majority doesn't include all the layers in the trigger decision, an unbiased sample of
data can be used to calculate the efficiency for a given detector layer, excluding that layer from the
trigger decision. For example if the trigger is requiring a coincidence of at least 3 RPC layers
out of the 4 in the low-p$_T$ (BM) station one can have an unbiased efficiency measurement for layer 1,
by using only events where layers 2, 3 and 4 had a hit. This method has been applied in the analysis
and in Fig.~\ref{fig:rpcperf:effBM} the RPC efficiency is estimated extrapolating muon tracks
reconstructed from MDT 
hits and requiring the presence of an RPC hit within $\pm$ 70 mm. Corrections of the high voltage for temperature
and pressure variations \cite{Abb}, \cite{Cam} are not applied.

\begin{figure}
\begin{center}
\begin{minipage}{14pc}
	\includegraphics[width=14pc]{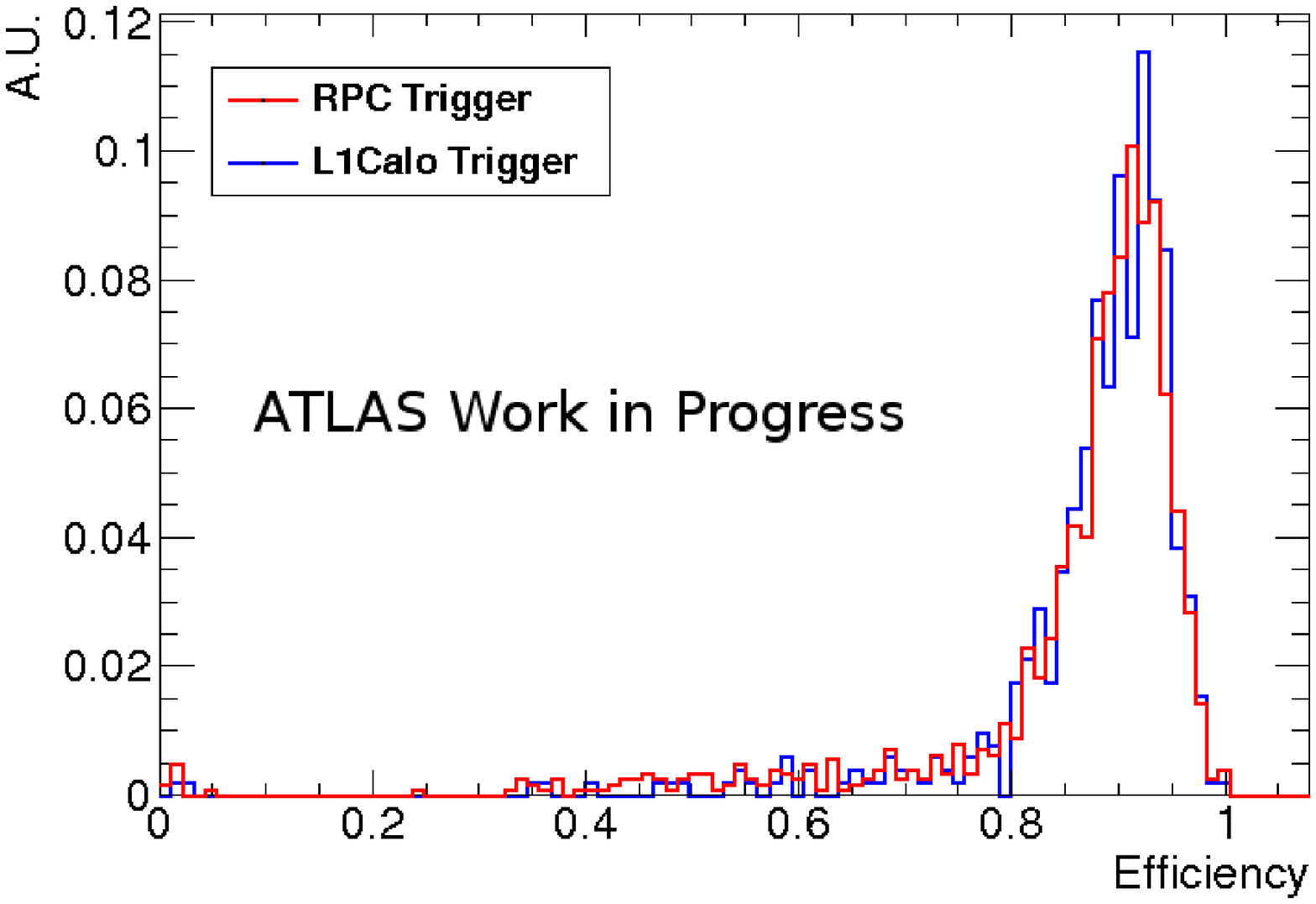}
\end{minipage}
\begin{minipage}{20pc}
	\caption{\label{fig:rpcperf:effBM}Distribution of average efficiency for RPC BM panels
at working point (high voltage 9.6 kV). The two histograms are referred to two different trigger
sources: RPC trigger and calorimeter-based trigger (L1Calo). The effectiveness of the algorithm to remove
offline the trigger bias is clearly visible. The histograms are normalized to unity.}
\end{minipage}
\end{center}
\end{figure}
In Fig.~\ref{fig:rpcperf:POk} the dependency of the efficiency on the applied high voltage for a given
read-out panel is shown. Analogous plots are produced for each read­-out panel.
The study of efficiency dependence on the high voltage is fundamental to determine the working point of the detector.
Indeed to operate in stable detector conditions the operating high voltage is choosen at the plateau
value extracted from this kind of plots.
\begin{figure}
\begin{center}
\begin{minipage}{14pc}
	\includegraphics[width=14pc]{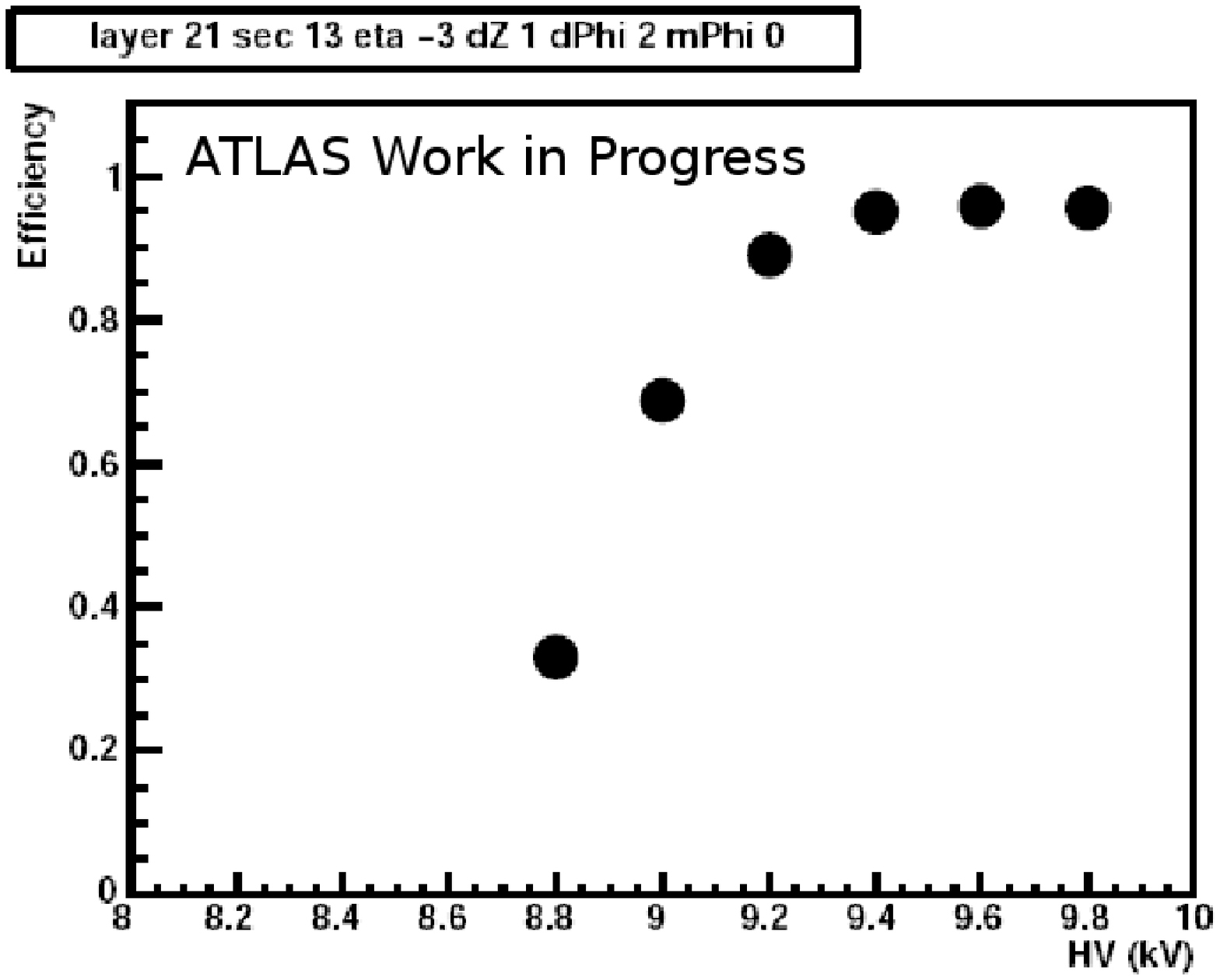}
\end{minipage}
\begin{minipage}{20pc}
	\caption{\label{fig:rpcperf:POk}Dependence of the efficiency on the high voltage applied to
the resistive plates for a given read-out panel.}
\end{minipage}
\end{center}
\end{figure}

Fig.~\ref{fig:rpcperf:effDiff} shows the difference in the RPCs efficiency measurement distribution at
different high voltage values. The efficiency plateau is already reached for 9.6 kV.
\begin{figure}
\begin{center}
\begin{minipage}{14pc}
	\includegraphics[width=14pc]{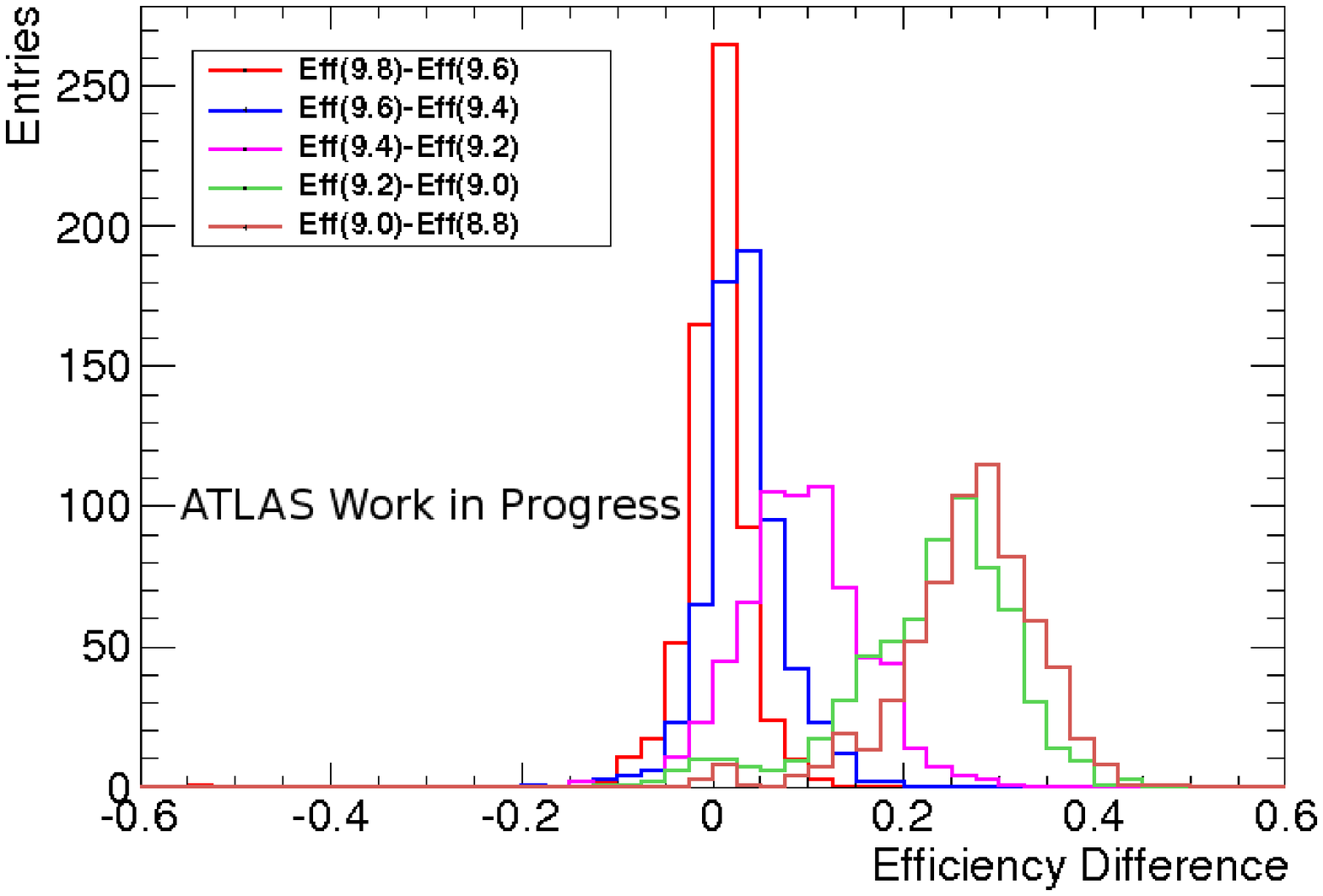}
\end{minipage}
\begin{minipage}{20pc}
\caption{\label{fig:rpcperf:effDiff} Distribution of efficiency gradients between different high voltage values.
Each panel gives an entry in each one of the distributions. The red plot for example shows that panels do not
change significantly their efficiency between 9.8 kV and 9.6 kV. Brown and green curves show instead the
effect of the rise of the efficiency plateaus.}
\end{minipage}
\end{center}
\end{figure}
Another important test is the comparison of the efficiency distribution in in the two cases with
magnetic field on and off. As shown in Fig.~\ref{fig:rpcperf:effmag} the efficiency is not affected by
turning on the field, as one would expect.
\begin{figure}
\begin{center}
\begin{minipage}{14pc}
	\includegraphics[width=14pc]{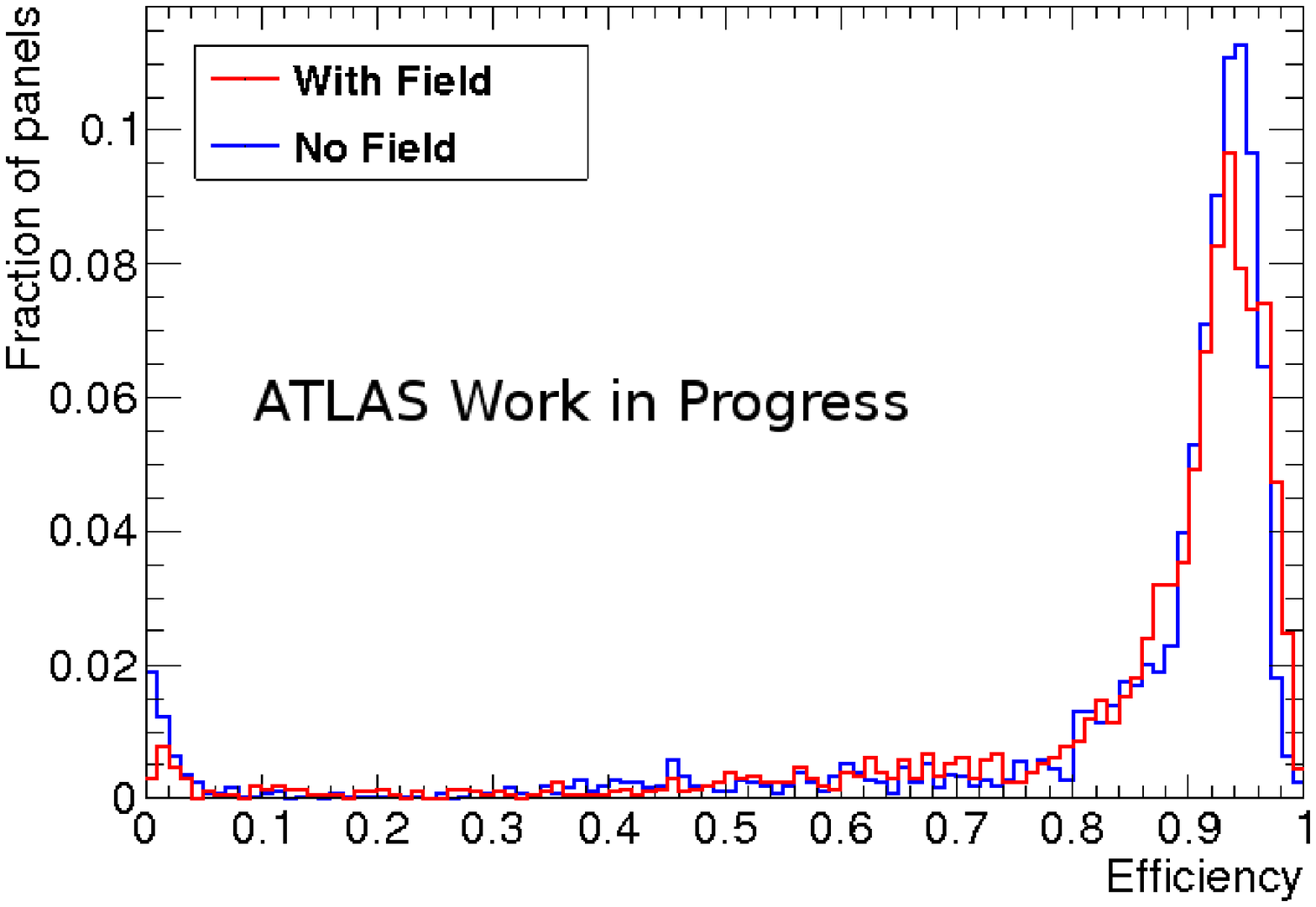}
\end{minipage}
\begin{minipage}{20pc}
	\caption{\label{fig:rpcperf:effmag}Distribution of average efficiency for RPC BM panels
at working point (high voltage 9.6 kV). The two histograms are referred to the case of magnetic field on
and off. The histograms are normalized to unity.}
\end{minipage}
\end{center}
\end{figure}

\subsection{\label{sec:clustersize}Cluster Size}

The signal induced on a strip and read out by the front-end electronics is defined as a hit.
To study the RPC performance and to check the RPC effciency a clustering algorithm
has been implemented. The algorithm is based on the following rules:

\begin{itemize}

  \item for each event a cluster is defined as a group of adjacent strips hit at the
  same time or within 15 ns. This time range is estimated as the maximum
  time for the signal induction/cross talk;
  \item in order not to count hits due to after pulses, only the first hit of each strip is considered
  by the cluster algorithm;
  \item the first hit in time defines the cluster time;
  \item the number of strips forming the cluster is the \emph{cluster size};
  \item the cluster centre is given by the geometrical centre of all the strips belonging
  to the same cluster.
  \item in a read-out panel more than one cluster can occour.

\end{itemize}
Fig.~\ref{fig:rpcperf:CSEtaPhi} shows the distribution of cluster size for $\eta$ and $\phi$ panels.
$\eta$ view cluster size is a little bit lower with respect to $\phi$ view.
This is as expected due to the chamber construction geometry. Indeed only $\eta$ read-out panels are
separated from the gas volume by a PET foil.
\begin{figure}
\begin{center}
\begin{minipage}{14pc}
	\includegraphics[width=14pc]{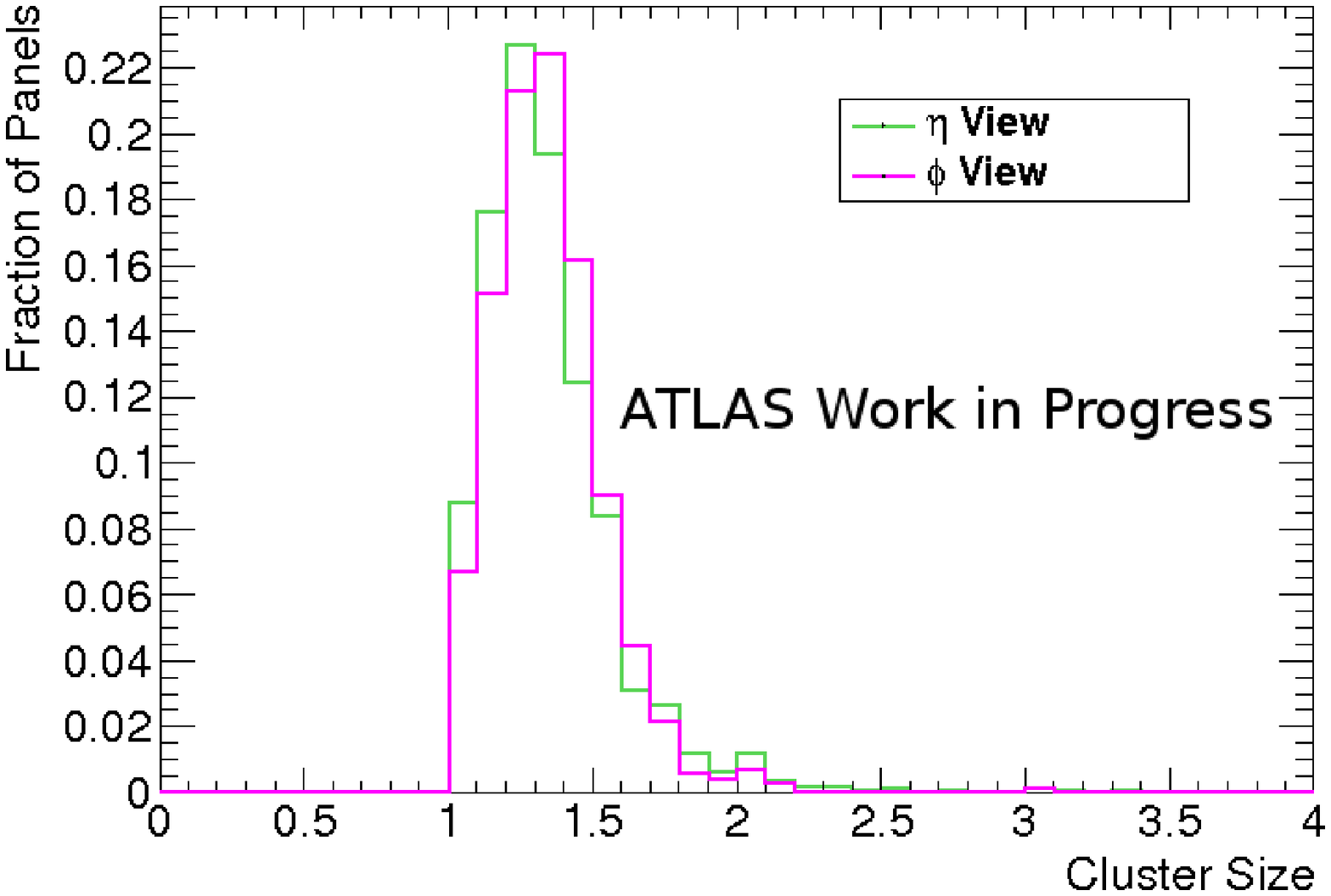}
\end{minipage}
\begin{minipage}{20pc}
	\caption{\label{fig:rpcperf:CSEtaPhi}Distribution of cluster size for $\eta$ and $\phi$ panels}
\end{minipage}
\end{center}
\end{figure}
In Fig.~\ref{fig:rpcperf:CSBMBO} the distribution of cluster size for BM and BO panels is shown. No relevant
difference between the two distribution is visible, as expected.
\begin{figure}
\begin{center}
\begin{minipage}{14pc}
	\includegraphics[width=14pc]{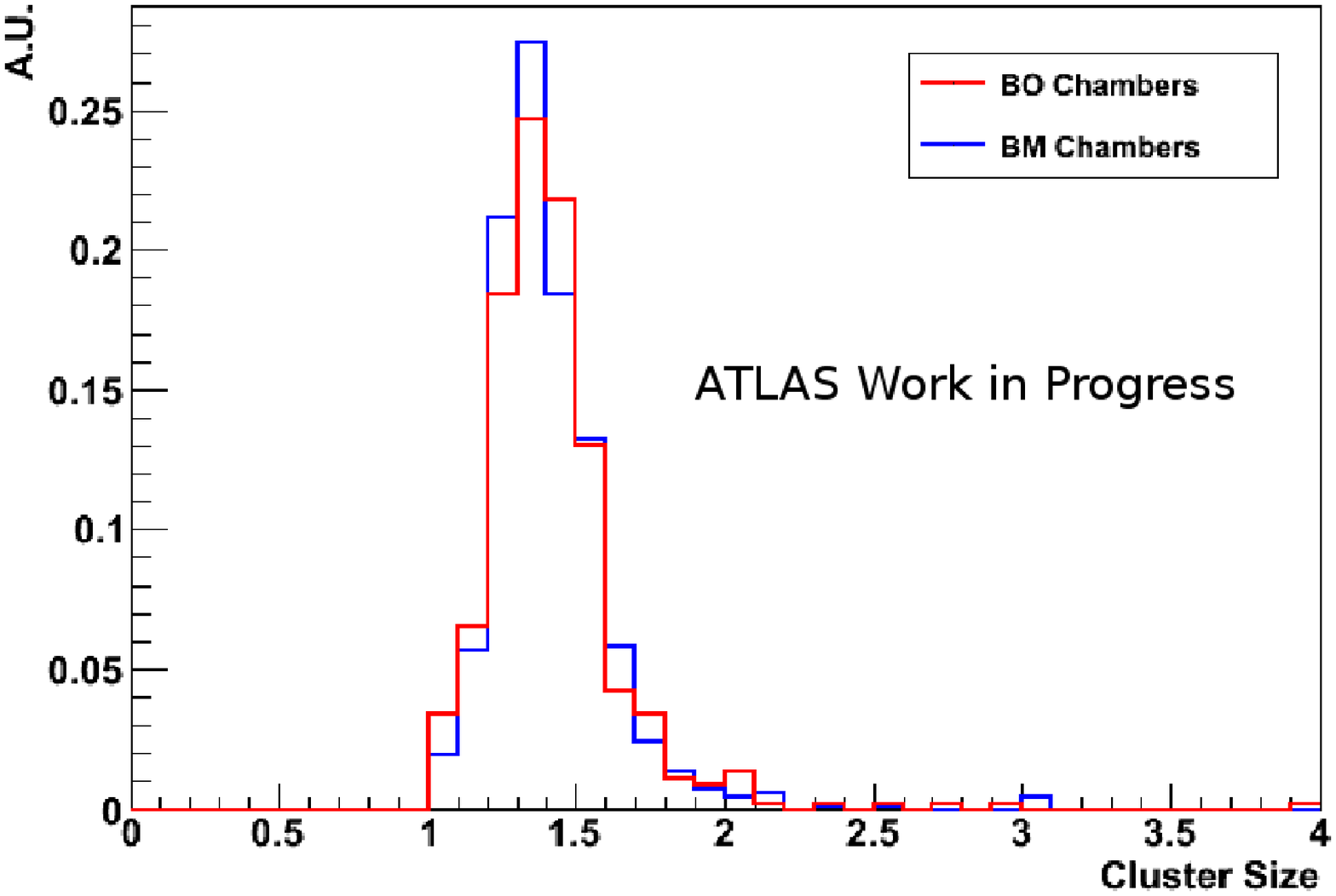}
\end{minipage}
\begin{minipage}{20pc}
	\caption{\label{fig:rpcperf:CSBMBO}Distribution of cluster size for BM and BO panels}
\end{minipage}
\end{center}
\end{figure}

\subsection{\label{sec:spresolution}Spatial Resolution}

Fig.~\ref{fig:rpcperf:Sigma12BM} shows the spatial resolutions as measured on $\eta$ panels, for clusters
of size 1 and 2. The residual~\footnote{Residuals are defined as the position of the RPC cluster respect to
the position of the MDT track extrapolation on RPC plane.} distributions are fitted with a Gaussian, and the resulting standard deviation
is divided by the strip pitch, to allow comparison between different panels.
As expected, clusters of size 2 give an improved spatial resolution due to the fact that the particle has
crossed a very narrow region near the border of two adjacent strips.
\begin{figure}[h]
\begin{center}
\begin{minipage}{14pc}
	\includegraphics[width=14pc]{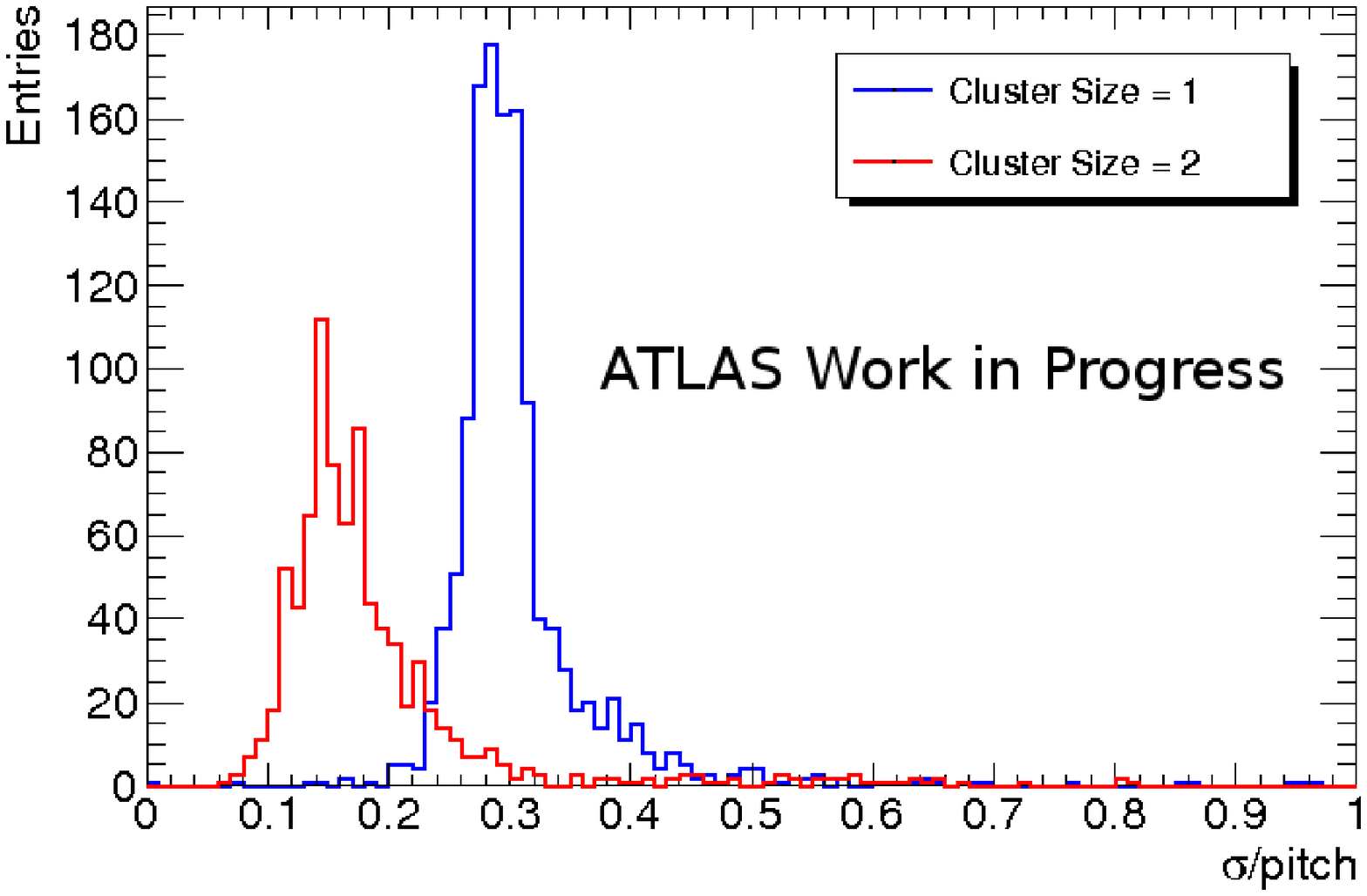}
\end{minipage}
\begin{minipage}{20pc}
	\caption{\label{fig:rpcperf:Sigma12BM}Distribution of spatial resolution divided by
the strip pitch for clusters of size 1 (blue line) and size 2 (red line).
Only $\eta$ RPC panels of BM chambers are shown.}
\end{minipage}
\end{center}
\end{figure}

Fig.~11 (12) shows the comparison of spatial
resolution with magnetic field on and off for clusters of size 1 (2).
No relevant difference between the two distributions is visible as expected.
\begin{figure}[h]
\begin{center}
\begin{minipage}{14pc}
	\includegraphics[width=14pc]{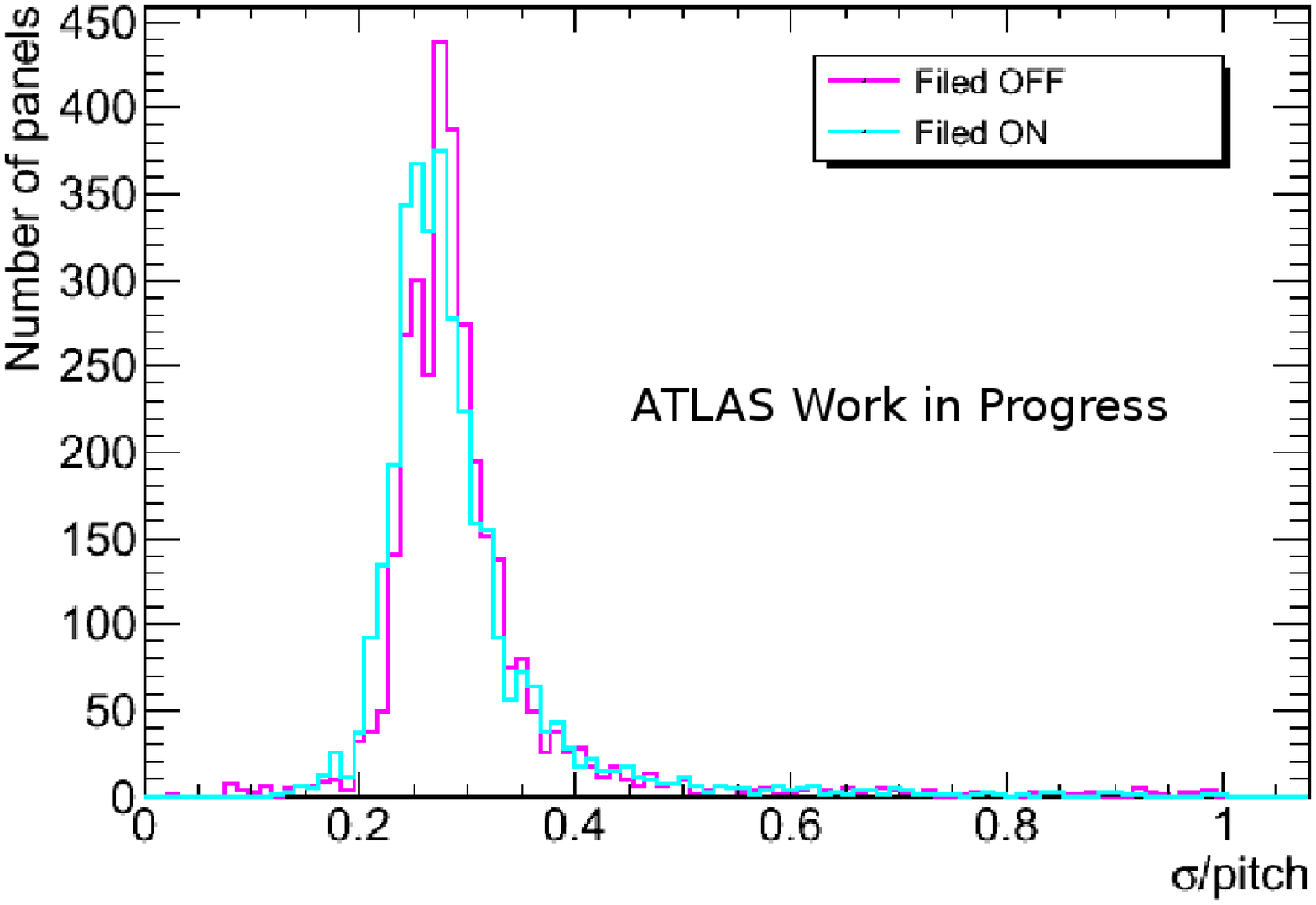}
\end{minipage}
\begin{minipage}{20pc}
	\caption{\label{fig:rpcperf:Sigma11ONOFF}Distribution of spatial resolution divided by
the strip pitch for clusters of size 1 in the case with magnetic field on and off.
Only $\eta$ RPC panels of BM chambers are shown.}
\end{minipage}
\end{center}
\end{figure}
\begin{figure}[h]
\begin{center}
\begin{minipage}{14pc}
	\includegraphics[width=14pc]{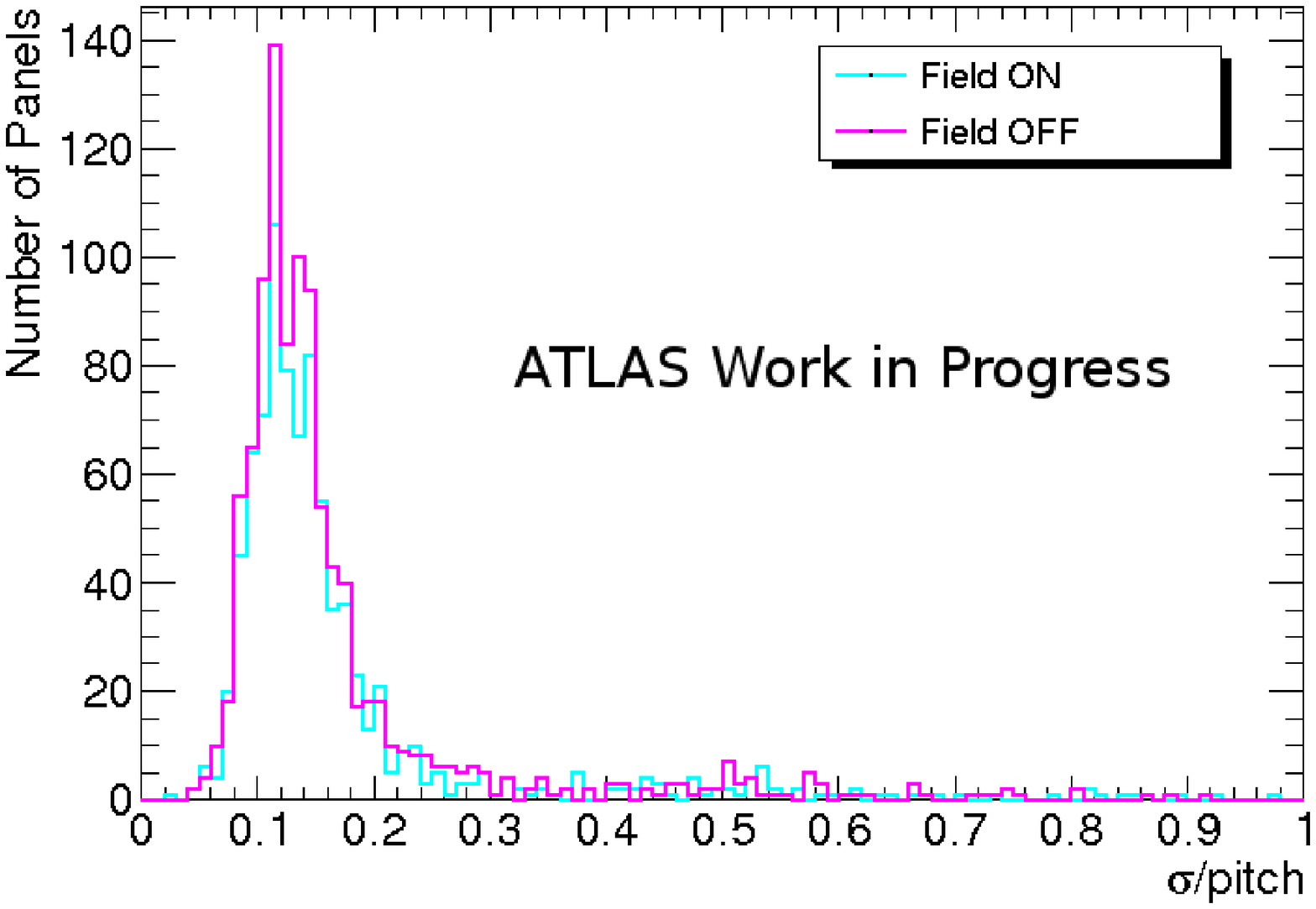}
\end{minipage}
\begin{minipage}{20pc}
	\caption{\label{fig:rpcperf:Sigma21ONOFF}Distribution of spatial resolution divided by
the strip pitch for clusters of size 2 in the case with magnetic field on and off.
Only $\eta$ RPC panels of BM chambers are shown.}
\end{minipage}
\end{center}
\end{figure}

\section{\label{sec:conlusions}Conclusions}

The ATLAS detector is completely installed in the cavern and is ready for the beams.

Resistive Plate Chambers undergone extensive tests to verify their performance and the results shown here
represent a part of the effort to bring the ATLAS Muon Spectrometer to life with cosmics.



\newpage

\lhead[\fancyplain{}{}]{\itshape \small{Anomalous U(1)' Phenomelogy: LHC and Dark Matter}}
\rhead[\itshape \large{1$^\mathrm{st}$ RYRM Proceedings}]{\fancyplain{}{A. Racioppi}}

\def\bea{\begin{eqnarray}}
\def\eea{\end{eqnarray}}

\def\bei{\begin{itemize}}
\def\eei{\end{itemize}}
\def\bs{\begin{slide}}
\def\es{\end{slide}}
\def\nn{\nonumber}
\def\half{\frac{1}{2}}
\def\pd{\partial}
\def\L{\mathcal{L}}
\def\M{\mathcal{M}}
\def\a{\alpha}
\def\ad{{\dot\alpha}}
\def\b{\beta}
\def\bd{{\dot\beta}}
\def\g{\gamma}
\def\G{\Gamma}
\def\d{\delta}
\def\D{\Delta}
\def\m{\mu}
\def\n{\nu}
\def\t{\tau}
\def\r{\rho}
\def\th{\theta}
\def\w{\omega}
\def\thb{{\bar\theta}}
\def\l{\lambda}
\def\lb{{\bar\l}}
\def\Lb{{\bar\Lambda}}
\def\s{\sigma}
\def\sm{\frac{\vec{\s}}{2}}
\def\sb{\bar{\s}}
\def\psib{{\bar{\psi}}}
\def\e{\epsilon}
\def\f{\phi}
\def\z{\zeta}
\def\A{\vec{A}}
\def\MZp{M_{Z'}}
\def\MZO{M_{Z_0}}
\def\MW{M_W}
\def\p{\not{p}}
\def\({\left(}
\def\){\right)}
\def\[{\left[}
\def\]{\right]}
\def\ft{\tilde{f}}
\def\gt{\tilde{g}}
\def\qt{\tilde{q}}
\def\ut{\tilde{u}}
\def\dt{\tilde{d}}
\def\topt{\tilde{t}}
\def\bt{\tilde{b}}
\def\hti{\tilde{h}}
\def\utcs{{{\tilde{u}^{c \dag}}}}
\def\dtcs{{{\tilde{d}^{c \dag}}}}
\def\lt{\tilde{l}}
\def\et{\tilde{e}}
\def\mt{\tilde{\mu}}
\def\taut{\tilde{\tau}}
\def\nt{\tilde{\nu}}
\def\Nt{\tilde{N}}
\def\etcs{{{\tilde{e}^{c \dag}}}}
\def\dpd{\overleftrFightarrow{\pd}}
\def\vf{\varphi}
\def\la{\langle}
\def\ra{\rangle}
\def\Tr{\textnormal{Tr}}
\def\STr{\textnormal{STr}}
\def\Q{Q_{CS}}
\def\i{\textnormal{i}}
\def\Eps{\epsilon^{\mu\nu\rho\sigma}}
\def\emn{\eta_{\m\n}}
\def\dmn{\d_{\m\n}}
\def\ithth{\int d^2\theta d^2\bar\theta\;}
\def\thth{\theta^2 \bar\theta^2}
\def\cD{\mathcal{D}}
\def\QQ{{Q_Q}}
\def\QU{{Q_{U^c}}}
\def\QD{{Q_{D^c}}}
\def\QL{{Q_L}}
\def\QE{{Q_{E^c}}}
\def\QHu{{Q_{H_u}}}
\def\QHd{{Q_{H_d}}}
\def\cA{\mathcal{A}}
\def\cw{\cos\th_W}
\def\sw{\sin\th_W}
\def\cscw{\text{cosec}\th_W}
\def\cotw{\cot\th_W}
\newcommand{\bth}{{\bf 3}}
\newcommand{\btw}{{\bf 2}}
\newcommand{\bon}{{\bf 1}}
\def\slashed{\ds}
\def\as{\slashed a}
\def\bs{\slashed b}
\def\cs{\slashed c}
\def\conj{{{\rm h.c.}}}
\def\MPlanck{M_{\rm P}}
\def\gu{\underline{g}}
\def\Baryon{{\rm B}}
\def\Lepton{{\rm L}}
\def\Stuckelberg{{St\"uckelberg }}
\def\axion{\text{axion}}
\def\chiral{\text{chiral}}
\def\GS{\text{GS}}
\def\GCS{\text{GCS}}
\def\NG{\text{NG}}
\def\GeV{\text{GeV}}
\def\BR{\text{BR}}
\def\numeq{n^\text{eq}}
\def\hri#1#2{\href{http://arxiv.org/abs/#1}{arXiv:#1 #2}}
\def\hre#1#2{\href{http://arxiv.org/abs/#1/#2}{arXiv:#1/#2}}
\def\ds#1{#1\kern-1ex\hbox{/}}
\def\sla{\raise.15ex\hbox{$/$}\kern-.57em}
\begin{center}

\setcounter{section}{8}
\section{\large{Anomalous $U(1)'$ Phenomelogy: LHC and Dark Matter}}
\label{racioppi}
\normalsize{Antonio Racioppi$^*$}

\emph{\small{Dipartimento di Fisica dell'Universit\`a di Roma , ``Tor Vergata" and\\
I.N.F.N.~ -~ Sezione di Roma ~ ``Tor Vergata''\\
\mbox{Via della Ricerca  Scientifica, 1 - 00133 ~ Roma,~ ITALY}\\ }}

%
%
%
\vspace{.5cm}

\begin{minipage}{.8\textwidth}
\small{We study an extension of the MSSM by an anomalous abelian vector multiplet and a \Stuckelberg multiplet.
The anomalies are cancelled by the Green-Schwarz mechanism.
We compute the decays $Z' \to Z_0 \gamma$ and $Z' \to Z_0 Z_0$ and find that the expected number of events per year at
LHC is at most of the order of 10. Then we compute the relic density predicted by our model with a new dark matter
candidate, the axino, which is the LSP of the theory. We find agreement with experimental data considering
coannihilations with a wino-like neutralino NLSP and a wino-like chargino NNLSP.}

\vspace{.25cm}
$*$ e-mail: \href{mailto:Antonio.Racioppi@roma2.infn.it}{Antonio.Racioppi@roma2.infn.it}
\end{minipage}
\end{center}
\setcounter {section} {0}
\setcounter {subsection} {0}
\setcounter {figure} {0}
\setcounter{equation}{0}

\section{Introduction}
A great deal of work has been done recently to embed the standard model of particle physics (SM) into a brane construction.
Such brane constructions naturally lead to extra anomalous $U(1)$'s in the four dimensional low energy theory and,
in turn, to the presence of possible heavy $Z^\prime$ particles in the spectrum. These particles should be among the
early findings of LHC and besides for the above cited models they are also a prediction of many other theoretical
models of the unification of forces. It is then of some interest
to know if these $Z^\prime$ particles contribute to the cancellation of the gauge anomaly in the way predicted from
string theory or not~\footnote{For references on these topics, see \cite{myPhD}, \cite{Langacker:2008yv} and the bibliography therein.}.
In \cite{myPhD} the present author studied a supersymmetric (SUSY) extension
of the minimal supersymmetric standard model (MSSM) in which the anomaly is cancelled {\it \`a la} Green-Schwarz.
The advantage of this choice over the standard one is that it allows for arbitrary values of the quantum numbers of the extra U(1).
The model is only string-inspired in order to explore the phenomenology of these models keeping a high degree of flexibility.
As a first step towards the study of hadron
annihilations producing four leptons in the final state (a clean signal which might be studied at LHC) we then
compute the decays $Z' \to Z_0 \gamma$ and $Z' \to Z_0 Z_0$. We find that the expected number of events per year at
LHC is at most of the order of 10. Then we compute the relic density predicted by our model with a new dark matter
candidate, the axino, which is the LSP of the theory. We find agreement with experimental data considering
coannihilations with a wino-like neutralino NLSP and a wino-like chargino NNLSP.

\section{Model Setup} \label{sect:Model Setup}
In this section we briefly discuss our theoretical framework. We assume an extension of the MSSM with an additional abelian
vector multiplet $V^{(0)}$ with arbitrary charges. The anomalies are cancelled with the Green-Schwarz (GS) mechanism
and with the Generalized Chern-Simons (GCS) terms. All the details can be found in~\cite{myPhD}.
All the MSSM fields are charged under the additional vector multiplet $V^{(0)}$, with charges which are
given in Table~\ref{QTable}, where $Q_i, L_i$ are the left handed quarks and leptons respectively while $U^c_i,
D^c_i, E^c_i$ are the right handed up and down quarks and the electrically charged leptons. 
The superscript $c$ stands
for charge conjugation. The index $i=1,2,3$ denotes the three different families. $H_{u,d}$ are the two Higgs scalars.
The key feature of this model is the mechanism of anomaly cancellation. As it is well known, the MSSM is anomaly free.
In our MSSM extension all the anomalies that involve only the $SU(3)$, $SU(2)$ and $U(1)_Y$ factors vanish identically.
However, triangles with $U(1)'$ in the external legs in general are potentially anomalous.
\vspace{3cm}
\begin{table}[h]
\caption{Charge assignment.}\label{QTable}
\begin{ruledtabular}
\begin{tabular}{ccccc}
            & $SU(3)_c$    &   $SU(2)_L$   & $U(1)_Y$ & ~$U(1)'~$\\
  \hline
    $Q_i$   & $\bth$       &  $\btw$       &  $1/6$   & $Q_{Q}$ \\
  $U^c_i$   & $\bar \bth$  &  $\bon$       &  $-2/3$  & $Q_{U^c}$\\
  $D^c_i$   & $\bar \bth$  &  $\bon$       &  $1/3$   & $Q_{D^c}$\\
    $L_i$   & $\bon$       &  $\btw$       &  $-1/2$  & $Q_{L}$ \\
  $E^c_i$   & $\bon$       &  $\bon$       &  $1$     & $Q_{E^c}$\\
      $H_u$ & $\bon$       &  $\btw$       &  $1/2$   & $Q_{H_u}$\\
      $H_d$ & $\bon$       &  $\btw$       &  $-1/2$  & $Q_{H_d}$ \\
\end{tabular}
\end{ruledtabular}
\end{table}

These anomalies are~\footnote{We are working in an effective field theory framework and we ignore all the gravitational effects. In particular, we do not consider the gravitational anomalies which, however, could be cancelled by the Green-Schwarz mechanism.}
\bea
   U(1)'-U(1)'-U(1)':     &&\ \cA^{(0)}  \\
   U(1)'-U(1)_Y - U(1)_Y: &&\ \cA^{(1)} \\
   U(1)'-SU(2)-SU(2):     &&\ \cA^{(2)} \\
   U(1)'-SU(3)-SU(3):     &&\ \cA^{(3)} \\
   U(1)'-U(1)'-U(1)_Y:    &&\ \cA^{(4)}
\eea
All the remaining anomalies that involve $U(1)'$s vanish identically due to group theoretical arguments.
Consistency of the model is achieved by the contribution of a \Stuckelberg field $S$ and its appropriate couplings to the anomalous $U(1)'$.
The \Stuckelberg lagrangian written in terms of superfields is
\begin{widetext}
\be
 \L_S = {\frac{1}{4}} \left. \( S + S^\dagger + 4 b_3 V^{(0)} \)^2 \right|_{\thth}
      - {\frac{1}{2}} \left\{ \[\sum_{a=0}^3 b^{(a)}_2 S \Tr\( W^{(a)} W^{(a)} \) + b^{(4)}_2 S
 W^{(1)}W^{(0)}\]_{\th^2} +h.c. \right\}
   \label{Laxion}
\ee
\end{widetext}
where the index $a=0,\ldots,3$ runs over the $U(1)',\, U(1)_Y,\, SU(2)$ and $SU(3)$ gauge groups respectively.
The \Stuckelberg multiplet is a chiral superfield
\bea
 S &=&  s+ i\sqrt2 \th \psi_S + \th^2 F_S - i \th \s^\m \bar\th \pd_\m s+\nn\\
 &&{\frac{\sqrt2}{2}}  \th^2 \bar\th \bar\s^\m \pd_\m \psi_S-{\frac{1}{4}} \thth \Box s \label{Smult}
\eea
and transforms under $U(1)'$ as
\bea
   V^{(0)} &\to& V^{(0)} + i \( \Lambda - \Lambda^\dag \) \nn\\
   S  &\to& S - 4 i ~b_3 ~\Lambda \label{U1'}
  \label{U1Trans}
\eea
where $b_3$ is a constant related to the $Z'$ mass.
As it was pointed out in \cite{Anastasopoulos:2006cz}, the \Stuckelberg mechanism is not enough to cancel all the
anomalies: non gauge invariant GCS terms must be added.
\bea
 &&\!\!\!\!\!\!
   \L_{GCS} =- d_4 \[ \( V^{(1)} D^\a V^{(0)} \! - \! (0 \leftrightarrow 1) \) W^{(0)}_\a \! + h.c. \]_{\thth} \!\!\! +\nn\\
 &&\phantom{\!\!\!\!\!\!\!  \L_{GCS} =}
            +  d_5 \[ \( V^{(1)} D^\a V^{(0)} \! - \! (0 \leftrightarrow 1) \) W^{(1)}_\a \! + h.c. \]_{\thth} \!\!\! +\nn\\
       &&\!\!\!\!\!\!\!
+  d_6 \Tr \bigg[\!\!\! \( V^{(2)} D^\a V^{(0)} \! - \!  (0 \leftrightarrow 2)\) W^{(2)}_\a \!+ n.a.c\! + h.c. \bigg]_{\thth} \nn\\
   \label{GCS_1}
\eea
where $n.a.c.$ refers to non abelian completion terms.
The $b$ constants in (\ref{Laxion}) and the $d$ constants in (\ref{GCS_1}) are fixed by the anomaly cancellation procedure (for details see~\cite{myPhD}).

For a symmetric distribution of the anomaly, we have
\bea
 && \qquad b^{(0)}_2 b_3 =-\frac{\cA^{(0)}}{384\pi^2} ~ \qquad b^{(1)}_2 b_3 = - \frac{\cA^{(1)}}{128 \pi^2} \label{bsds} \\
 && \qquad b^{(2)}_2 b_3 = -\frac{\cA^{(2)}}{64 \pi^2} \quad \qquad b^{(4)}_2 b_3 = -\frac{\cA^{(4)}}{128 \pi^2}\nn\\
 && d_4 =- \frac{\cA^{(4)}}{384 \pi^2}~~~~~ \qquad d_5 = \frac{\cA^{(1)}}{192\pi^2}~~~~~ \qquad d_6= \frac{\cA^{(2)}}{96 \pi^2} \nn
\eea
It is worth noting that the GCS coefficients $d_{4,5,6}$ are fully determined in terms of the $\cA$'s by the gauge invariance, while the
$b_2^{(a)}$'s depend only on the free parameter $b_3$, which is related to the mass of the anomalous $U(1)$.

The soft breaking sector of the model is given by
\be
 \L_{soft}=\L_{soft}^{MSSM}- {\frac{1}{2}}  \(M_0 \l^{(0)} \l^{(0)} + \frac{M_S}{2} \psi_S \psi_S  + h.c. \) \label{Lsoft}
\ee
where $\L_{soft}^{MSSM}$ is the usual soft susy breaking lagrangian while  $\l^{(0)}$ is the gaugino of the added $U(1)'$ and $\psi_S$ is the axino.
The axino soft mass term deserves some comment.
Based on~\cite{Girardello:1981wz} one could naively expect that such a term is not allowed since the \Stuckelberg multiplet is a chiral multiplet.
This turns out not to be correct because the \Stuckelberg multiplet couplings are not of the Yukawa type.
In fact in our model both the axino and the axion couple only through GS interactions.
It is worth noting that a mass term for the axion $\f$ is instead not allowed since it
transforms non trivially under the anomalous $U(1)'$ gauge transformation~(\ref{U1Trans}).

\section{Neutral Gauge Boson Sector} 
\label{Gauge}
It is well known that usually the $Z'$ and the Standard Model $Z_0$ mix with each other (see for instance see \cite{Langacker:2008yv}).
Anyway nobody forbids a priori the absence of such mixing. Then, as a first step, we consider the case where the SM and $U(1)'$ gauge sector are decoupled.
This can be achieved at tree level with the choice
\be
 \QHu=\QHd=0 \label{QHuchoice}
\ee
The mass terms for the neutral gauge fields become
\be
 \L_M = \frac{1}{2} \(V^{(0)}_\m \ V^{(1)}_\m \ V^{(2)}_{3\m} \) M^2
         \( \begin{array}{c} V^{(0)\m}\\
                             V^{(1)\m}\\
                             V^{(2)\m}_{3}
             \end{array} \)
     \ee
with $M^2$ being the gauge boson mass matrix
\be
 M^2= \( \begin{array}{ccc} M_{V^{(0)}} & ~~~0& ~~~0  \\
            ... & g_1^2 \frac{v^2}{4} & -g_1 g_2 \frac{v^2}{4}   \\
            ... & ...  & g_2^2 \frac{v^2}{4}  \\
         \end{array} \)
\label{BosonMasses}
\ee
where $M_{V^{(0)}}=4 b_3 g_0$  is the \Stuckelberg mass parameter for the anomalous $U(1)$ and it is assumed to be in the TeV range.
The lower dots denote the obvious terms under symmetrization. After diagonalization, we obtain the following eigenstates and eigenvalues
\bea
 &&\!\!\!\!\!\!\!\!\!\!\!\!\!\!\!\!\!\!\!\!
 A_\m    = \frac{g_2 V^{(1)}_\m + g_1 V^{(2)}_{3\m}}{\sqrt{g_1^2+g_2^2}} \, , \qquad ~M^2_{\g}=0
 \label{photon}\\
 &&\!\!\!\!\!\!\!\!\!\!\!\!\!\!\!\!\!\!\!\!\!\!
 Z_{0\m} = \frac{g_2 V^{(2)}_{3\m} - g_1 V^{(1)}_\m}{\sqrt{g_1^2+g_2^2}} \, , \qquad M^2_{Z_0}=\frac{1}{4} \(g_1^2+g_2^2\) v^2
 \label{Z0}\\
 &&\!\!\!\!\!\!\!\!\!\!\!\!\!\!\!\!\!\!\!\!
 Z'_\m  = V^{(0)}_\m \, , \qquad \qquad \qquad \quad ~M^2_{Z'}=M_{V^{(0)}}^2
 \label{Zprime}
\eea

To estimate the number of the anomalous decays that can be observed at LHC we shall use the
narrow width approximation,
\be
 N_{Z' \to \text{particles}}= N_{Z'} \ \text{BR} (Z' \to \text{particles})
\ee
where $N_{Z'}=\s_{Z'} \, \L \, t \ $ is the total number of $Z'$, $\text{BR} (Z' \to \text{particles})$ is the branching ratio,
$\L=10^{34} {\rm \,\,cm^{-2} s^{-1}}$ the luminosity and $t=$1 year. Finally $\s_{Z'}$ is the $Z'$ production cross section \cite{Langacker:2008yv}.
To estimate a rough upper bound for the anomalous BR, we assume that the LSP has a mass higher than 500 GeV (see Section~\ref{Neutralinos}),
so that we can ignore sparticles contribution in the computation of the branching ratios.
We integrate numerically the PDFs using a Mathematica package \cite{PDF}. In Fig.~\ref{NZp} we show the result for $N_{Z'}$ at $\sqrt s = 14$ TeV.
We can see that the number of the $Z'$ produced falls off exponentially with $M_{Z'}$, so we shall focus on the case $M_{Z'} \sim 1$ TeV.
We compute the two decay rates $\Gamma (Z'\to Z_0 \gamma)$ and $\Gamma (Z'\to Z_0 Z_0)$ for $M_{Z'} \sim 1$ TeV.
They depend on the remaining free parameters of the model, i.e. the charges $Q_Q$, $Q_L$ \cite{myPhD}.
We remember that $\QHu=\QHd=0$ and we choose $g_0=0.1$.
We show our results in Fig.~\ref{Contourplot 1 TeV} in the form of contour plots in the plane $Q_Q, Q_L$.
Our choices for $g_0$, $Q_Q$, $Q_L$ and $\MZp$ are in agreement with the current experimental bounds \cite{CDF}.
The darker shaded regions correspond to larger decay rates.
The white region corresponds to the value $10^{-6}$ GeV that can be considered as a rough lower limit for the detection of the corresponding process.
We see that the most favorite decay is $Z'\to Z_0 Z_0$ and we focus on it.
In Fig.~\ref{NZpZZ}, we estimate the number of decays for 1 year of integrated luminosity which turns out to be
$N_{Z'\to Z_0 Z_0} \sim 10$ for large values of the charges $\QL$ and $\QQ$.

\section{Neutralino Sector}
\label{Neutralinos}
Assuming the conservation of R-parity the LSP is a good weak interacting massive particle (WIMP) dark matter candidate.
As in the MSSM the LSP is given by a linear combination of fields in the neutralino sector.
The decoupling in the gauge bosons sector caused by (\ref{QHuchoice}) implies also a decoupling in the neutralino sector.
Written in the interaction eigenstate basis
$(\psi^{0})^T= (\psi_S, \ \l^{(0)},\ \l^{(1)} ,\ \l^{(2)}_3,\ \tilde h_d^0,\ \tilde h_u^0)$
the neutralino mass matrix is a six-by-six matrix given by
\be
 {\bf M}_{\tilde N}
  =   \(\begin{array}{cccccc}
       \frac{M_S}{2} & \frac{M_{V^{(0)}}}{\sqrt2} &   0   &   0   & 0  & 0 \\
               \dots &       M_0                  &   0   &   0   & 0  & 0   \\
               \dots &      \dots                 &  M_1  &   0   & -\frac{g_1 v_d}{2}& \frac{g_1 v_u}{2} \\
                  \dots &      \dots                 & \dots &  M_2  & \frac{g_2 v_d}{2} & -\frac{g_2 v_u}{2} \\
                  \dots &      \dots                 & \dots & \dots &      0     & -\m  \\
                  \dots &      \dots                 & \dots & \dots &    \dots     & 0
         \end{array}\) \label{massmatrix} ~~~~
   \ee
where $M_S,~M_0,~M_1,~M_2$ are the soft masses coming from the soft breaking terms (\ref{Lsoft}) while $M_{V^{(0)}}$ is given in eq.~(\ref{BosonMasses}).
Moreover we make the assumption that
\be
M_0\gg M_S,M_{V^{(0)}}
\label{axinodecoupling}
\ee
so that the axino is the LSP \cite{myPhD}.

As a rule of thumb~\cite{Jungman:1995df} a first order estimate of the relic density is given by
\begin{equation}
 \Omega_\chi h^2\simeq \frac{10^{-27}\,{\rm cm^3\, s^{-1}}}{\langle \s_{eff} v \rangle}
\label{oh2vssigmaeff}
\end{equation}
where $\langle \s_{eff} v \rangle$ thermal average of the effective cross section of the axino annihilations.
Since the axino is an XWIMP, $\langle \s_{eff} v \rangle$ cannot give a relic density in the WMAP preferred range
$0.0913\le\Omega h^2\le 0.1285$~\cite{wmapdata}.
Thus we are forced to consider a scenario in which the NLSP is close in mass to the LSP and
coannihilations between the axino and the NSLP are activated.
We suppose the NLSP to be a wino-like neutralino.
In this case a third particle comes to play: the wino-like chargino. This is due to the mass degeneracy between the two wino-like states.
In this case the effective cross section can be approximated as \cite{myPhD,Feldman:2006wd}
\be
 \la \s_\text{eff}^{(3)} v \ra \simeq \[ \frac{3 Q}{1+ 3 Q} \]^2 \la \s_\text{eff}^{(2)} v \ra
\ee
where $Q = (1 + \D_2)^{3/2} e^{-x_f \D_2}$ with $x_f\simeq 20$~\cite{Edsjo:1997bg}, \mbox{$\D_2=(M_2-M_S)/M_S$}
and $\la \s_\text{eff}^{(2)} v \ra$ is the thermal average of the effective cross section for a thermal bath composed by only
a wino-like neutralino and a wino-like chargino as particle species.
The rescaling factor between the three and two particle species relic density is given by the following relation
\be
 \( \Omega h^2\)^{(3)} \simeq \[ \frac{1+ 3 Q}{3 Q} \]^2 \( \Omega h^2\)^{(2)}\label{rescrelic2to3}
\ee
We performed a random sampling of MSSM models with wino-chargino coannihilations.
This situation is naturally realized in anomaly mediated supersymmetry breaking scenarios.
For each model we computed the relic density $( \Omega h^2)^{(2)}$ for the two coannihilating species with the DarkSUSY package~\cite{Gondolo:2004sc}.
We finally computed $( \Omega h^2)^{(3)}$ using~(\ref{rescrelic2to3}).
Models which satisfy the WMAP constraints are shown in Fig.~\ref{DMwino-chargino} for four reference values of $\D_2$.
The space of parameters with $\D_2\lesssim 5\%$ and an axino mass $M_S\gtrsim 700$ GeV is favored while as the mass gap increases lower
axino masses become favored, e.g. $100\; \text{GeV} \lesssim M_S<200$ GeV ($\D_2\simeq 20\%$).
The range $M_S\gtrsim 700$ GeV is compatible with the LSP mass range assumed in Section ~\ref{Gauge}.
\begin{figure*}
 \centering
 \includegraphics[scale=0.45]{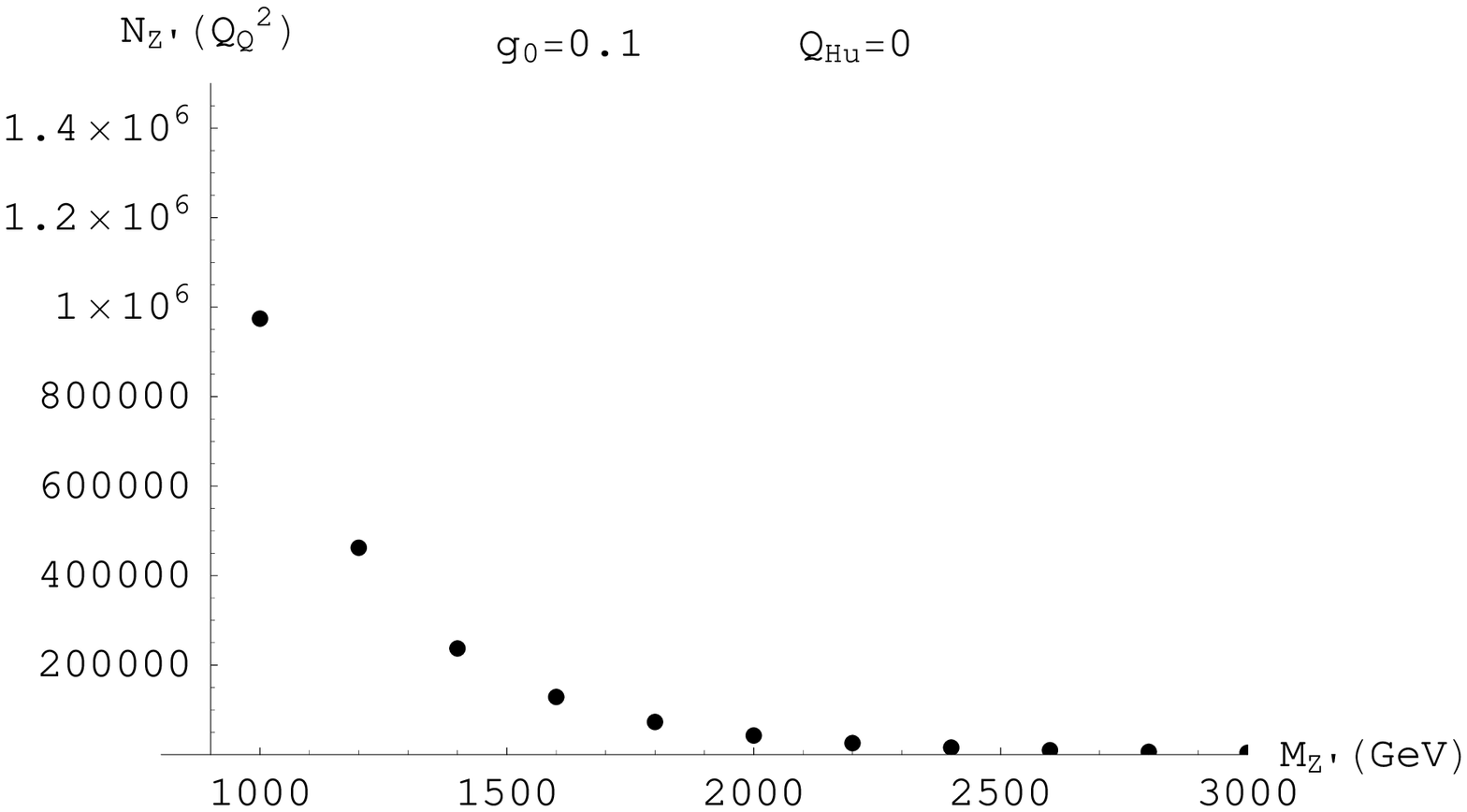}
 \caption{Number of $Z'$ produced at LHC in 1 year for $\L=10^{34} cm^{-2} s^{-1}$ and $\sqrt s = 14$ TeV,
          in units of $Q_Q^2$, in function of the mass of the $Z'$.} 
          \label{NZp}
\end{figure*}
\begin{figure*}
 \centering
 \includegraphics[scale=0.45]{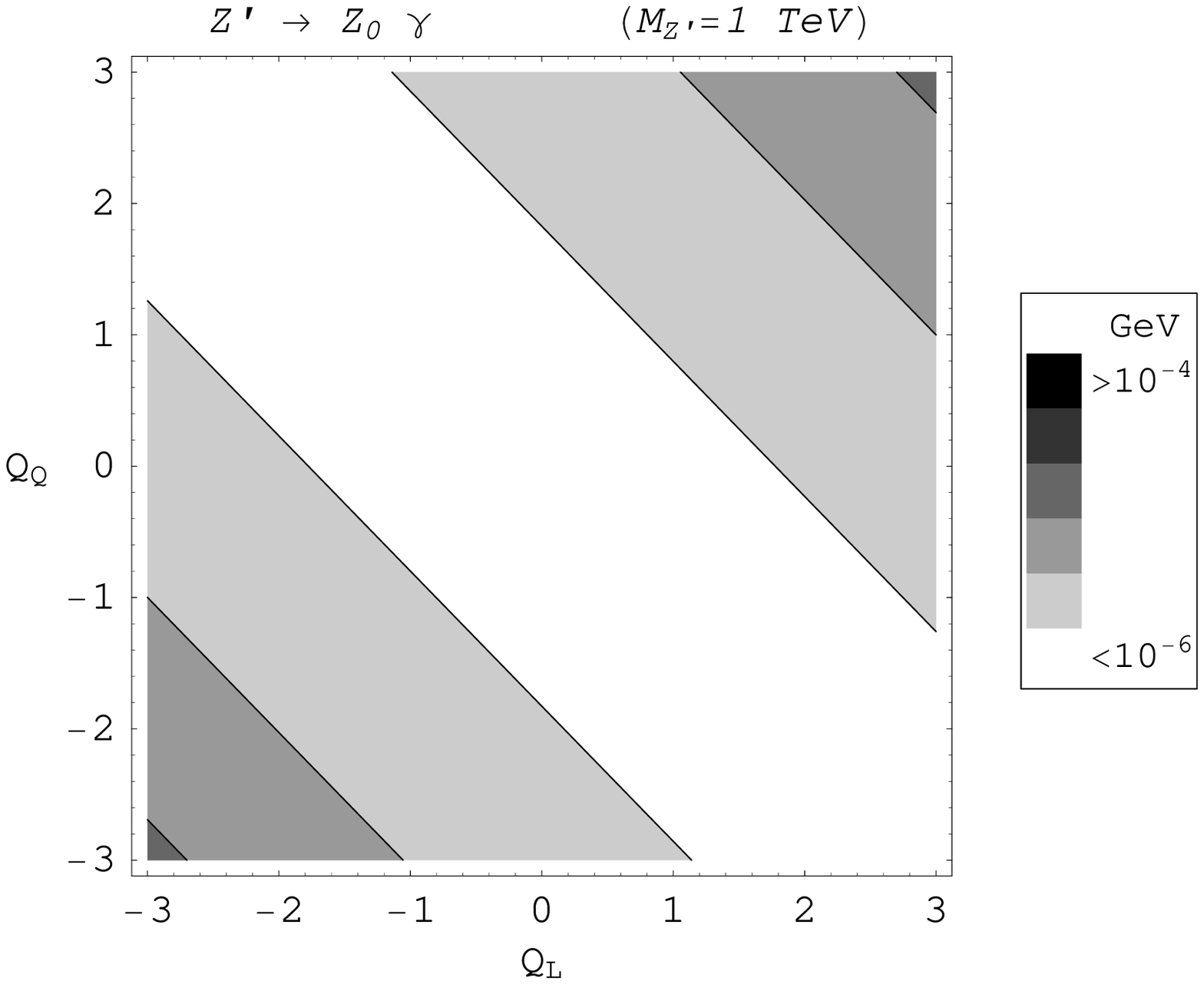}
 \includegraphics[scale=0.45]{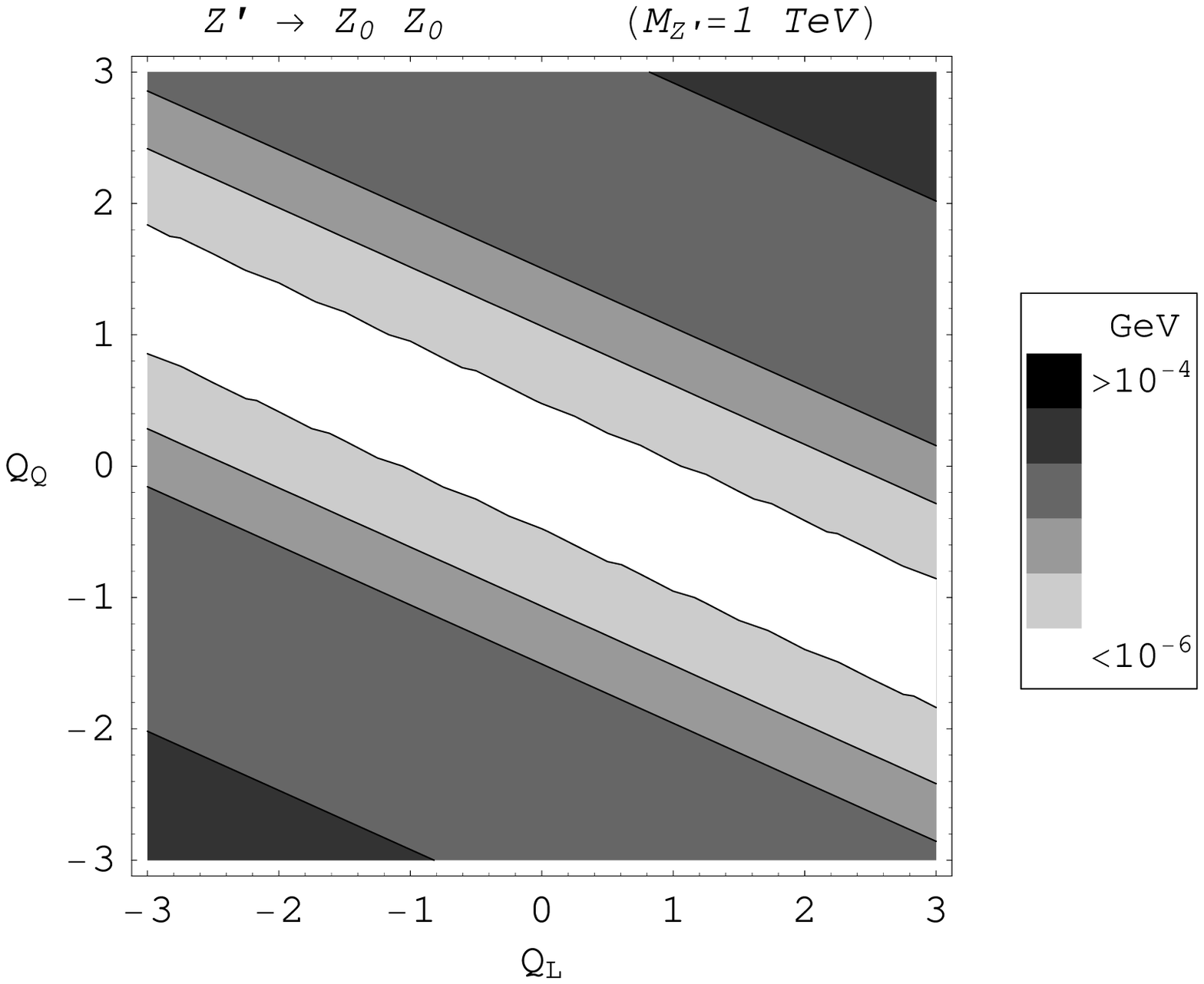}
 \caption{Decay rates for $Z'\to Z_0 \g$ (left) and $Z'\to Z_0 Z_0$ (right) for $\MZp=1$ TeV.}
  \label{Contourplot 1 TeV}
\end{figure*}
\begin{figure*}
 \centering
 \includegraphics[scale=0.45]{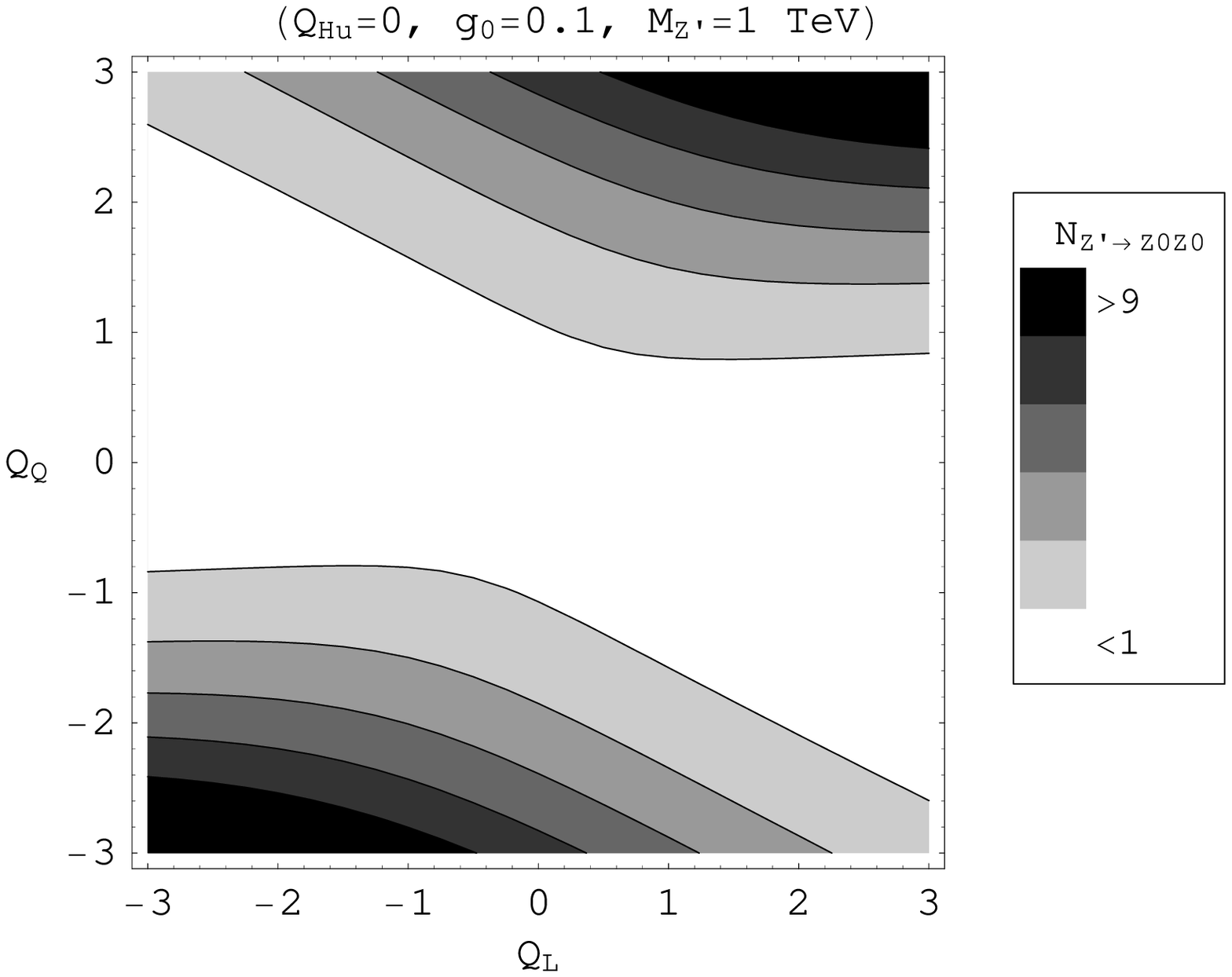}
 \caption{Number of $Z'\to Z_0 Z_0$ at LHC in 1 year for $\L=10^{34} cm^{-2} s^{-1}$, $\sqrt s = 14$ TeV and
          $\MZp=1$ TeV.}
           \label{NZpZZ}
\end{figure*}
\begin{figure*}
      \begin{center}
      \includegraphics[width=.45\textwidth]{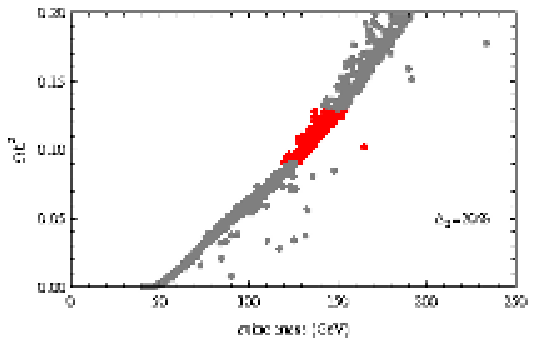}
      \includegraphics[width=.45\textwidth]{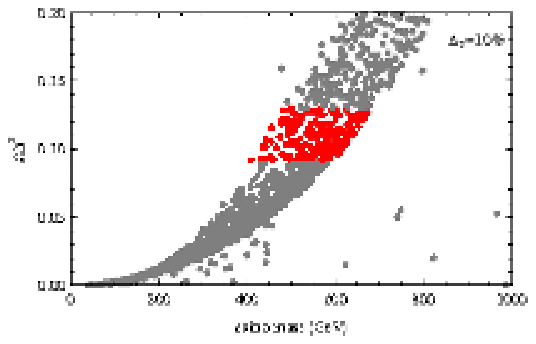}
      \includegraphics[width=.45\textwidth]{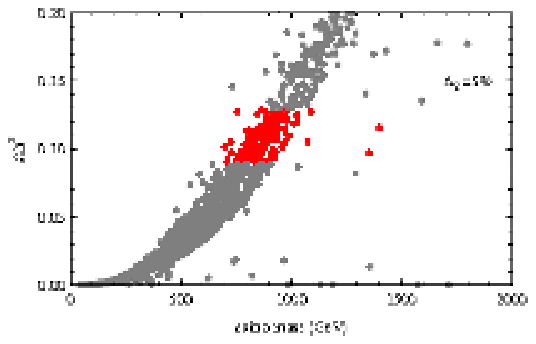}
      \includegraphics[width=.45\textwidth]{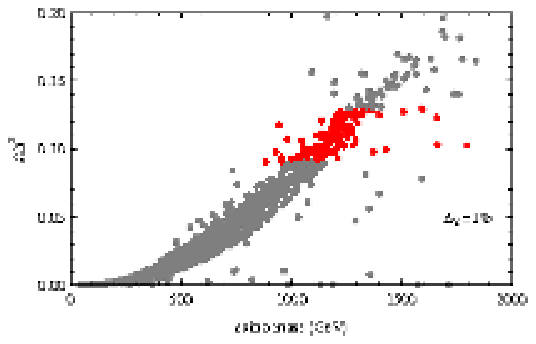}
     \caption{Axino relic density in the case in which the NSLP is a wino while the NNLSP is the lightest chargino.
              Red (darker) points denote models which satisfy WMAP data.
              Upper left panel: $\D_2=20\%$. Upper right panel:  $\D_2=10\%$.
              Lower left panel: $\D_2=5\%$. Lower right panel: $\D_2=1\%$.}
\label{DMwino-chargino}
\end{center}
\end{figure*}

\section{Conclusions}
We have briefly reviewed an extension of the MSSM by the addition of an anomalous abelian vector multiplet and showed
some original results concerning the phenomenology of an anomalous $Z'$.
We presented the first phenomenological results of our model for the choice $\QHu=\QHd=0$.
We estimated the number of anomalous decays for a $Z'$ with $M_{Z'} \sim 1$ TeV for 1 year of integrated luminosity in LHC.
The maximum number of $Z' \to Z_0 Z_0$ decays is of the order of 10.
Then we studied a possible dark matter candidate in our model.
In the decoupling limit $\QHu=\QHd=0$ and under the assumption $M_0\gg M_S,M_{V^{(0)}}$ the axino turns out to be the LSP.
Since the axino is an XWIMP, we find agreement with WMAP data considering coannihilations
with a wino-like neutralino NLSP and a wino-like chargino NNLSP.

\begin{acknowledgments}
A. R. would like to thank Prof.~Massimo Bianchi, Prof.~Francesco Fucito, Prof.~Gianfranco Pradisi, Prof.~Yassen
Stanev, Dr.~Pascal Anastasopoulos, Dr.~Andrea Lionetto, Andrea Mammarella and Daniel Ricci Pacifici for useful
discussions and collaborations.
\end{acknowledgments}


\newpage

\lhead[\fancyplain{}{}]{\itshape \small{Flavour mixing in an expanding universe }}
\rhead[\itshape \large{1$^\mathrm{st}$ RYRM Proceedings}]{\fancyplain{}{W. Tarantino}}



\newcommand{\we}{\omega(\eta)}
\newcommand{\intw}{\int{\we d\eta}}
\newcommand{\C}{\mathcal C}
\newcommand{\Ce}{\mathcal{C}(\eta)}
\newcommand{\Cpe}{\mathcal{C}'(\eta)}
\newcommand{\Cp}{\mathcal{C}'}
\newcommand{\rfv}{\mid 0 \rangle_f}
\newcommand{\lfv}{_f\langle 0 \mid}
\newcommand{\rmv}{\mid 0 \rangle}
\newcommand{\lmv}{\langle 0 \mid}
\newcommand{\bpsi}{\bar{\psi}}
\newcommand{\vp}{\vec{p}}
\newcommand{\vk}{\vec{k}}
\newcommand{\vq}{\vec{q}}
\newcommand{\vx}{\vec{x}}
\newcommand{\lag}{\mathcal L}
\newcommand{\gmn}{g^{\mu \nu}}
\newcommand{\st}{\sin \theta}
\newcommand{\sst}{\sin^2 \theta}
\newcommand{\ssst}{\sin^3 \theta}
\newcommand{\ct}{\cos \theta}
\newcommand{\cct}{\cos^2 \theta}
\newcommand{\ccct}{\cos^3 \theta}
\newcommand{\mfA}{\mathfrak{A}}
\newcommand{\mfB}{\mathfrak{B}}
\newcommand{\mfC}{\mathfrak{C}}
\newcommand{\mfD}{\mathfrak{D}}
\newcommand{\mfa}{\mathfrak{a}}
\newcommand{\mfb}{\mathfrak{b}}
\newcommand{\mfc}{\mathfrak{c}}
\newcommand{\mfd}{\mathfrak{d}}

\newcommand{\mfArs}{\mathfrak{A}_{rs}}
\newcommand{\mfBrs}{\mathfrak{B}_{rs}}
\newcommand{\mfCrs}{\mathfrak{C}_{rs}}
\newcommand{\mfDrs}{\mathfrak{D}_{rs}}
\newcommand{\mfars}{\mathfrak{a}_{rs}}
\newcommand{\mfbrs}{\mathfrak{b}_{rs}}
\newcommand{\mfcrs}{\mathfrak{c}_{rs}}
\newcommand{\mfdrs}{\mathfrak{d}_{rs}}

\newcommand{\Ku}{\mathfrak{A}_{rs}}
\newcommand{\Kd}{\mathfrak{B}_{rs}}
\newcommand{\Kt}{\mathfrak{C}_{rs}}
\newcommand{\Kq}{\mathfrak{D}_{rs}}
\newcommand{\ku}{\mathfrak{a}_{rs}}
\newcommand{\kd}{\mathfrak{b}_{rs}}
\newcommand{\kt}{\mathfrak{c}_{rs}}
\newcommand{\kq}{\mathfrak{d}_{rs}}

\newcommand{\uad}{u^{(a,d)}(\vp)}
\newcommand{\uadd}{u^{(a,d)}(\vp)^{\dagger}}
\newcommand{\uade}{u^{(a,d)}(\vp,\eta)}
\newcommand{\uadde}{u^{(a,d)}(\vp,\eta)^{\dagger}}

\newcommand{\eti}{\tilde{\eta}_{i}}

\newcommand{\VN}{V^2_N(p)}
\newcommand{\VM}{V^2_M(p)}

\begin{center}

\setcounter{section}{9}
\section{\large{Flavour mixing in an expanding universe}}
\label{tarantino}


\normalsize{Walter Tarantino$^*$}

\emph{\small{King's College London}}



\vspace{.5cm}


\begin{minipage}{.8\textwidth}
\small{Motivated by a microscopic model of string-inspired foam, in which foamy structures are provided by brany point-like defects (D-particles) in space-time, 
we discuss flavour mixing in curved space-time for spin-0 and spin-1/2 particles. This can be view as the low energy limit of flavour non-preserving interactions of stringy matter excitations with the defects,
and non-trivial space-time background induced by quantum fluctuations of the D-particles.
We show, at late epochs of the Universe, that both the fermionic and the bosonic vacuum condensate behaves as a fluid with negative pressure and positive energy; however the equation of state has $w_{\rm fermion} > -1/3$ and so the contribution of fermion-fluid flavour vacuum  alone could not yield accelerating Universes. On the other hand, for the boson fluid
the equation of state is, for late eras, close to $w_{\rm boson} \to -1$, and hence overall the D-foam universe appears accelerating at late eras.}

\vspace{.25cm}
$*$ e-mail: \href{mailto:walter.tarantino@kcl.ac.uk}{walter.tarantino@kcl.ac.uk}
\end{minipage}

\end{center}

\setcounter {section} {0}
\setcounter {subsection} {0}
\setcounter {figure} {0}
\setcounter{equation}{0}

\section{Introduction}

Flavour mixing is an interesting topic in quantum field theory, both for phenomenological (quarks, mesons, neutrinos present the mixing) and theoretical reasons:
flavour mixing requires massive neutrinos, whereas in the Standard Model they are treated as massless particles. Massive neutrino mixing can thus provide a sight
on the physics beyond the Standard Model. This work is based on the papers \cite{tarantino} and \cite{mavrosarkar}.

During recent years it has been suggested that the neutrino physics might play an important r\^ole in the understanding of the nowadays accelerated expansion of the universe
\cite{dark energy};
in particular, it has been motivated that a mathematically correct treatment of the flavour mixing in a free theory requires the introduction of a Fock space different from the one
usually considered in QFT (the irreducible representation of the Poincar\'e group, where the states have well defined mass).
According with \cite{vitiello}, introducing the mixing \textit{\`a la} Pontecorvo:
\be
\left\{\begin{array}{rcl}
\psi_e(x)&=&\psi_1(x) \ct+\psi_2(x)\st\\
\psi_\mu(x)&=&-\psi_1(x) \st+\psi_2(x)\ct
\end{array}
\right.
\ee
where $\psi_{\iota}(x)$ (with $\iota=e,\mu$) are the flavoured fields, and $\psi_{i}(x)$ (with $i=1,2$) are the fields
with well defined mass ($m_1$ and $m_2$, respectively),
the Fock space for the flavour states can be built using the ladder operators defined by
\be
\left\{\begin{array}{rcl}
\hat{a}_e^{(a,b)}(\vp,t)&=&\hat{G}_\theta^{\dagger}(t)\hat{a}_1^{(a,b)}(\vp)\hat{G}_\theta(t)\\
\hat{a}_\mu^{(a,b)}(\vp,t)&=&\hat{G}_\theta^\dagger(t)\hat{a}_2^{(a,b)}(\vp)\hat{G}_\theta(t)
\end{array}
\right.
\ee
where 
\be\label{G}
\hat{G}_\theta(t)=\exp\left[\theta\! \int\! d\vx \left( \hat{\psi}^\dagger_1(x)\hat{\psi}_2(x)-\hat{\psi}_2^\dagger \hat{\psi}_1(x)\right)\right]~.
\ee
The \textit{flavour vacuum} state is therefore defined as
\be\label{fv}
\rfv\equiv \hat{G}_\theta^\dagger(t)\rmv
\ee
with $\rmv$ the usual mass eigenstate vacuum.
It has been claimed~\cite{giunti} that the flavour-vacuum formalism, although mathematically consistent, nevertheless leads to no physically different predictions from the conventional formalism.
However, the authors of \cite{henning} have argued that probability conservation is only realised within the flavour state vacuum in quantum field theories with mixing; moreover the oscillation probability among flavours is modified, compared to the traditional formalism, by extra terms which, although small, nevertheless are in principle experimentally detectable. In this sense they postulated that in such cases the flavour vacuum is the physical vacuum.
It has then been claimed~\cite{dark energy} that the vacuum condensate due to fermion particle mixing, evaluated in the (physical) flavour vacuum, seems to behave as a source of Dark Energy, in the sense of yielding a non-trivial flavour-vacuum energy density, leading to an acceleration of the expansion of the Universe, at late eras.

The so defined \textit{flavour vacuum} seems to merge naturally as the \textit{physical} vacuum from a brane model that has been object of studies during the last years \cite{Dfoam}.
In this model a D3-brane with strings attached on it moves into a ``foam'' of D-particles, topological defects that puncture the bulk in which the D3-brane is embedded.
The interaction of the foam with the brane and the strings is highly non-trivial: from one side the foam can change the flavour of the strings that
interact with it; on the other hand the foam induces a deformation of the metric of the brane itself.
In the usual interpretation of the low energy limit,
our universe is identified with the D3-brane and the attaching point of the strings on it represents a particle.
Therefore, the effect of the interaction with the foam can be represented with the flavour mixing mechanism and an induced curvature of the spacetime.

\section{Description of the model}
In our work we have considered a mixing of two fields (first two spin-0 and then two spin-1/2 fields) on a curved background as the low energy limit of
the D-particle model.
The flavour vacuum expectation value of the stress-energy tensor has been evaluated in both bosonic and fermionic case.
In this effective model, few (but crucial) differences with respect to the analysis already present in the literature occur.
First of all, all other calculations have been performed in the context of the usual QFT in a flat spacetime, considering negligible all the contributions
coming from the non trivial particle production due to the expansion of the universe, induced by the \textit{flavour vacuum}.
In our model, the curvature of the spacetime plays a fundamental r\^ole, being deeply connected with the flavour mixing, at the string scale.
Therefore in the QFT model that we consider as the low energy limit of the brane model, the curvature is taken into account
as a classical background that appears in the Lagrangian, in the spirit of QFT in Curved Spacetime, largely studied in literature \cite{birrell}.
In this framework the curvature of the background
is considered to be classic and to depend only from points of the spacetime (so no back reaction of the fields \textit{on} the curvature and no quantization of the curvature itself).
An important feature of this formulation is the creation of particles due to the external gravitational field. The importance of this effect in our
analysis is related with the fact that our theory normalized in such a way to cut off contributions for the vacuum condensate that do not come from the particle creation mechanism.
The basic formalism of quantum field theory is generalized to curved space-times
in a straightforward way \cite{ford}: 
consider the covariant generalization of the action,
choose a foliation of the spacetime into spacelike hypersurfaces, let $\Sigma$ be a particular hypersurface with unit normal vector $n^{\mu}$ 
labeled by a constant value of the time coordinate $t$, choose the canonical (anti)commutation relations
in such a way that the integration over $\Sigma$ of the (anti)commutator of the field and its conjugate momentum gives 1.
As it this well known, when the Poincar\'e group is no longer a symmetry of the theory, the definition of particle becomes ambiguous.
Creation and annihilation operators are defined choosing a specific decomposition in modes of the generic solution of the equation of motion,
but there is no unique choice for the set of modes: different decomposition will lead to different Fock spaces.
This is true also in Minkowksi space-time, but the Poincar\'e symmetry leads to the unique choice of a set of modes that allows to
define a vacuum state that is invariant under the transformations of the group and therefore is the agreed vacuum for all inertial
observers in Minkowksi space-time.
However, in many problems of interest the curved space-time can be treated as
asymptotically Minkowskian in the remote past and future (the ``in'' and the ``out'' region).
In these regions there is clearly a ``natural'' choice for the vacua, corresponding to the physical states of no particle detected by all inertial observers.
Working in the Heisenberg picture, if we choose the state of the quantum field in the in region to be the
vacuum state, then it will remain in that state during its subsequent evolution.
However, at later times, outside the in region, freely falling particle
detectors may still register particles in this ``vacuum'' state. In particular, in the out region,
then the in vacuum may not coincide with the
out vacuum. Therefore we can say that a ``particle production'' by the external gravitational field occurred.
This phenomenon was first carefully investigated by Parker \cite{parker}, \cite{parkerboson}, in the case of an FRW-universe.
Working in the Heisenberg picture, in his pioneer works he introduced expressions valid at all times for the fields in terms of the ladder operators defined in the in region.
For instance, in the fermionic case we have
\begin{widetext}
\be\label{psi}
\hat{\psi}(\eta,\vec{x})=\left(\frac{1}{L \sqrt{\\C(\eta)}}\right)^{\frac{3}{2}} \!\!\!\!\!\!\!\!
\sum_{\mbox{\tiny{$\begin{array}{c} \vp  \\ a,b,c=\pm 1 \end{array}$}}}  \!\!\!\!\!\!\!\!
\hat{A}^{(a,b)}(\vp)D^{a}_{c}(p,\eta) v^{(c,abc)}(ac\vp,\eta) e^{ia\vp\cdot\vx-ic\!\int\!\!\w(\eta) d\eta}.
\ee
\end{widetext}
where $L$ is the parameter that enters the periodic boundary condition $\psi(\eta,\vx+\vec{n}L)=\psi(\eta,\vx)$ ($\vec{n}$ being a vector with integer Cartesian components),
 $\w(\eta)\equiv\sqrt{p^2+m^2 \C(\eta)}$, $\C(\eta)$ being the conformal scale factor,
$v^{(a,b)}(\vp,\eta)$ is a spinor defined by
\be
\left\{
\begin{array}{l}
v^{(a,b)}(\vp,\eta)\equiv v^{(a,b)}(\vp/\sqrt{\C(\eta)})\\
(-i a\sqrt{p^2+m^2 } \gamma^0+i a \vec{\gamma}\cdot\vp+m)v^{(a,b)}(\vp)=0\\
v^{(a,b)\dagger}(\vp)v^{(a',b')}(\vp)=\delta_{a,a'}\delta_{b,b'}
\end{array}
\right.
\ee
$\hat{A}^{(a,b)}$ are operators such that $\{\hat{A}^{(a,b)}(\vp), \hat{A}^{(a',b')\dagger}(\vq)\}=\delta_{a,a'}\delta_{b,b'}\delta_{\vp,\vq}$
with $a,b=\pm1$ and
\begin{widetext}
\bea
\label{Deffectivemass}      D_{(a')}^{(a)}(p,\eta)&=&\delta_{a'}^{a}+a'\!\!\int^{\eta}_{-\infty}\!\!\!d\eta'      \Big(\sqrt{\C(\eta')}\frac{p}{2 m}\frac{\w'(\eta')}{\w(\eta')}\, e^{2 i a' \!\!\!\int\!\!\!\,\w(\eta')d\eta'}D_{(\!-a'\!)}^{(a)}(p,\eta')\Big)
\eea
\end{widetext}
again with $a,a'=\pm-1,1$.

This formulation of the theory allows us to build a \textit{flavour vacuum} at early times by combining equation (\ref{G}) and equation (\ref{psi}),
by saying:
\be
\rfv \equiv \hat{G}^{\dagger}_{\theta}(-\infty)\rmv.
\ee

The so defined model is divergent (just like its Minkowskian counterpart), therefore it needs to be renormalized. The second important difference with the previous works,
is that the normal ordering and the cutoff, used in the renormalization procedure, are now dictated by the microscopic underlying brane model;
the subtraction procedure, in particular, has to be such that in the limit of the absence of D-particles and their fluctuations the mixing phenomenon and its effects should
disappear. 
Moreover in our D-particle foam model one needs to distinguish two effects, as far as the structure of the underlying space-time is concerned.
The first effect concerns a \emph{background} space-time, over which propagation of low-energy matter excitations (fermions or bosons) takes place.
The background space-time in our case of D-particle foam has been argued to be obtained from quantum fluctuations of individual D-particles,
which in the case of first quantised string framework are due to a summation over world-sheet topologies~\cite{szabo},
upon (statistically) averaging over populations of D-particles on the D3-brane world. This leads to a \emph{background} space-time metric of the form $g_{\mu\nu}=\C(\eta)\eta_{\mu\nu}$.
The individual MSW interactions of the flavoured matter excitations with the D-foam background~\cite{barenboim},
produce extra back-reaction local fluctuations on the space-time structure. They do not cause metric distortions,
but affect the particle mode's energy-momentum dispersion relations.
This parallels and is in a similar spirit to the standard result when one considers particle production at the end of inflation~\cite{chung}.
Hence in the effective QFT we have to take into account interactions of the neutrinos and the D-particle medium.
In this respect, the scale factor $\C(\eta)$ appearing in the above formulae for the fermions, should be considered as representing the background space-time.
The energies $\omega$ on the other hand will contain an ``\emph{effective}'' scale factor
\begin{equation}
\C_{\rm eff} (\eta) = \C(\eta) + \Delta \C(\eta) \equiv \C(\eta) \left(1 + \frac{\Delta \C(\eta)}{\C(\eta)}\right).
\label{effscale}
\end{equation}
In the dispersion relation
\bea\label{oeff}
\omega_{\rm eff} &=& \sqrt{p^2 + m^2\C_{\rm eff}(\eta)}\nn\\
& \simeq & \sqrt{p^2 + m^2\C(\eta)} + \frac{m^2\,\Delta \C(\eta)}{2\sqrt{p^2 + m^2\C(\eta)}}~,\\
\eea
to leading order in the approximation  $|\Delta \C| \ll 1 $. The $\Delta \C$ is the MSW contribution.
Therefore our subtraction (normal ordering) procedure
is  \emph{defined} in such a way that in the \emph{absence} of any MSW interaction of the fermion matter with D-particles, the stress tensor should \emph{vanish}.

Further divergences are removed by a cutoff that takes into account the fact that the capture process is likely to be most efficient for stringy matter
with small momenta\footnote{The reader's attention is called at this point to the fact that in string theory there are no actual ultraviolet momentum divergences.
These are artifacts of the low-energy local effective field theory which is defined up to energies of the order of the string scale, $M_s$,
or better the Planck scale (defined as the ratio of $M_s/g_s$, with $g_s$ the string coupling, assumed weak $g_s < 1$).
Thus, by definition any momentum integral will be cut-off at that scale automatically.}.
One can get an estimation of it by considering the scale inherent in the momentum dependence of the single particle momentum distributions in the flavour vacuum.

On recalling that for a relativistic fluid $T_{00}/\C(\eta)$ represents the \textit{energy density}, and $T_{ii}/\C(\eta)$ the \textit{pressure},
one can easily derive that fermionic vacuum condensate behaves as a fluid with positive energy density and negative pressure, with an equation of state that satisfies $-1/3<w<0$,
with varying the cutoff; on the other hand the bosonic condensate behaves as a fluid with positive energy density and negative pressure, equal in modulus to the energy density
independently from the cutoff, leading to an equation of state $w=-1$.

\section{Conclusions}
Therefore we can see that in our model only the bosonic fluid provides a mechanism for an accelerating universe, whereas the fermionic condensate, albeit source
of dark energy, does not have pressure low enough:
to ensure that the flavour vacuum contributions to the dark energy leads to \emph{accelerating} Universes at late epochs, as the current phenomenology indicates, one should have bosons simultaneously present with fermions.
In the context of supersymmetric low-energy field theories, such as those derived in the low-energy limit of superstrings, the relevant bosons may be the \emph{sneutrinos}, 
the supersymmetric partners of neutrinos, which have large masses due to target-space supersymmetry breaking.
However, the construction of a flavour vacuum in a supersymmetric theory, that typically is by contraction a interacting theory, is not a trivial task
and we hope to clarify this point in a forthcoming publication.

\begin{acknowledgments} 
I would like to thank N.E. Mavromatos and S. Sarkar for their contribution to this work. My work, as well as my participation to RYRM was supported by a King's College London (UK)
graduate scholarship.
\end{acknowledgments}



%

\end{document}